\newcommand{\hi}{H{\sc i}~}
\newcommand{\HI}{H{\sc i}}

\documentclass[preprint2]{pp7}

\usepackage{xcolor}

\input{pp7.h}

\usepackage{hyperref}
\usepackage[normalem]{ulem}

\begin{document}

\title{\textbf{\LARGE Initial Conditions for Star Formation:\\
A Physical Description of the Filamentary ISM }}

\author {\textbf{\large Alvaro Hacar}}
\affil{\small\it Department of Astrophysics, University of Vienna, T{\"u}rkenschanzstrasse 17, 1180, Vienna (Austria)}
\author {\textbf{\large Susan E. Clark}}
\affil{\small\it Department of Physics, Stanford University, Stanford, California 94305, USA \\
Kavli Institute for Particle Astrophysics \& Cosmology, P. O. Box 2450, Stanford University, Stanford, CA 94305, USA}
\author {\textbf{\large Fabian Heitsch}}
\affil{\small\it Department of Physics \& Astronomy, University of North Carolina - Chapel Hill, Chapel Hill, North Carolina 27599, USA}
\author {\textbf{\large Jouni Kainulainen}}
\affil{\small\it Chalmers University of Technology, Department of Space, Earth and Environment, SE-412 93, Gothenburg, Sweden }
\author {\textbf{\large Georgia V. Panopoulou}}
\affil{\small\it California Institute of Technology, MC350-17, 1200 East California Boulevard, Pasadena, CA 91125, USA}
\author {\textbf{\large Daniel Seifried}}
\affil{\small\it University of Cologne, I. Physical Institute, Z\"ulpicher Str. 77, 50937 Cologne, Germany}
\author {\textbf{\large Rowan Smith}}
\affil{\small\it Jodrell Bank Centre for Astrophysics, Department of Physics and Astronomy, University of Manchester, Oxford Road, Manchester M13 9PL, UK}

\begin{abstract}
\baselineskip = 11pt
\leftskip = 1.5cm 
\rightskip = 1.5cm
\parindent=1pc
{\small The interstellar medium contains filamentary structure over a wide range of scales. 
Understanding the role of this structure, both as a conduit of gas across the scales and a diagnostic tool of local physics, is a major focus of star formation studies. 
We review recent progress in studying 
filamentary structure in the ISM, interpreting its properties in terms of physical processes, and exploring formation and evolution scenarios. We include structures from galactic-scale filaments to tenth-of-a-parsec scale filaments, comprising both molecular and atomic structures, from both observational and theoretical perspectives. 
In addition to the literature overview, we assemble a large amount of catalogue data from different surveys and provide the most comprehensive census of filamentary structures to date. Our census consists of 22\,803 filamentary structures, facilitating a holistic perspective and new insights.  
We use our census to conduct a meta-analysis, leading to a description of filament properties over four orders of magnitudes in length and eight in mass. Our analysis emphasises the hierarchical and dynamical nature of filamentary structures. Filaments do not live in isolation, nor they generally resemble static structures close to equilibrium.  
We propose that accretion during filament formation and evolution sets some of the key scaling properties of filaments. This highlights the role of accretion during filament formation and evolution and also in setting the initial conditions for star formation. 
Overall, the study of filamentary structures during the past decade has been observationally driven. 
While great progress has been made on measuring the basic properties of filaments, our understanding of their formation and evolution is clearly lacking. In this context, we identify a number of directions and questions we consider most pressing for the field. 
\\~\\~\\~}
 %leave this in to get the correct vertical space after the abstract
\end{abstract}  

%\maketitle

%%% Section rules
%%%%%%%%%%%\section{\textbf{LEVEL ONE HEAD (all caps, bold)}}
%%%%%%%%%%%\subsection{\textbf{Level Two Head (upper and lower case, flush left, bold)}}
%%%%%%%%%%%\subsubsection{\textbf{Level three head (italic)}}

\section{THE FILAMENTARY ISM: \\ A CENTURY OF DISCOVERIES}

% Barnard
{\it ``Among the most surprising things in connection with these nebula-filled holes are the vacant lanes that so frequently run from them for great distances. These lanes undoubtedly have had something to do with the formation of the holes and with the nebula in them.''} \citep{Barnard1907}. With these words, E.~E. Barnard first reported the direct connection between filaments (dark lanes), dense cores (holes), and stars (nebulae). 
This description precedes observations of molecular line emission \citep{Wilson1970}, and hints at a connection between the geometry of interstellar matter and the star formation process. More than a century and many groundbreaking observations later, we are still trying to unravel the multiscale physics of star formation, and the meaning of a ubiquitous ISM geometry: filaments.

A series of theoretical and observational works investigated the basic physical properties of filaments in molecular clouds in the following decades. The geometrical simplicity of idealized filaments allowed semi-analytic derivations of hydrostatic equilibrium solutions \citep{Stodolkiewicz1963,Ostriker1964} and quasi-static gravitational fragmentation \citep{Larson1985}. %and 
These were later extended to include the influence of magnetic fields \citep{Nagasawa1987,Hanawa1993} and external pressure \citep{Fiege2000}. Extinction and infrared (IR) observations illustrated the filamentary nature of clouds, showing signatures of regular Jeans-like fragmentation \citep{Schneider1979} and promoting the formation of cores and stars at high efficiencies \citep{Hartmann2002}. Polarization measurements quantified the relative orientation between magnetic fields and filaments \citep{Hall55,Vrba1976,Goodman1990}. In parallel, large-scale molecular maps first explored the internal kinematics of filaments \citep{Loren1989b} typically exhibiting multiple velocity components \citep{Duvert1986}. 
Limited in resolution and sensitivity, most of these early works targeted ``prototype'' filaments in the solar neighbourhood, such as the B213-L1495 filament in Taurus \citep{Mizuno1995}, the Ophiuchus Streamers \citep{Loren1989a}, or the Integral Shape Filament (ISF) in Orion \citep{Bally1987}. Later, these studies were extended to other filaments in our Galaxy, with particular attention paid to filaments with extraordinary mass \citep{Schneider2010} or length \citep{Jackson2010}.

% Herschel and PPVI
The detailed study of filaments was revolutionized by the \textit{Herschel Space Telescope} wide-field far-infrared (FIR) continuum maps \citep{Pilbratt2010}. The unprecedented dynamic range of \textit{Herschel} highlighted the high degree of filamentary organization of the gas in molecular clouds \citep{Andre2010,Molinari2010}. 
Its enhanced sensitivity provided the first homogeneous measurements of filament masses, radial profiles, and characteristic radii across entire clouds \citep{Arzoumanian2011,Arzoumanian2019,Palmeirim2013,Konyves2015}. 
The filaments exhibited a characteristic density profile with a nearly constant diameter of 0.1 pc \citep{Arzoumanian2011}.
The strong influence of these initial \textit{Herschel} results merited a chapter in the previous Protostars \& Planets VI \citep[PPVI,][]{Andre2014}, becoming one of the most cited papers of this series. 

% after PPVI
After PPVI, a new generation of studies is transforming our description of filaments. Continuum and line observations have now identified filaments over an enormous range of scales and environments, from sub-parsec structures within clouds \citep{Hacar2013} up to kpc-sized objects associated with spiral arms \citep{Zucker2015}, and from the densest high-mass star-forming regions \citep{Trevino-Morales2019} to the atomic ISM \citep{Clark2014}. High-resolution ALMA observations have revealed complex networks of filaments related to massive star formation \citep{Peretto2013,Hacar2018} and Infrared-Dark Clouds \citep[IRDCs,][]{Henshaw2014,Barnes2021}.
Novel theoretical works explored different formation mechanisms for filaments at different scales \citep{Hennebelle2013,Inoue2013,Inoue2018,Chen2016,Duarte2017,Abe2021}.
In contrast to a classical static description, modern simulations show how filaments dynamically evolve and interact with their environments over time \citep[e.g.][]{Smith2014b}. Zoom-in simulations illustrate filaments with a complex substructure that greatly depart from idealized cylinders \citep[e.g.][]{LiKlein2019}.
The widespread detection of filaments in many types of simulations and observations suggests that while the physical processes that underlie filament formation must be anisotropic, they are not necessarily unique. 

% This paper
This chapter reviews the state-of-the-art observational and theoretical research on the filamentary nature of the ISM by the time of the Protostars \& Planets VII conference in 2022.
By synthesizing the substantial efforts of the star formation community over the last 10~years, we aim to provide an updated description of filaments at different scales and environments. 
This chapter introduces a novel meta-analysis of multiple filament surveys presented in the literature.
Our goal is to articulate what is currently known, and to illuminate the path forward for future studies of ISM filaments. 
We restrict our discussion to filamentary structures with linear scales between $\sim$~500 pc and $\sim$~0.01 pc, 
a range of scales that excludes larger-scale structures like Galactic spiral arms, and structures below the scale of protostellar core collapse.
Our aim is to describe the initial conditions for the formation of individual stars and clusters in the Milky Way, setting the scene for further discussions on the origin and evolution of stars and disks 
(see PPVII reviews by Pineda et al and Pattle et al). 

%%%%%%%%%%%%%%%%%%%%%%%%%%%%%%%%%%%%%%%%%%%%%%%%%%%%%%%%%%%
\section{DEFINITIONS AND HYDROSTATIC EQUILIBRIUM}\label{sec:basicprop}

% Basic observables
 The total mass, $M$, and length, $L$, are two properties often derived from observations of filaments. 
 Both $M$ and $L$ values are typically obtained from the analysis of maps of the column density, $N$, derived either from continuum (given a gas-to-dust ratio) or molecular line (assuming a molecular abundance) observations. Mass and length define a filament's 
 line mass $m$ (mass per unit length), typically in units of M$_\odot$~pc$^{-1}$:
 \begin{equation}
 m = \frac{M}{L}.
 \end{equation}

% Dynamics:
The gas dynamics inside filaments are inferred from the direct measurement of the line full-width-half-maximun ($\Delta v$) and line centroids ($v_\mathrm{LSR}$) in molecular line spectra. The total gas velocity dispersion ($\sigma_\mathrm{tot}$) along the line-of-sight (LOS) can be derived from the observed velocity dispersion of a given molecular tracer $i$ ($\sigma_{\mathrm{obs,} i}=\Delta v_i/\sqrt{8 \, \mathrm{ln}\, 2}$) via the non-thermal gas component
\begin{equation}
    \sigma_\mathrm{nt}^2=\sigma_{\mathrm{obs,} i}^2-\sigma_{\mathrm{th,} i}^2\, ,
\end{equation}
with $\sigma_{\mathrm{th,} i}=\sqrt{k_\mathrm{B}T/\mu_i m_\mathrm{p}}$, where  $\mu_i$ is the molecular weight of the observed species (e.g. \mbox{$\mu$ ($^{13}$CO) = 29)}, $m_\mathrm{p}$ the proton mass, $k_\mathrm{B}$ the Boltzmann constant and $T$ the (independently-determined) gas kinetic temperature. 
Using the thermal velocity dispersion of the gas
\begin{equation}
    c_\textrm{s} = \sigma_\mathrm{th, gas} = \sqrt{\frac{k_\mathrm{B}T}{\mu m_\mathrm{p}}} \, ,
\end{equation}
with \mbox{$\mu$ = 2.33} or 1.27 for molecular or atomic gas, respectively, and combining it with $\sigma_\mathrm{nt}$
results in the total gas velocity dispersion
\begin{equation}
    \sigma_\mathrm{tot}^2=\sigma_\mathrm{nt}^2+c_\mathrm{s}^2 \, .
    \label{eq:sigmatot_definition}
\end{equation}
Additional parameters describing the global gas motions, such as the velocity gradient at scale $L$ ($\nabla v_\mathrm{LSR}=\Delta v_\mathrm{LSR}/\Delta L$) and centroid dispersions ($\sigma (v_\mathrm{LSR})$), can be obtained from the statistics of the line centroids across filaments. The comparison between $\sigma_\mathrm{nt}$ (or $\sigma v_\mathrm{LSR}$) and $c_\mathrm{s}$ indicates whether the non-thermal motions inside filaments are subsonic ($\sigma_\mathrm{nt}< c_\mathrm{s}$), transonic ($c_\mathrm{s}\le\sigma_\mathrm{nt}\lesssim 2 c_\mathrm{s}$), or supersonic ($\sigma_\mathrm{nt}>2 c_\mathrm{s}$).

% Stability
The critical line mass of a filament is derived from a hydrostatic, isothermal cylinder model \citep{Stodolkiewicz1963,Ostriker1964}:
\begin{equation}
    m_\mathrm{crit}(T)=\frac{2 c_\mathrm{s}^2}{G} \sim 16.6 \left(\frac{T}{10\, \mathrm{K}}\right)\, \mathrm{M}_\odot \, \mathrm{pc}^{-1} \, .
    \label{eq:mcrit_therm}
\end{equation}
Filaments can be categorized in terms of gravitational stability, by considering the ratio of the line mass to a critical line mass \citep{Ostriker1964,Fischera2012a},  
\begin{equation}
  f\equiv\frac{m}{m_\mathrm{crit}}\, .\label{eqn:deff}
\end{equation}
Filaments with $f>1$ (supercritical) become radially unstable and must collapse under their own gravity, while only those with $f<1$ (subcritical) can remain in hydrostatic equilibrium. 
This critical line mass plays the same role as the isothermal Jeans mass in early studies of molecular clouds  \citep[e.g.][]{Klessen2000}: it relies on a series of assumptions (hydrostatic equilibrium, isolation, isothermality) that may or may not be applicable to observed filaments. This is readily apparent from the fact that many filaments have line masses well in excess of the critical value ($f\gg1$).  
To include the contribution from non-thermal motions to the filament's stability, the term $c_\textrm{s}$ can be replaced by the total velocity dispersion to obtain the virial line mass
\begin{equation}
    m_\mathrm{vir}(\sigma_\mathrm{tot})=\frac{2 \sigma_\mathrm{tot}^2}{G}\sim 465\left(\frac{\sigma_\mathrm{tot}}{1\,{\rm km \, s}^{-1}}\right)^2 \, \mathrm{M}_\odot \, \mathrm{pc}^{-1} \, .
\label{eq:Mcrit_tot}
\end{equation}
This effectively increases the line mass compared to Eq.~\ref{eq:mcrit_therm}.

Furthermore, the stabilising contribution of the magnetic field can -- approximately -- be accounted for by adding the square of the Alfv\'en speed
%\begin{equation}
$v_\textrm{A} = \frac{B}{\sqrt{4 \pi \rho}} \,$,
%\label{eq:alfven}
%\end{equation}
i.e. the magnetic pressure supporting the filament against radial collapse as exerted by a field oriented parallel to the filament to Eq.~\ref{eq:sigmatot_definition}.
If we instead assume that the magnetic field is oriented perpendicular to the filament, then in the limit of very strong field
%\begin{equation}
    $m_\textrm{vir} \propto \phi_\textrm{cl}$
%\label{eq:mvir_bfield}    
%\end{equation} 
\citep{Tomisaka2014,Kashawagi2021}, where $\phi_\textrm{cl}$ is the magnetic flux through the filament per unit length. This results in less support against gravity compared to a parallel field configuration \citep[see also][]{Seifried2015}.

The column density profiles of filaments are often parameterized by Gaussian functions (see \S\ref{sec:radialprof}) or by Plummer-like curves \citep{Nutter2008,Arzoumanian2011},
\begin{equation}
  N(x)=A_p\frac{n_0 R_\mathrm{flat}}{(1+(x/R_\mathrm{flat})^2)^{(p-1)/2}} \, ,
\end{equation}
 where $x$ is the projected distance from the filament spine, and $A_p$ is a proportionality constant. The corresponding three-dimensional density structure
\begin{equation}\label{eq:density_profile}
  n(r)=\frac{n_0}{(1+(r/R_\mathrm{flat})^2)^{p/2}}
\end{equation}
can be inferred as function of the central volume density $n_0$ and radius $r$. The filament's (inner) flat radius is related to its central column density \mbox{$N(x = 0) = N_0$} as $R_\mathrm{flat} = N_0/(A_p n_0)$.
The 
%column density
full-width-half-maximum FWHM of the column density distribution is 
\begin{equation}
  FWHM=2R_\mathrm{flat}\left(2^{2/(p-1)}-1\right)^{1/2} \, .
\end{equation}
This definition assumes a filament in isolation, i.e. the half-maximum refers to a background column density of \mbox{$N = 0$}. 

The isothermal Ostriker filament corresponds to $p=4$, $A_p=\pi/2$, and
\begin{equation}
R^2_\mathrm{flat} = \frac{2 c_\mathrm{s}^2}{\pi G\rho_0} \, ,
\label{eq:rflat}
\end{equation}
with \mbox{$\rho_0 = \mu m_\textrm{p} n_0$}, which leads to \mbox{$N_0 = \frac{\pi}{2} n_0 R_\mathrm{flat}$} and \mbox{$FWHM \approx 1.53 R_\mathrm{flat}$}. This solution can be generalized to include the effects of turbulence, magnetic fields, and rotation \citep{Nakamura1993,Hanawa1993}.

According to a linear stability analysis, hydrostatic filaments can fragment under the influence of gravity if perturbations are larger than their critical value 
\begin{equation}
    \lambda_\mathrm{crit}=3.93\cdot R_\mathrm{flat} \, ,
\label{eq:lambdacrit}
\end{equation}
showing a maximum growth rate for \mbox{$\lambda_\mathrm{max}\sim 2\times \lambda_\mathrm{crit}$} \citep{Stodolkiewicz1963,Nagasawa1987,Inutsuka1992}. 

%%%%%%%%%%%%%%%%%%%%%%%%%%%%%%%%%%%%%%%%%%%%%%%%%%%%%%%%%%%

\section{A NEW ERA OF GALACTIC SURVEYS}\label{sec:surveys}

%\subsection{Recent observations: Filaments in the literature}

\begin{figure*}[t]
    \centering
    \includegraphics[width=0.97\textwidth]{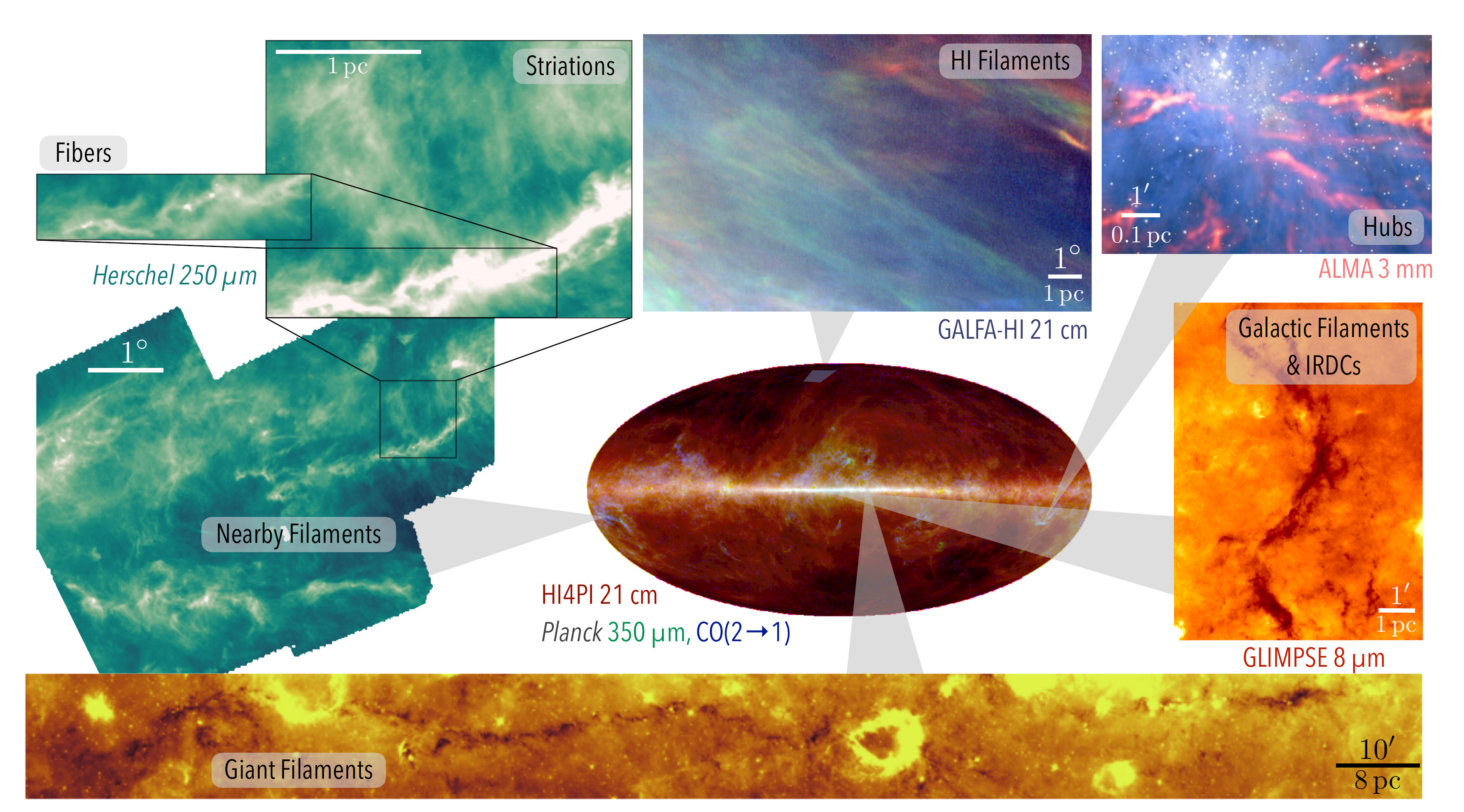}
    \caption{\small Observations exemplifying each of the filament families identified in section \ref{sec:surveys}. Clockwise from top: (1) \hi filaments (\S\ref{sec:hi}) shown in a three-color GALFA-\hi intensity map from \citet{Clark2018confproc}, where red, green, and blue represent the \hi brightness temperature in sequential velocity channel maps with $\Delta v \sim 3 \mathrm{km} \mathrm{s}^{-1}$. (2) Hub-filament structures (\S\ref{sec:hubs}) in OMC-1 as observed by ALMA \citep[Credit: ESO/H. Drass/ALMA (ESO/NAOJ/NRAO)/A. Hacar: http://eso.org/public/news/eso1809/ ; see also][]{Hacar2018}. (3) IRDC G11.11-0.12 from GLIMPSE \citep{kainulainen2013g11}. (4) ``Nessie", the prototypical giant filament (\S\ref{sec:giantfilaments}) \citep{Jackson2010,Mattern2018Nessie}. (5) \textit{Herschel} 250 $\mu$m maps show nearby filaments in Taurus, with two levels of inset figures: a zoom-in to a region containing striations (\S\ref{sec:striations}) and a region with dense fibers (\S\ref{sec:fibers}).   }
\label{fig:filfamilies}
\end{figure*}

We present an overview of observational studies of filamentary structures on various scales ($0.01$--$500$~pc) in the ISM. A classification into distinct, well-separated groups of filaments is difficult or even impossible. For this reason, we present results sorted by {\it filament families} (\S\ref{sec:nearbyfilaments}--\ref{sec:hubs}). These categories are
commonly presented in the literature and can sometimes depend on the technique, resolution, and sensitivity of the observations used. Consequently, filament families can partly overlap and individual filaments might belong to several families.
We illustrate some characteristic examples of each of these families in Fig.~\ref{fig:filfamilies}.

%----------------------------------------------------------
\subsection{Filaments in nearby molecular clouds}
\label{sec:nearbyfilaments}
%----------------------------------------------------------

% What are they; How are they identified?

The filamentary, parsec-scale substructure of molecular clouds closer than roughly 500 pc has been a subject of intensive study using the % Several nearby clouds have been studied in detail by the
\emph{Herschel} satellite data \citep[][]{Arzoumanian2011,Palmeirim2013,Benedettini2015, Arzoumanian2019,Konyves2020,Pezzuto2021}, and techniques from dust emission to extinction and molecular line emission \citep[][]{Malinen2012,Hacar2013,Seo2015,Kainulainen2016,Arzoumanian2018,Orkisz2019,Suri2019}. These works employ a variety of algorithms and techniques to identify %, or ``extract'',
filaments from the cloud maps (see \S\ref{sec:algorithms}). %A complex network of structures is commonly identified,
Typical methods identify complex networks of structures, comprising tens or even hundreds of individual, but often connected, (sub-) parsec-scale filaments. Filaments defined with these methods make up 2-11\% of the area within the cloud maps \citep[][]{Arzoumanian2019}. However, the filamentary network dominates the mass budget at high column densities \mbox{($A_\mathrm{V} > 7$ mag)} and harbors most of the star-forming cores in the clouds \citep{Konyves2015,Konyves2020}. The filaments are analysed individually, leading to a census of their properties. Overall, these works have now built a detailed view of the parsec-scale filamentary networks within the clouds at a resolution of roughly \mbox{10\ 000 AU} (0.05 pc). 

% What are their properties?
\citet[][]{Arzoumanian2019} conducted a homogenous, systematic study of the properties of filaments in dust-emission-derived column density maps of eight nearby molecular clouds. 
This census focuses on the basic parameters of the filaments and their distributions, i.e., line masses, lengths, and radial profiles. Typically, these filaments span a range of line masses of 5-17 M$_\odot$ pc$^{-1}$ and lengths of 0.3-0.8 pc. They reach peak column densities of 3-9 $\times$ 10$^{21}$ cm$^{-2}$ and dust-based temperatures of 14-16 K. 
The radial density profiles of the filaments have received ample attention because of their %due to the
connection to the basic physics of hydrostatic cylinders (see \S\ref{sec:radialprof}). Studies based on dust emission maps from \textit{Herschel} measure the distribution of inner widths to have the mean value of about 0.1~pc with a typical spread of a factor of 2 \citep{Arzoumanian2011,Arzoumanian2019}. %Such values are also sometimes found when using C$^{18}$O as a tracer
Similar values have been measured in C$^{18}$O filaments \citep[][]{Orkisz2019, Suri2019}, while observations of denser gas tracers indicate smaller widths at least for a subset of structures \citep[][see also \S\ref{sec:fibers} and \S\ref{sec:radialprof}]{Hacar2018}. 
The robustness of these results and whether they indicate a characteristic scale for filament widths continue to be debated \citep[][see also \S\ref{sec:widths}]{Panopoulou2017,Ossenkopf-Okada2019,Panopoulou2021}. The radial profiles of the filaments tend to be well-described by Plummer-like profiles with exponents $p$ around 1.5-2.5.  Observations of polarized dust emission reveal correlations between the orientation of the magnetic field and the long axis of filaments: filaments at column densities below $\sim \rm 10^{21} ~cm^{-2}$ are preferentially parallel to the local magnetic field, while at $N \gtrsim \rm 10^{22}~ cm^{-2}$ they are oriented preferentially perpendicular to the magnetic field \citep[][and \S\ref{sec:bfields}]{Planck2016XXXVSoler,Jow2018}.

%----------------------------------------------------------
\subsection{Filaments from Galactic plane surveys}
\label{sec:surveyfilaments}
%----------------------------------------------------------

% What are they; How are they identified?
The most substantial Galactic plane filament surveys to date have been the dust-emission based HiGal \citep[][]{Molinari2010,Schisano2020}, ATLASGAL surveys \citep[][]{Schuller2009,Li2016}, and the $^{13}$CO and C$^{18}$O survey SEDIGISM \citep[][]{Schuller2017,Mattern2018}. These surveys covered large, continuous sections of the Galactic plane, resulting in large samples that enable statistical, Galaxy-wide studies. For example, about \mbox{18,400} individual filaments were characterized using the HiGal survey \citep{Schisano2020}. However, survey results may be biased in ways difficult to quantify, due to limited resolution and sensitivity; for example, typical resolutions of the surveys are roughly 15-30\arcsec, translating to about 0.25-0.5~pc at 3.5~kpc distance. The mass sensitivities are tracer- and distance-dependent, but are typically on the order of a few 100 M$_\odot$. Thus, filaments detected in Galactic plane surveys likely include \emph{some} population of filaments similar to those detected within nearby clouds (\S\ref{sec:nearbyfilaments}), but also larger-scale filaments, substructure of which is not resolved at distances of several kpc. 

The properties of the filaments detected in the Galactic plane surveys vary with survey sensitivity. 
The most sensitive survey to date \citep[HiGal, see][]{Schisano2020} finds filaments with typical lengths of about 5-10~pc, masses of 500-1\ 000~M$_\odot$, line masses of 20-200 ~M$_\odot$~pc$^{-1}$, and temperatures of 10-35~K. For comparison, the typical filaments from the ATLASGAL survey are almost an order of magnitude more massive, reflecting the poorer sensitivity of ATLASGAL.
The majority of ATLASGAL survey emission \citep[$\sim 70\%$,][]{Li2016} is in filamentary structures. This is analogous to the result from nearby clouds that most of the dense gas is in the filamentary network (\S\ref{sec:nearbyfilaments}), however, a direct comparison is difficult because of the different spatial scales the survey probes. So far, only ATLASGAL-identified filaments have been systematically studied with molecular lines,
with measured $^{13}$CO velocity dispersion around \mbox{0.6-2 km s$^{-1}$} \citep[][]{Mattern2018}. A high fraction of ATLASGAL filaments are detected in $^{13}$CO (2-1), but only a small minority in C$^{18}$O (2-1) \citep{Mattern2018}. Ongoing surveys in CO will enable substantial improvements in coverage and statistics \citep{Su2019}.

%----------------------------------------------------------
\subsection{IRDC filaments}\label{sec:irdc}
%----------------------------------------------------------

Infrared Dark Clouds (IRDCs) commonly show filamentary morphologies, either overall, or in parts. Their properties have been investigated mostly in case studies \citep{Peretto2014, Henshaw2014,Busquet2016,Henshaw2016,Williams2018,Chen2019,Sokolov2019,Kong2019outflows,Baug2020,SLi2021}. The filamentary IRDCs are commonly identified from the \emph{Spitzer}/GLIMPSE survey by eye as prominent absorption features. Due to selection effects, the studied objects are typically located at distances of a few kiloparsecs \citep{Rathborne2016,Rygl2010}, reach high column densities \mbox{($N$ $\gtrsim 50 \times 10^{21}$ cm$^{-2}$)} and high line masses \mbox{($m \gtrsim 100$ M$_\odot$ pc$^{-1}$)}, and have lengths between roughly one and a few tens of parsecs. 
Detailed studies of the column density structure have uncovered filamentary networks and hub-filament structures within them \citep{Busquet2016,Ohashi2016,Henshaw2017}. 

While IRDC filaments have been extensively studied in the continuum, studies of their kinematics have been more sparse until recently \citep[e.g.][]{Peretto2014,Henshaw2016,Williams2018,Chen2019,SLi2021,Arzoumanian2022}. A homogeneous analysis of the kinematics of these structures would greatly enhance our ability to categorize these structures in terms of their evolutionary/environmental stage. Even less well studied are the magnetic field properties of IRDCs, with only a few clouds having dedicated investigations \citep[][see \S\ref{fig:kinematics} for further details]{Pillai2015,Santos2016,Busquet2016,Anez-Lopez2020,Wang2020c,Tang2019,Liu2018}.

%----------------------------------------------------------
\subsection{Giant Filaments}
\label{sec:giantfilaments}
%----------------------------------------------------------

% What are they; How are they identified?

A recent focus in filament studies has been the search for the longest, most massive filamentary structures in the Milky Way. The general driver of these works is to understand the galaxy-scale distribution and organisation of dense gas that is thought to correspond to kiloparsec-scale continuous structures commonly seen in external galaxies \citep{Elmegreen2018}. For brevity, we refer to these structures as 'Giant Filaments', noting that further sub-categories have been proposed based on their physical properties \citep{Zucker2018} and that their properties may depend on the galactocentric radius \citep{Colombo2021}.
%
% What are their properties?
%
The first systematic works have identified tens of Giant Filaments \citep{Ragan2014, Abreu-Vicente2016, Zucker2015, Wang2015, Wang2016, Li2016, Colombo2021}, varying from tens to hundreds of parsecs in length, the longest potentially being up to 2~kpc long \citep{Veena2021}. They cover a wide range of masses up to some 10$^6$~M$_\odot$ and their aspect ratios range from about 4 to 100. Typical line masses are around \mbox{1000 M$_\odot$ pc$^{-1}$}. The studies use a wide variety of approaches to identify structures based on dust extinction and emission signatures and CO line emission. In the abscence of homogeneous criteria, possible biases arising from different identification mechanisms are difficult to quantify. Reflecting the variety, the structures are commonly referred to with various names, e.g., 'Bones', 'Galactic filaments', 'Giant Molecular Filaments', or 'Large-scale filaments'. Most of these structures are detected in the galactic plane surveys (at least partially), however, they are usually broken down into smaller structures by the filament finding algorithms of those studies. Generally, the Giant Filaments harbor rich sub-structure consisting of a hierarchy of clumps and cores and smaller-scale filaments \citep[e.g.,][]{Jackson2010,Mattern2018Nessie, Wang2020densegas}. 

% What do we think about them? What are the key open questions.
\cite{Zucker2018} analysed the properties of filaments from different surveys homogeneously, finding qualitative differences between filament sub-samples.
%They showed that they trace different types of objects. 
\citet{Zucker2018} identified three distinct categories within the parameter space of dense gas mass fraction versus aspect ratio: high aspect ratio giant molecular clouds, networks of dense compact sources, and highly elongated, high column-density filaments ('the Bone candidates'). These categories may have differing physical origin and they may be dominated by different physical processes. Specifically, the high aspect ratio, high column-density filaments were suggested to be the best candidates to trace the spiral arms of the Milky Way. For a subset of their objects, \citet{Zucker2018} derived steeper radial profiles than what is obtained for nearby filaments (\S\ref{sec:nearbyfilaments}).

\citet{Zhang2019} also presented a homogeneous analysis of the physical properties of Giant Filaments, specifically focusing on scaling relations and star formation activity. Giant filaments follow scaling relations similar to Larson's relations and dense gas vs. star formation rate relations similar to those found in molecular clouds in general. Thus, their study promotes a view that long filaments are not special in terms of their star formation activity, but follow the same general scaling behavior as other clouds despite their elongated morphology.

%----------------------------------------------------------
\subsection{Dense fibers}\label{sec:fibers}
%----------------------------------------------------------

% Discovery
The analysis of the gas kinematics in nearby clouds such as B213-L1495 in Taurus, or the Integral Shape Filament in Orion revealed a new type of dense molecular filaments at small scales, named dense fibers. First reported by \citet{Hacar2013}, fibers are distinguished as velocity-coherent structures in Position-Position-Velocity space. They have 
(tran-)sonic velocity dispersions \mbox{($\sigma_\textrm{tot} \leq 2 \times c_\textrm{s}$)} and smooth oscillatory velocity profiles with local velocity variations ($\delta v_{lsr}\lesssim$~$c_\textrm{s}$). Fibers are commonly observed in diffuse (e.g. C$^{18}$O) and dense (e.g. N$_2$H$^+$) molecular tracers, and rank among the lowest-mass molecular filaments (\mbox{$M$ $\sim$ 5-10 M$_\odot$}). They typically show sub-parsec lengths ($L\lesssim$ 1pc), and have line masses close to critical \citep[\mbox{$m \sim m_\textrm{crit}$}; see ][for a statistical description]{Hacar2018}. Fibers also exhibit high central gas densities \mbox{$n_0 > 10^4$ cm$^{-3}$}, and small characteristic widths \citep[FWHM=0.02-0.1 pc; see][]{Fernandez-Lopez2014,Hacar2018,Monsch2018,Schmiedeke2021}, making them narrower than most of the \textit{Herschel} filaments identified in dust continuum. 
%The first polarization measurements within these objects indicate magnetic field orientations perpendicular to their main axis \citep{Doi2020}. The magnetic field orientation may, however, change between neighbouring fibers and seems to be different from the magnetic field detected at scales $>$~1pc \citep{Arzoumanian2019,Doi2020}.
The first polarization measurements within fibers indicate random magnetic field orientations with respect to their main axis \citep{Doi2020}. The magnetic field orientation also changes between neighbouring fibers and appears different from the magnetic field detected at scales $>$~1pc \citep{Arzoumanian2019,Doi2020}.

Fibers correspond to the fine substructure within larger and more massive objects identified in the continuum \citep[e.g.][]{Andre2014}. When observed at high enough resolution, an increasing number of works report a rich fibrous substructure in low-mass clouds \citep{Arzoumanian2013,Feher2016}, intermediate mass clusters \citep{Fernandez-Lopez2014,Hacar2017b}, IRDCs \citep{Henshaw2014,Chen2019,Sokolov2019}, and high-mass star-forming regions \citep{Hacar2018,Trevino-Morales2019,Shimajiri2019,SLi2021,Cao2021}. %The recently reported increased surface density of (sub-)critical fibers ($\Sigma_\textrm{fibers}$) in filamentary regions such as Taurus, Perseus, and Orion might explain the substructure of the apparently highly supercritical filaments (\mbox{$m > m_\textrm{crit}$}) at larger scales \citep[i.e. $m \propto \Sigma_\textrm{fibers}$; ][]{Hacar2018}.
The presence of an ensemble of of (sub-)critical fibers ($\Sigma_\textrm{fibers}$) in filamentary regions such as Taurus, Perseus, and Orion might explain the substructure of what are apparently highly supercritical filaments (\mbox{$m > m_\textrm{crit}$}) at larger scales \citep[i.e. $m \propto \Sigma_\textrm{fibers}$; ][]{Hacar2018}.

% Simulations
 Simulations with high levels of refinement \citep{Smith2014b,Moeckel2015,Kirk2015,LiKlein2019} show filament bundles with a rich substructure of fibers. 
Some of the fibers extracted in observations may not correspond to real gas structures but rather to artefacts produced in the line intensity profiles of low-density tracers \citep[e.g. CO; ][]{Clarke2017,Zamora2017}. Nonetheless, the excellent correspondence between most fibers and the column density distribution within filaments \citep{Andre2014}, as well as their detection in high-density tracers \citep[e.g. N$_2$H$^+$;][]{Hacar2018}, suggests that artifacts are neglibible in large samples. 

Some fibers harbour individual or small groups of cores regularly spaced at distances consistent with their corresponding Jeans length \citep{Tafalla2015}. Combined with their (tran-)sonic velocity dispersions %\mbox{($\sigma_\textrm{tot} \leq 2 \times c_\textrm{s}$)}
up to parsec-scales, periodic velocity oscillations are found to relate the positions of these cores with the streaming motions produced by gravitational fragmentation along their main axis \citep{Hacar2011,Hacar2017b,Heigl2018a}. 
Fibers are proposed to be the first sonic-like structures formed at the end of the turbulent cascade, which then impart their sonic-like properties to the embedded cores \citep{Hacar2011}.

\subsection{Striations}\label{sec:striations}

The diffuse parts of several nearby molecular clouds feature elongated structures termed `striations' \citep[see \textit{Herschel} images of Taurus, Chamaeleon-Musca, Polaris Flare, L1642 in][respectively] {Kirk2013,Cox2016,Miville2010,Malinen2016}. Striations were first identified by \citet{Goldsmith2008} in a large dynamic range image derived from CO observations of the Taurus molecular cloud. Their distinctive characteristics are that, (a) they appear quasi-periodically spaced and, (b) they are parallel to the magnetic field \citep{Chapman2011,Panopoulou2016a,Malinen2016}. Striations are typically found at column densities of $N(H_2) \sim 10^{20}-10^{21}$ cm$^{-2}$. While sometimes observed to be connected with dense, star-forming filaments \citep{Palmeirim2013}, striations are also found in areas devoid of denser material \citep{Goldsmith2008,Miville2010}.

Striations may be related to flows parallel to the magnetic field that channel material onto denser filaments \citep{Palmeirim2013,Cox2016}. 
\citet{Heyer2016} studied the velocity structure of striations and found oscillatory behavior in cuts perpendicular to their axis. They proposed that this might be a signature of Kelvin-Helmholtz instability or of magnetohydrodynamical (MHD) waves. 

In a dedicated numerical study, \citet[][]{Tritsis2016a} quantified the observational properties of molecular cloud striations in Taurus. While flows along/perpendicular to the field or Kelvin-Helmholtz instabilities were unable to reproduce the observed column density contrast ($\sim$ 25\%), the propagation of MHD waves was able to match the observed properties. The MHD wave model for the formation of striations has made two predictions that have so far been confirmed: the existence of normal modes in environments where the waves are trapped \citep{Tritsis2018} and the correspondence of the velocity and column density power spectra which follow the dispersion relation of MHD waves \citep[found in HI data,][]{Tritsis2019}. 
\citet{Chen2017} proposed that corrugations of sheets can be caused by the thin shell instability, yet,
this mechanism over-predicts the column density contrast of observed striations. Striation-like structures are found in numerous simulations when considering strongly magnetized (trans- to sub-Alfv\'enic) media \citep[e.g][see also \S\ref{sec:formation_bfield}]{Beattie2020}.

There is evidence that striations are not a feature of the molecular phase only, but may also exist in the more diffuse atomic phase of the ISM \citep[e.g.][]{Tritsis2019}. \citet{Wareing2016} performed simulations to investigate the role of the thermal instability in forming filamentary structures and found striations in their strongly magnetized models which reproduced the column density contrast of molecular cloud striations, but formed in the diffuse atomic medium.

Numerous works identify diffuse filaments parallel to magnetic field lines \citep{PlanckCollaborationXXXII2016,Clark2014,Inutsuka2015,LiKlein2019}, however care must be taken to determine which of these show the quasi-periodic spacing, low column density contrast and oscillatory velocity profiles that are characteristic of striations. We note that the term `striations' has sometimes been used to refer generally to diffuse filaments parallel to the magnetic field \citep[e.g.][]{Busquet2013,Miettinen2020}, however this choice could lead to confusion. More work is needed to quantify the properties of striations in a larger sample of clouds.

\subsection{\hi Filaments}\label{sec:hi}

Sensitive, high-resolution observations of diffuse \hi emission reveal ubiquitous filamentarity: the sky is patterned with slender, linear \hi filaments, or \hi ``fibers" \citep{Clark2014}. Similarly, fine \hi filaments are seen in absorption toward the Galactic Center \citep{McClure-Griffiths2006}. In both cases typical column densities are $\lesssim 10^{20}\,\mathrm{cm}^{-2}$ \citep[see also]{Kalberla2016}.  %\textcolor{red}{typical column densities are around xxxxx cm$^{-2}$}.
An even more striking property of the \hi filaments is their excellent parallel alignment with the local magnetic field orientation, as probed first by starlight polarization \citep{McClure-Griffiths2006, Clark2014} and later by polarized thermal dust emission \citep{Clark2015, Martin2015, Kalberla2016}. The \hi filaments are well-aligned with the measured plane-of-sky magnetic field orientation on average, with some measurable misalignment that may be a useful probe of turbulence in the nearby ISM \citep{Huffenberger2020,Clark2021}. 

The \hi filaments have very high aspect ratios, particularly as seen in the $4'$-emission measured in the Galactic Arecibo L-Band Feed Array Survey \citep[GALFA-\HI;][see Fig.~\ref{fig:filfamilies}]{Peek2018}. 
However, their magnetic alignment is still distinctly measurable in lower-angular resolution data like \HI4PI \citep{HI4PICollaboration2016}, an all-sky \hi map made by combining the $9'$ EBHIS data \citep{Winkel2016} with the $16'$ GASS survey \citep{McClure-Griffiths2009}. 
Polarization maps ``predicted" from \hi geometry via the assumption that \hi filaments are perfectly aligned with the magnetic field bear a striking resemblance to real measurements of the polarized dust emission \citep{Clark2018,ClarkHensley2019}.

Because the \hi filaments studied by \citet{Clark2014} are particularly prominent in narrow velocity channels, \citet{Lazarian2018} argued that these structures were ``velocity caustics": an imprint of the turbulent velocity field uncorrelated with the underlying density field \citep[e.g.][]{Lazarian2000}. This interpretation has been ruled out by a number of independent analyses. 
\citet{Clark2019} tested the velocity caustics picture in the diffuse \hi by computing the correlation between velocity channel emission structure and dust emission traced in broadband measurements of the FIR, which is insensitive to the gas velocity field. This analysis finds no measurable velocity caustic effect. 
Furthermore, \citet{Clark2019} find that the ratio of FIR/$N_\textrm{HI}$ increases toward sightlines that contain small-scale structure in the \hi channel map emission. 
Indeed the FIR/$N_\textrm{HI}$ ratio is positively correlated with the cold neutral medium (CNM) fraction in the diffuse \hi \citep{Murray2020}, and with the intensity of small-scale structure \citep{Kalberla2020}. Moreover, \citet{Peek2019} showed that the equivalent width of Na{\sc i} absorption is more sensitive to the \hi column density in small-scale channel map structure than to the total \hi column density. 
Together these results support the interpretation that the \hi channel map filaments are real density structures that occupy a colder, denser phase of the \hi gas than the warmer, more diffuse medium.

Although a prominent feature of the high Galactic latitude sky, \hi filaments are not confined to diffuse regions of the ISM. The Riegel-Crutcher cloud, backlit by bright radio emission from the Galactic center, contains many magnetically aligned \hi filaments that are detected in absorption \citep{McClure-Griffiths2006}. These filaments are reminiscent of the high Galactic latitude \hi filaments detected in emission, with similar aspect ratios and somewhat higher column densities \citep[c.f.][]{Clark2014}.
Surveys of the Galactic plane reveal ubiquitous filamentary structure in \HI, including extremely massive structures that are apparently coherent both spatially and spectrally. \citet{Soler2020} and \citet{Syed2022} report a particularly extended \hi filament that, assuming the circular rotation distance estimate of 17 kpc is accurate, would be over 1 kpc long. A similar giant \hi filament (length 1.1 kpc) was recently reported by \citet{Li2021}. These structures are qualitatively different than the diffuse \hi filaments for which properties considered in this work (e.g. Table \ref{tab:filprop1}) are derived. 

\subsection{Hubs, ridges, and networks}\label{sec:hubs}

% hubs
Filaments can also form complex associations. In the so-called hub-filament structures (HFS), or simply hubs, multiple filaments extend radially up to several parsecs away from a central parsec-size clump with column densities of $>~10^{22}$~cm$^{-2}$ and several hundreds of M$_\odot$ in mass \citep{Myers2009a}. Line and continuum observations show converging filaments in a multitude of IRDCs \citep[e.g.][]{Peretto2010,Peretto2013,Busquet2013} and Galactic plane clumps \citep{Kumar2020}. Elongated HFS, sometimes referred to as ridges, have central FWHM up to $\sim$~0.5~pc \citep{Hennemann2012,Russeil2013} and are among the most massive filaments in the Galaxy. Observed line masses \mbox{$m$ $\sim$ 300 M$_\odot$ pc$^{-1}$} significantly exceed the critical hydrostatic limit \mbox{$f \gg 1$} \citep{Schneider2010,Hill2011}. Nearby examples of these HFS are found in regions such as NGC~1333 and OMC-1 \citep{Myers2009a}, DR21 \citep{Schneider2010}, or Mon-R2 \citep{Trevino-Morales2019}.

% clusters
Hubs and ridges are associated with the earliest phases of high-mass and cluster formation \citep[see][and references therein]{Motte2018}.
All nearby young stellar associations exceeding 25~stars~pc$^{-2}$ \citep{Myers2009a} as well as a large fraction of the massive galactic pre- and protostellar clumps \citep{Kumar2020} are found in these filamentary systems. Many HFS appear to be highly dynamic objects. 
Longitudinal velocity gradients ($\nabla v_{LSR}>$~1~km~s$^{-1}$~pc$^{-1}$) and filamentary accretion flows ($>$~30~M$_\odot$~Myr$^{-1}$) are observed along filaments feeding their central hub \citep{Kirk2013,Peretto2013,Peretto2014,Hacar2017a,Baug2018,Williams2018,Chen2019,Saajasto2019,Dewangan2020,Liu2021,Ren2021}. Evidence for gas acceleration following free-fall velocity profiles \citep{Hacar2017b,Williams2018}, spiral patterns \citep{Trevino-Morales2019}, and hour-glass shaped and close-to-critical magnetic fields \citep{Pattle2017}, denote gravity as the main driver of these inflow motions at parsec-scales (see also \S\ref{sec:dyn-acc}). 

% networks
At sub-parsec scales, densely populated fiber networks are also observed {\it within} hubs such as OMC-1 \citep{Wiseman1998,Hacar2018}, Serpens \citep{Fernandez-Lopez2014}, NGC~6334 \citep{Arzoumanian2021,SLi2021} or DR21 \citep{Cao2021}, among others. Fibers in these systems show large longitudinal \citep{Hacar2018} and lateral \citep{Dhabal2018,Chen2020a} velocity gradients, as well as a variable magnetic field morphology \citep{Arzoumanian2021,Pattle2021}, sometimes exhibiting rapid changes with respect to the magnetic field at cloud scales \citep{Doi2020}.
The complexity of these fiber arrangements appears to increase with total mass of the host clump \citep{Hacar2018}. The most massive (supra-Jeans) cores in these regions are typically located at the network nodes (local hubs) \citep{Hacar2017a,Zhang2019} suggesting that junctions, collisions, and gravitational focusing effects between fibers may become important in dense systems. Simulations indicate that cores formed in these nodes exhibit larger masses than those formed during the fragmentation of individual fibers \citep{Clarke2020,Hoemann2021}. 
Additional filamentary accretion into these nodes may contribute to the rapid assembly of massive stars and disks \citep{Banerjee2006,Smith2011,Kirk2015,Smith2016}. High resolution observations of massive clumps precursors reinforce this interpretation \citep{Beuther2020}.

\subsection{Other filaments}

The study of filaments is necessarily restricted to data with which we can resolve filaments. Naturally, filaments also exist in other galaxies, and large-scale filaments have been identified in the Magellanic Clouds \citep{Fukui2018,Tokuda2019} and M100 \citep{Elmegreen2018}. Another frontier in the study of ISM filaments is the Galactic center, an extreme environment relative to the rest of the Galactic disk (see PPVII review by Henshaw et al). The low star formation rate in the Galactic center relative to its high surface density makes it an intriguing case study for the theory of star formation. G0.253+0.016 (``The Brick") is a famous IRDC that has an unusually high mass, but very little evidence of ongoing star formation \citep{Kauffmann2013,Pillai2015,Henshaw2019}. For most of this work, we focus on filaments in environments more typical of the Galactic disk, but we note that the Galactic center is an excellent testbed for theories of filament formation and evolution \citep[e.g.][]{Kruijssen2019}. 

The ISM is home to a wealth of other ``filaments" beyond those considered here. These include the non-thermal radio filaments at the Galactic center \citep{Yusef-Zadeh1987,Heywood2022}, radio polarimetric filaments and depolarization canals \citep{Haverkorn2000},  filaments in the ionized medium traced in H$\alpha$ \citep{Planck2016_XXV}, polarization gradient filaments \cite{Gaensler2011,Campbell2021}, and filamentary structures associated with ISM shocks or supernova remnants \citep{McCullough2001, Bracco2020}, including large angular scale structures associated with supershells.

\subsection{Filament identification algorithms}\label{sec:algorithms}

New families of filaments have proliferated concomitantly with new filament identification algorithms. The techniques in use range from specialized, physically-motivated algorithms to general-purpose feature identification codes. Each carries particular biases that inform the filament classification and subsequent analysis. Each algorithm optimizes for the detection of a certain class of structures: this is the \textit{de facto} definition of a filament.  

\textsc{DisPerSe} \citep{Sousbie2011} is a widely-used algorithm for identifying filaments, especially in continuum data. Originally developed for quantifying cosmic web structures, it was quickly applied to \textit{Herschel} observations of Galactic filaments \citep{Arzoumanian2011, Peretto2012, Hill2011}. \textsc{DisPerSe}'s provenance as a code for identifying structures in the distribution of large-scale structures informs what it identifies as ``filaments". The cosmic web is structured by gravitational forces, and cosmic web filaments connect dark matter halos. This property motivates \textsc{DisPerSe}'s definition of filaments as ascending 1-manifolds: one-dimensional structures connecting overdensities in a Morse decomposition of data. Thus, the application of \textsc{DisPerSe} to filament detection in ISM image data carries the implicit definition of a filament as a linear structure connecting image overdensities. For example, \textsc{DisPerSe} identifies filaments on images consisting entirely of randomly placed cores \citep{Panopoulou2014}. The output of \textsc{DisPerSe} is one-dimensional: a filament ``spine" along which properties of the structure may be analyzed. 

Other algorithms similarly define filaments by their skeletons via some quantifiable aspect of image morphology. The code \textsc{getfilaments} and its successor \textsc{getsf} \citep{Men'shchikov2013, Men'shchikov2021}, which identify filaments via the application of a series of Gaussian smoothing kernels, define filaments as structures that are persistent over multiple scales and substantially anisotropic (elongated along one axis). \textsc{FilFinder} employs a multi-step preprocessing scheme before defining a filament spine via a medial axis transform \citep{Koch2015}. These are, in essence, matched filter algorithms that detect a known signal in the presence of noise by convolving a measurement (in this case an image of the ISM) with a filter designed to maximize the signal-to-noise in the presence of the desired signal. An important caveat is that the definition of a good skeleton has been entirely subjective. However, recently \citet{Green2017} introduced a measure of skeleton ``goodness-of-fit'', the mean structural similarity index, to remove visual bias in filament identification. \citet{Jaffa2018} introduce the method of $J$-plots, which is based on the moment of inertia, to characterise the morphology of (already identified) objects in 2D. It differentiates single filaments, circular or ring-like structures and condensed cores, at least for simple object geometries.  

Identifying filaments in molecular line emission is an inherently three-dimensional problem, and one that must contend with the ambiguities inherent in inferring 3D spatial information from position-position-velocity (PPV) data \citep[e.g.,][]{Beaumont2013}. Friends In VElocity (\textsc{five}) approaches this problem via a friends-of-friends search in PPV space, seeded by high signal-to-noise spectral components \citep{Hacar2013,Hacar2018}. \textsc{five} then trawls the PPV cube, defining velocity-coherent structures that meet some additional criteria. Another approach is to construct 3D filaments from stacks of 2D filaments identified in velocity channel maps. 
\textsc{Fil3D} is a recently-developed algorithm that takes 2D channel map filaments (as delineated by \textsc{FilFinder} or a similar algorithm) and identifies 3D filaments where such structures are substantially overlapping and continuous in velocity space (Putman et al. in prep). \textsc{Fil3D} was motivated by application to filaments in \hi channel maps (Kim et al. in prep). 

%RHT, Hessian matrices, gradients, etc.
Other techniques focus on quantifying filamentarity in image data, rather than identifying filaments as discrete objects. Many of these algorithms borrow from edge detection methods in machine vision. As filaments can be understood as sharp discontinuities in image space, they are highlighted by spatial gradients \citep{Koch2013, Soler2013, Planck2016XXXVSoler, Orkisz2019}. The direction of the maximum gradient is a simple metric for the orientation of a filament, or any other ``edge" in the image plane.
One can measure the second derivative of image intensity by computing the Hessian matrix, which computes the local curvature of structures \citep{Molinari2011,Polychroni2013,PlanckCollaborationXXXII2016}. A filament can then be defined as an image feature with negative curvature along one dimension. Each of these methods identifies filaments at a preferential scale: image gradients and higher-order derivatives are local operators, and thus high-pass filters. These methods are most sensitive to filaments with widths that are similar to the scale of the derivative kernel. Other methods effectively set a minimum spatial scale for filamentary structure.  

Some authors have studied the filamentary ISM by simply applying a high-pass filter or unsharp mask to image data \citep{Kalberla2016, Clark2019}. This method does not explicitly parameterize filaments, but removes smoothly varying emission at large angular scales. To the extent that the emission studied is inherently filamentary, selecting the small-scale structure via an unsharp mask has the effect of highlighting filaments. An unsharp mask is the first step in the Rolling Hough Transform (RHT), which parameterizes the linearity of image data as a function of orientation \citep{Clark2014}. The RHT quantifies the probability that any given image pixel is part of a line in image space at any given orientation. Thus, while not explicitly a filament finder, it can be used to identify linear features, or to measure the orientation of filaments identified via other algorithms.  

With such a variety of approaches to filament detection, cross-comparison between various algorithms is a useful way to identify sources of systematic discrepancy between derived filament properties. \citet{Chira2018} compare \textsc{DisPerSe}, \textsc{FilFinder}, and minimal spanning tree algorithms \citep[e.g.,][]{Wang2016} applied to dendrograms \citep[e.g.,][]{Goodman2009}. \textsc{DisPerSe} returns consistently higher line masses than the other two methods. 

One defining characteristic of many of the algorithms described here is that they are maximally sensitive to filaments at a particular angular scale. Some algorithms have instead adopted explicit descriptions of multi-scale structure \citep[e.g.][]{Ossenkopf-Okada2019,Robitaille-2019, Robitaille2020, Allys2019}. Such treatments describe structure as a function of angular scale, and are thus capable of quantifying hierarchical morphology.

%%%%%%%%%%%%%%%%%%%%%%%%%%%%%%%%%%%%%%%%%%%%%%%%%%%%%%%%%%%
\section{A CENSUS OF MILKY WAY FILAMENTS}\label{sec:census}

\begin{deluxetable}{l|c|c|c|c|c|c|c}
\scriptsize
%\tablewidth{0pt}
\tablewidth{59em}
\tablecaption{FILAMENT FAMILIES (I): FUNDAMENTAL PROPERTIES$^{(1)}$}
\renewcommand{\arraystretch}{.7}
\tablehead{Family$^{(\#)}$ & Targets & $M$ & $L$ & FWHM & $N_0$ & $T$\\
 & & (M$_\odot$) & (pc) &  (pc) & ( 10$^{21}$ cm$^{-2}$)  & (K) }
\startdata
Nearby Filaments & 707 & 2.1 - 13.4 & 0.4 - 1.0 & 0.07 - 0.16 & 3.2 - 9.0 & 13 - 16 &  \\ 
Galactic Survey & 18854 & 98 - 2.3E3 & 3.6 - 14 & 0.3 - 1.3(?) & 5.9 - 14.4 & 15 - 19 &  \\ 
IRDCs & 115 & 46 - 239 & 1.2 - 2.8 & $\sim$~0.3(?) & 13 - 40 & 16 - 25 &  \\ 
Giant Filaments & 153 & 7.8E3-1.5E5 & 32 - 83 & $\sim$~1(?) & 6.0 - 8.4 & 15 - 19 &  \\ 
Dense fibers & 127 & 1.5 - 15.0 & 0.2 - 0.5 & 0.03 - 0.10 & 12 - 69 & 10 - 22 &  \\ 
Striations & 2$^{(*)}$ & 2.5 - 5.6 & 1.0 & 0.2 - 0.4(?) & 0.5 - 1.0 & 15 - 20$^{(2)}$ &  \\ 
HI filaments & 2823 & 0.1 - 0.5 & 1.4 - 3.6 & ? & $<1$(?) & 100$^{(3)}$ &  \\
\enddata
\tablenotetext{(1)}{\small Values correspond to the interquartile [Q25\%,Q75\%] range for each measured quantity. (2) Typical molecular gas temperatures at low column densities \citep[e.g.][]{Planck2011}. (3) Typical temperature for the CNM. (?) Unknown  or  poorly  constrained. (*) Limited statistics.}
\tablenotetext{(\#)}{\small References: 
{\bf Nearby Filaments:} \citet{Arzoumanian2013,Arzoumanian2019,Palmeirim2013,Nagahama1998,Kainulainen2016,Hacar2016b,Chung2021}.
{\bf Galactic Plane Surveys:} \citet{Schisano2020,Li2016,Wang2016,Mattern2018,Xiong2019}.
{\bf IRDCs:} \citet{Schneider2010,Hennemann2012,Kainulainen2013,Peretto2014,Lu2014,Beuther2015,Pillai2015,Chen2019,Busquet2016,Santos2016,Trevino-Morales2019,Leurini2019,Arzoumanian2021,Arzoumanian2022}.
{\bf Giant Filaments:} \citet{Contreras2013,Ragan2014,Zucker2018,Zhang2019,Colombo2021}.
{\bf Dense fibers:} \citet{Hacar2011,Hacar2013,Hacar2016a,Hacar2017a,Hacar2018,Lee2014,Tafalla2015,Seo2015,Dhabal2019,Eswaraiah2021,Schmiedeke2021,Dhabal2018,SLi2021}.
{\bf Striations:} \citet{Chapman2011,Panopoulou2016a}.
{\bf HI Filaments:} \citet{Clark2014}, Putman et al. in prep.
}
\label{tab:filprop1}
\normalsize
\end{deluxetable}

% Large dataset
The increasing number of observational results describing the filamentary properties of the ISM provide a panchromatic view of these gas structures across our Galaxy. As illustrated in \S\ref{sec:surveys}, however, the physical characteristics of these filaments may significantly differ depending on the family, observational technique, or scales considered in each case. For the first time, the maturity of this field allows a direct analysis of filament populations in the ISM beyond studies of individual targets or regions.

% How do we construct it\ how to covert its properties
We have created a census of filaments across the Milky Way to perform a meta-analysis of their physical characteristics.
We restrict ourselves to the main physical properties of these filaments typically reported in the literature, namely (and when available), their total mass ($M$), length ($L$), central column density ($N_0$), width (FWHM), gas kinetic temperature ($T$), total ($\sigma_\textrm{tot}$) and non-thermal ($\sigma_\textrm{nt}$) velocity dispersion, internal gradients ($\nabla v_\textrm{LSR}$), and magnetic field strength ($B_0$). The assembly of this catalog requires a homogenization of distinct measurements, including standard conversions into total gas masses and velocity dispersion from different molecular datasets. 
Our catalog includes information of 49 individual observational works, including continuum, molecular line, and polarization surveys, together with dedicated works on specific targets. In multiple occasions we combine several independent but complementary studies describing distinct properties in well investigated targets (e.g. B213-L1495). 
Our catalog includes a total of 22803 filaments (entries) across the Galaxy. The mean statistical properties classified by families, including all references, are listed in Tables \ref{tab:filprop1} \& \ref{tab:filprop2}. 

% Caveats
Several caveats should be considered during the analysis of our filament sample. By construction, our sample is limited by the targets and surveys included in individual studies in the literature. As a result, the number of targets per filament family significantly varies across our sample, from the handful of striations characterized in the literature to the thousands of filaments in nearby clouds and Galactic plane surveys. Our filament catalog is also largely inhomogeneous including multiple entries with only partial measurements. Similarly, some entries may not be unique as the same targets may be included in multiple catalogs. 
Due to the availability of these observations, our sample is dominated in number (83\%) by the mass and length estimates provided by large continuum surveys with a smaller fraction ($<$5\%) of targets including kinematic measurements and magnetic field estimates.

Not all quantities might be equally accurately determined. Continuum and line observations provide robust measurements of $M$, $L$, and $N_\textrm{0}$ of filaments. In contrast, parameters such as the FWHM are indirectly inferred from modeling or assuming some filament shape (e.g. uniform cylinder). Distance uncertainties are not always quoted in the original catalogs, and are therefore unaccounted for in the majority of the data. Gas kinetic temperatures are sometimes equated to the dust temperatures as these are easier to obtain in FIR observations (i.e. \mbox{$T \approx T_\textrm{dust}$}). Other parameters, such as $\sigma_\textrm{tot}$ or $\sigma_\textrm{nt}$, might depend on the molecular tracer used. Magnetic fields are inferred via the Davis-Chandrasekhar-Fermi method (\S\ref{sec:bfields}). 
While standard practice in the field, this inhomogeneity in determining parameters
should be considered during the interpretation of our results.

\begin{figure*}[ht!]
    \centering
    \includegraphics[width=\textwidth]{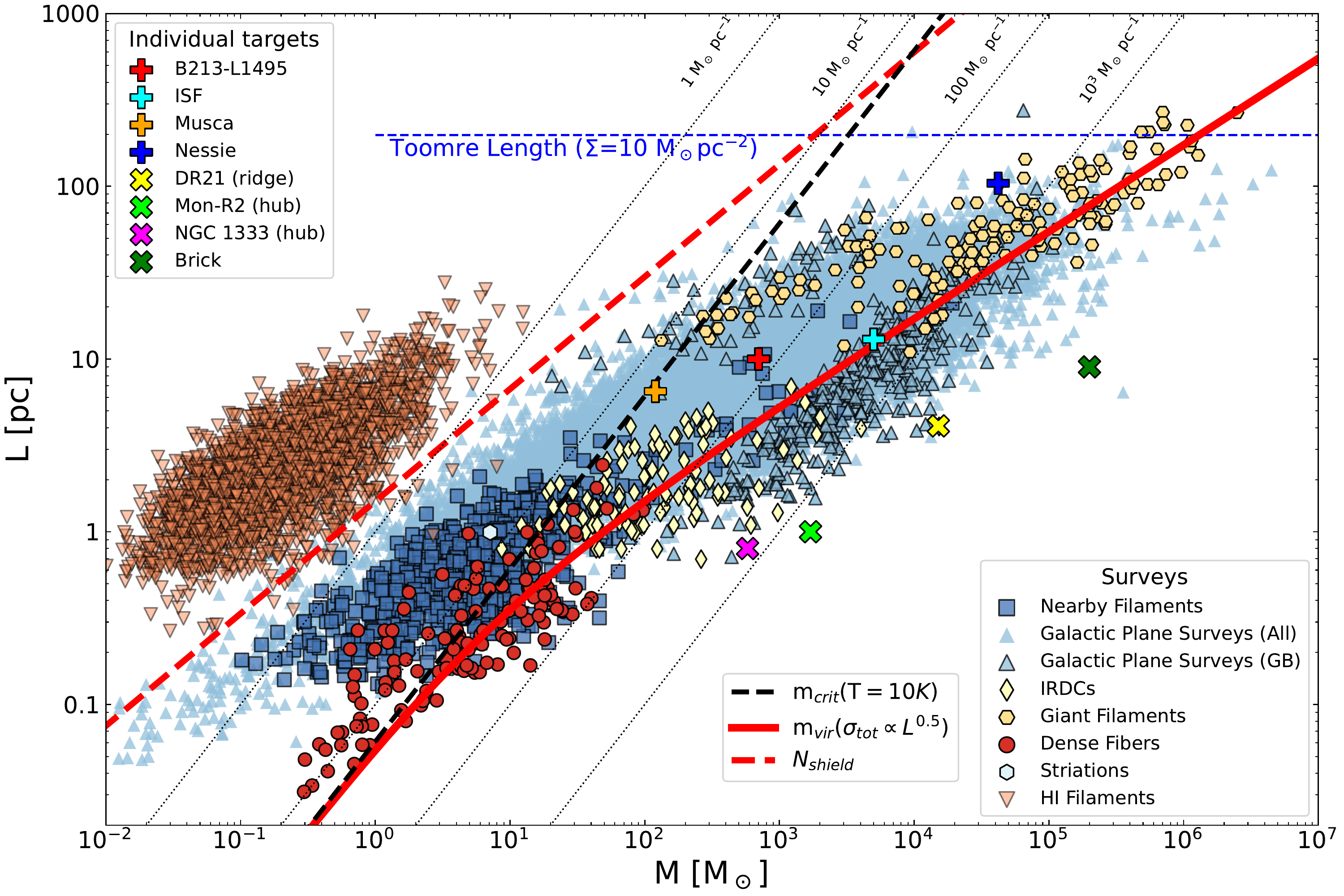}
    \caption{\small Mass and length measurements for all filaments in our catalog (\S\ref{sec:census}). Points are colour-coded according to the corresponding filament families described in \S\ref{sec:surveys}: Nearby Filaments, Galactic Plane Surveys, IRDCs, Giant Filaments, Dense Fibers, Striations, and \hi Filaments (see legend). In the case of Galactic Plane Surveys a subcategory identifies those detected by ground-based observations (GB).
    Some prototypical filaments are also included in this plot (see symbols in legends). Black dotted lines indicate constant $m$ = 1, 10, 100, 1000~M$_\odot$~pc$^{-1}$ values, the black dashed line corresponds to \mbox{$m_\textrm{crit}$(10~K) = 16.4 M$_\odot$ pc$^{-1}$}. The expected maximum filament size set by the Toomre length for the Galactic midplane at the Sun's radius is indicated by a blue dashed line. The red solid line shows $m_\textrm{vir}$ (Eq.~\ref{eq:MLscaling}) including turbulent motions. The red dashed line corresponds to the column density $N_\textrm{shield}$ required for self-shielding  (Eq.~\ref{eq:Nshield}).
    \label{fig:Main_prop}}
\end{figure*}

This meta-analysis provides a panchromatic view of the filamentary properties at different scales and environments. The compilation reflects not only the current state-of-the-art but also the large community effort and achievements describing the filamentary nature of the ISM combining different surveys, observational techniques, and tracers. We will use this sample to describe the statistical properties and global trends within and between filament families (\S\ref{sec:MandL} - \ref{sec:velfields}). We describe the dynamic properties of filaments (\S\ref{sec:theory}) to interpret the observed correlations (\S\ref{sec:phydes}) and we identify future avenues for the study of filaments in the ISM (\S\ref{sec:conclusions}).

\subsection{Filaments across scales: tracers of ISM physics}\label{sec:MandL}

% M-L figure
The most comprehensive overview of different filaments reported in the literature can be obtained from the description of their mass and length shown in Fig.~\ref{fig:Main_prop}. To facilitate their comparisons, different filament families and prototypes (see \S\ref{sec:surveys}) are marked in this plot. 

Observations report filaments across almost eight orders of magnitude in mass, from $\sim$~0.01 to 5$\times$10$^6$~M$_\odot$, and four orders of magnitude in length, from $\sim$~0.03 to 300~pc. 
The molecular filaments describe a continuous distribution in mass and length, including all the nearby filaments and fibers and extending towards the longest Galactic Plane and Giant Filaments, with an approximate scaling relation \mbox{$L \propto M^{0.5}$} (\S\ref{sec:mass_size}). Only the HI fibers clearly depart from this general trend. 

% Scale dependence
The large range in mass and length indicates that filaments possess different stability and dynamical properties depending on the scale. Most of the filaments detected at scales $L$~$\gtrsim$~10~pc show very high $m$ (up to \mbox{$\geq 100$ M$_\odot$~pc$^{-1}$}) and thus largely exceed the expected $m_\textrm{crit}$ for a hydrostatic filament %at \mbox{$T$ = 10 K} 
(Eq.~\ref{eq:mcrit_therm}). This indicates that these objects are dynamically evolving. 
Many of the shorter filaments are closer to, but still slightly above, $m_\textrm{crit}$ with 
line masses of 15-30 M$_\odot$~pc$^{-1}$. 
On the opposite side, a non negligible fraction of (sub-)parsec scale filaments show sub-critical $m$ suggesting them to be either evolving or transient structures. 
It is thus clear that filaments identified in the literature do not form a unique or homogeneous population. Instead, different filaments probe distinct ISM properties depending on the scale, from large cloud complexes (i.e. Giant Filaments) to small substructures within clouds (e.g. filaments and fibers in nearby clouds).

Filament prototypes require a special mention.
Targets such as B213-L1495, Musca, Nessie, or DR21 may define the typical (sometimes extreme) properties of other filaments within the same mass and length ranges. However, it is unlikely that objects with dissimilar properties are governed by the same physical processes (see also \S\ref{sec:theory}). Giant Filaments such as Nessie are likely affected by processes at galactic scales (e.g. shear or spiral arms) that have a small influence in the evolution of small scale filaments within clouds. On the other hand, it is expected that gravity plays a more dominant role in massive filaments such as DR21 than in \hi filaments, despite having comparable sizes (see also the discussion on formation mechanisms in \S\ref{sec:formation}). The extrapolation of these prototypes for the interpretation of filaments at different scales should be avoided or at least treated with extreme caution. 

% Observational biases
Several observational biases can be identified during the classification and study of different filaments (see mean values in Tables \ref{tab:filprop1} \& \ref{tab:filprop2}). First, distance appears as one of the primary discriminators between families. Using the same observational techniques (e.g. \textit{Herschel}), Galactic Plane surveys identify cloud-size filamentary structures much larger in size ($L~>$~10~pc) and mass ($M~>$~100~M$_\odot$) \citep[e.g.][]{Schisano2020} than the corresponding (sub-) parsec long ($L\sim$~0.5-5~pc) and low-mass ($M\sim$~1-10~M$_\odot$) filaments reported in nearby studies \citep[e.g.][]{Arzoumanian2019}. The latter describe the small-scale filaments within molecular clouds likely unresolved when observed at kpc distances across the Galactic Plane. Similarly, observational criteria 
%(e.g. \mbox{$L$ $>$ 3 FWHM}) 
artificially bias the study of smaller filaments in nearby clouds. For example, limiting the filament length to $L>3$~FWHM in combination with unresolved widths may suppress shorter filaments. 
The combination of these effects directly translate into systematic differences in the corresponding line mass derived in these studies.

Second, sensitivity limits ground-based Galactic surveys to the detection of filaments with extreme $m$ values (\mbox{$>$ 500 M$_\odot$ pc$^{-1}$}) \citep[e.g.][]{Li2016,Zucker2019}
within the general population of Galactic Plane targets \citep[see][for a discussion]{Schisano2020}. Similarly, the use of molecular tracers might restrict the detection of filaments to the densest targets at a given scale \citep[e.g. fibers, ][]{Hacar2018}.

Third, observational techniques may create artificial distinctions in the otherwise continuous distribution of filaments across scales. For example,
filamentary IRDCs are observationally identified by their high extinction in the mid-IR \citep[e.g.][]{Peretto2010}, while their mass and length cannot be distinguished from an average filament in the Galactic plane. 

% Physical interpretation: anisotropic ISM
The observed range of filament lengths is intimately connected to the fragmentation properties and characteristic timescales of the ISM. The maximum length of a filament in our Galaxy is limited to $\sim$~200~pc by the Toomre instability for a typical midplane gas surface density
of 10~M$_\odot$~pc$^{-2}$ \citep[e.g.][]{Li2017}.  At the opposite end, star formation occurs within parsec-size filaments at densities $>10^3$~cm$^{-3}$ where gravity dominates their fragmentation forming quasi-spherical cores (see \S\ref{sec:fragmentation}). These scale-dependent properties may change with the Galactocentric distance (with different Toomre values), elevation with respect to the midplane (e.g. \hi filaments), local environment (e.g. CMZ), and local gas density (e.g. high-mass clouds). Functioning as gas compasses, the detailed characterization of the filamentary properties of the ISM can thus provide unique insights of its most fundamental physical properties.

%---------------------------------------------
\subsection{Thermal Physics of Filaments}\label{sec:pressure}

\begin{figure}[t]
    \centering
    \includegraphics[width=\linewidth]{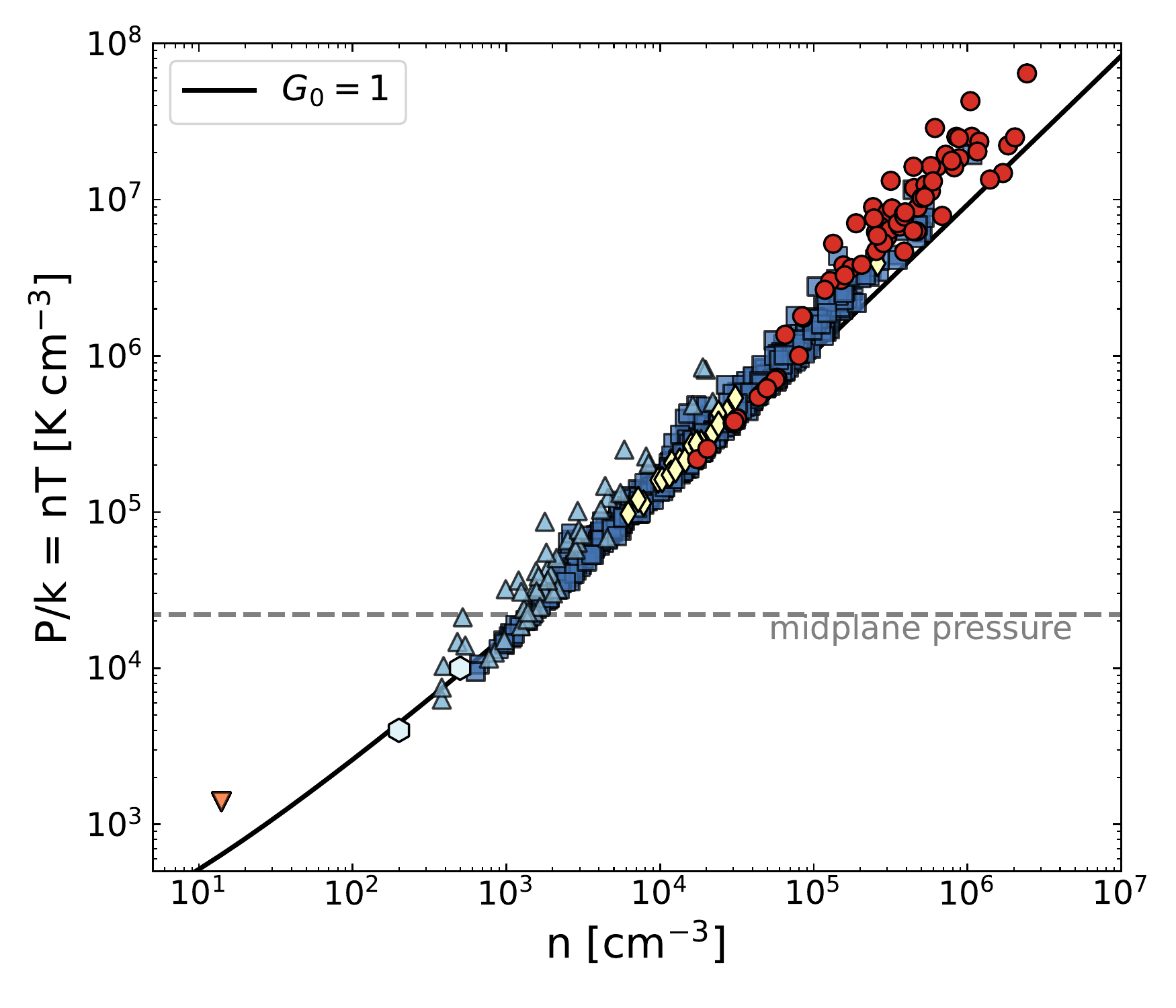}
    \caption{\small Thermal pressure as function of the filament density. The thermal equilibrium curve corresponds to a shielding column of \mbox{$N = 10^{21}$ cm$^{-2}$} and an interstellar radiation field strength of  \mbox{$G_0 = 1$} \citep{Wolfire1995}. Symbols are similar to Fig.~\ref{fig:Main_prop}. In the case of H{\small{I}} the temperature and density are assumed similar to the CNM.
    The horizontal line indicates the typical Galactic midplane pressure of $\sim2\times10^4$~K~cm$^{-3}$ \citep{Cox2005}.}
    \label{fig:thermal}
\end{figure}

Filaments cover not only a large parameter space in mass and length (Fig.~\ref{fig:Main_prop}), but also in thermal pressure (Fig.~\ref{fig:thermal}), suggesting that they exist in a wide variety of environments and evolutionary stages \citep{RiveraIngrahametal2017}. The thermal pressure of H{\small{I}} filaments is only about $10$\% of the Galactic midplane pressure ($2\times 10^4$~K~cm$^{-3}$, \citealp{Cox2005}).
Magnetic fields aligned with the H{\small{I}} filaments \citep{McClure-Griffiths2006,Clark2014} could provide the ``missing'' pressure component. Thermal pressures substantially above the Galactic midplane pressure may be set by the filament's self-gravity, by pressurization due to the ``weight'' of the ambient cloud \citep{Lada2008}, or by accretion (\S\ref{Sec:env-accretion}). Sustained flow compressions could also lead to overpressures, with the thermal pressure balancing the ram pressure of the inflowing gas \citep{VazquezSemadeni2006}.  
All filaments are close to thermal equilibrium curves \citep{Wolfire1995,Koyama2002}, suggesting that radiative cooling timescales are short compared to dynamical timescales \citep{Audit2005}. 

Average filament temperatures vary only slightly between filaments (Tab.~\ref{tab:filprop1}, see also Fig.~\ref{fig:thermal}). Temperature profiles inferred from dust temperatures \citep{Arzoumanian2011,Palmeirim2013,Arzoumanian2019} indicate a temperature drop of $\Delta T\sim 2-3$~K toward a filament's spine. For the Integral Shape Filament (ISF) in Orion A, \citet{Schuller2021} find temperature changes of $\Delta T\sim 20$~K corresponding to $100$\% in terms of the central temperature when compared to its environment. Along a filament gas and dust temperatures seem to vary mostly only within a few Kelvin \citep{Sokolov2017}. Simulations of individual filaments, including chemistry and radiation attenuation, \citep{Seifried2016,Seifried2017} are consistent with observed temperature profiles, though some studies show strong temperature increases toward the center \citep{Anathpindika2021}. Observed and modeled filament profiles do not support the assumption of isothermality, with temperatures increasing by at least $\approx 25$\% outwards. Such temperature profiles result in an effective adiabatic exponent of \mbox{$\gamma_\textrm{eff} < 1$}, consistent with thermal equilibrium models \citep{Koyama2002}. The effective equation of state \mbox{$P\propto \rho^{\gamma_\textrm{eff}}$} is relevant for the radial stability of a filament \citep{Toci2015}. Filaments with \mbox{$\gamma_\textrm{eff} > 1$} are unconditionally stable. Isothermal (Ostriker) filaments have a critical point at infinite overdensity, and
filaments with $\gamma_\textrm{eff}<1$ become critical at finite overdensities. \citet{Toci2015} state that observed temperature ranges are smaller than those predicted by polytropic models and thus motivate non-thermal support, yet this conclusion seems to depend on the specific value of $\gamma_\textrm{eff}$ and on the assumed filament width.

\subsection{Radial profiles}\label{sec:radialprof}

The radial density profiles of filaments have received broad attention \citep[e.g.][]{Arzoumanian2011} due to the connection to the basic physics in hydrostatic cylinders directly inferred from three main observables: the central column density ($N_0$), the radial dependence (via the power-law index $p$), and the full-width half maximum (FWHM) of the column density profile of filaments (see definitions in \S\ref{sec:basicprop}). 

Filaments cover more than three orders of magnitude in column density, from the atomic and diffuse \HI~filaments (\mbox{$N_0 < 10^{20}$ cm$^{-2}$}) to the densest filamentary IRDCs and fibers (\mbox{$N_0 > 5\times10^{22}$ cm$^{-2}$}) (see Table~\ref{tab:filprop1}).
Resolved nearby filaments typically show higher $N_0$ values \citep[$\sim 6-7\times 10^{21}$~cm$^{-2}$;][]{Arzoumanian2011,Arzoumanian2019} than those detected in Galactic Plane surveys \citep[$\sim 2-3\times 10^{21}$~cm$^{-2}$;][]{Schisano2014}.  
Positive correlations are found between $N_0$ and the star-forming properties of resolved filaments (e.g. star-forming clumps and lines mass $m$), as well as between $N_0$ and the filament column density background $N_\textrm{bg}$ \citep{Schisano2014,Arzoumanian2019}.

The radial dependence of the column density profile provides insights about the stability, equation of state, and pressure confinement of filaments (\S\ref{Sec:env-pressure}). An isothermal hydrostatic cylinder has a radial density profile of $n\propto r^{-p}$ with \mbox{$p$ = 4} \citep{Stodolkiewicz1963,Ostriker1964}, although this dependence is rarely observed in filaments \citep{Pineda2010,Hacar2011,Fischera2012b,Monsch2018,Zucker2018}. Instead, most filaments tend to be well-described with Plummer-like profiles with exponents around \mbox{$p \sim$ 1.5-2.5} \citep{Arzoumanian2011,Palmeirim2013} which can be interpreted as a departure from equilibrium \citep[e.g.][]{Kawachi1998}. 

Shallow (\mbox{$p <$ 4}) profiles are expected for magnetized \citep{Fiege2000,Tomisaka2014,Kashawagi2021,Kirk2015} and externally pressurized filaments \citep{Fischera2012a}. Non-isothermal \citep{Recchi2013,Smith2014b} or polytropic \citep{Gehman1996b} equations of state as well as rotation \citep{Recchi2014} will also flatten the profile. 
%Nonetheless, shallow (\mbox{$p <$ 4}) profiles are expected in the case of non-ideal filaments under the presence of magnetic fields \citep{Fiege2000,Tomisaka2014,Kashawagi2021,Kirk2015} or external pressure \citep{Fischera2012a}, as well as filaments exhibiting non-isothermal \citep{Recchi2013,Smith2014} or polytropic equations of state \citep{Gehman1996b}, and rotation \citep{Recchi2014}. 
The observational distinction between these scenarios is however hampered by the limited column density contrast (i.e. $N_0$/$N_\textrm{bg}$, see above) in most resolved filaments \citep[see][]{Arzoumanian2019}, and measured values of $p$ may vary with location in a filament \citep{Fischera2012b}.

\begin{figure*}[ht!]
    \centering
    \includegraphics[width=\linewidth]{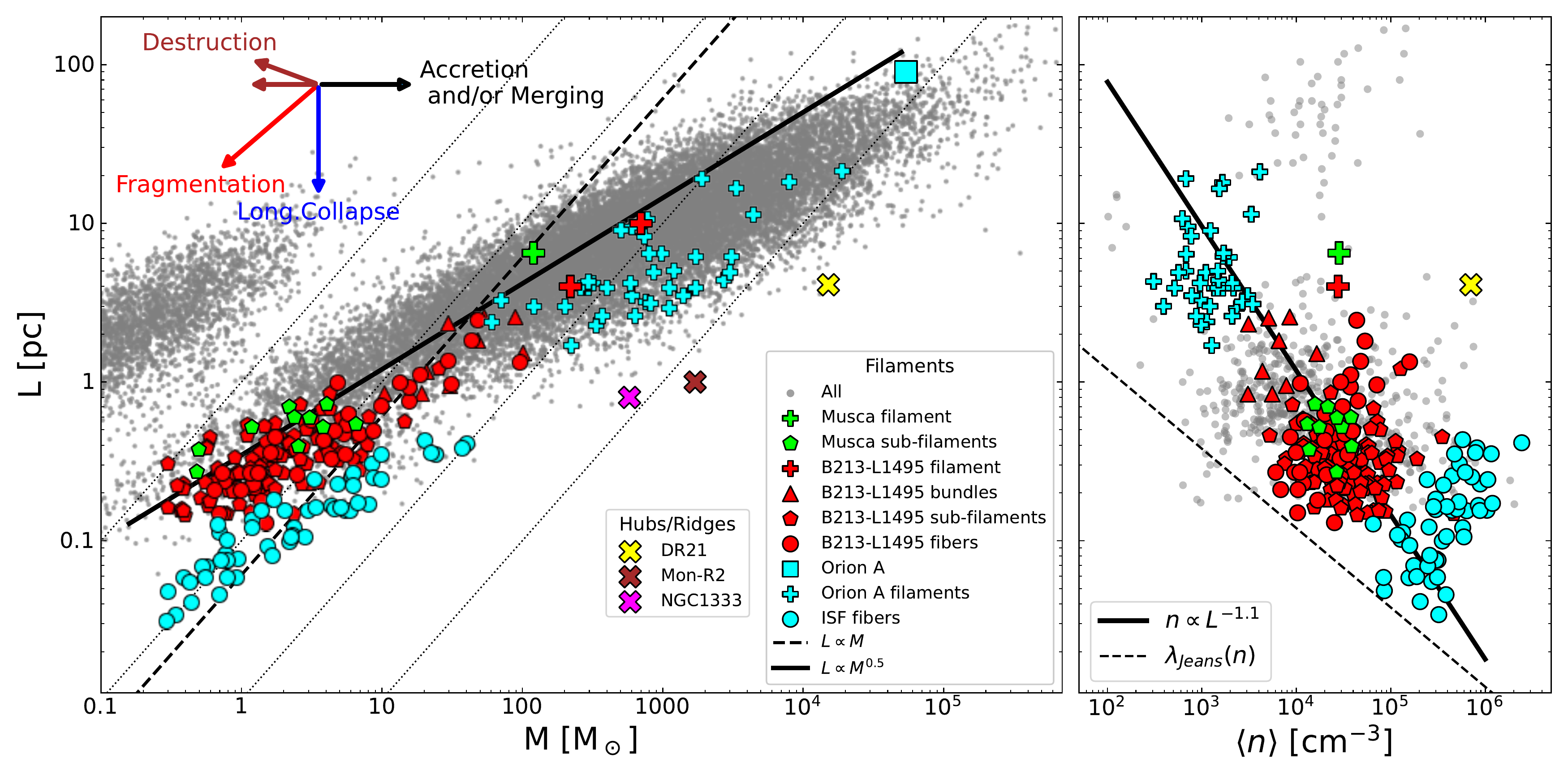}
    \caption{\small Mass-length {\bf (left panel)} and density-length correlations {\bf (right panel)} observed at different scales in Orion A (cyan), B213-L1495 (red), and Musca (green) in comparison to the entire filament population (grey; see also Fig.~\ref{fig:Main_prop}).
    Different symbols (see legend) denote the nested structure of these filamentary regions, from the largest to smallest being filamentary clouds (squares), parsec-size filaments (crosses), bundles (triangles), sub-filaments (pentagons), and fibers/sub-filaments (circles).
    The overall filament distribution follow the relations \mbox{$L \propto M^{0.5}$} and \mbox{$n \propto L^{-1.1}$} (black solid lines). The arrows in the left panel indicate the change in position under various processes.
    {\bf References \& tracers:}
    All: see Tables~\ref{tab:filprop1};
    Orion A -- cloud: FIR continuum \citep{Lombardi2014,Grossschedl2018}, filaments: $^{13}$CO \citep{Nagahama1998}, ISF fibers: N$_2$H$^+$ \citep{Hacar2018};
    B213-L1495 -- filament: C$^{18}$O \& FIR continuum \citep{Hacar2013,Palmeirim2013}, bundles \& fibers: C$^{18}$O \citep{Hacar2013}, sub-filaments: FIR continuum \citep{Arzoumanian2019};
    Musca -- cloud: extinction \citep{Kainulainen2016}, sub-filaments: FIR continuum \citep{Arzoumanian2019}; DR21 -- FIR continuum \citep{Schneider2010}; NGC1333 -- FIR continuum \citep{Hacar2013}; Mon-R2 -- FIR continuum \citep{Trevino-Morales2019}.
    }
\label{fig:hierarchy}
\end{figure*}

The dust-based \textit{Herschel} studies in local clouds suggest that
the distribution of inner widths has a mean {\it universal} value of about FWHM~$\simeq$~0.1~pc with a typical spread of a factor of 2 \citep{Arzoumanian2011,Arzoumanian2019,Palmeirim2013,Andre2014,Schuller2021}. Similar mean values, although showing larger dispersion, are also found in molecular observations of resolved filaments using tracers such as C$^{18}$O \citep[][]{Orkisz2019, Suri2019}. The measured filament widths may, however, depend on the filament scale and density (\S\ref{sec:widths}).
Larger widths of FWHM~$\sim$~0.26-0.34 pc are  reported in more distant ridges \citep[][]{Hennemann2012} and Galactic Plane filaments \citep[][]{Schisano2014} using \textit{Herschel}, and in nearby filaments observed in $^{13}$CO \citep{Panopoulou2014}.
At the opposite end, narrower widths of FWHM~$\lesssim$~0.05~pc are measured in denser sub-filaments and fibers using interferometric observations of dense molecular tracers \citep{Fernandez-Lopez2014,Hacar2018,Dhabal2018,Monsch2018}.
The inverse FWHM-$N_0$ relation predicted for pressurized filaments \citep{Fischera2012a,Heitsch2013a} has not been observed yet \citep{Arzoumanian2019,Suri2019}.
Resolution \citep{Schisano2014} and confusion \citep{Suri2019} may also affect the resulting FWHM measurements. 
The interpretation and robustness of a possible characteristic scale is thus a matter of on-going debate \citep[e.g. ][]{Panopoulou2017,Ossenkopf-Okada2019,Arzoumanian2019} and deserves a careful re-evaluation of both physical and observational arguments (see also \S ~\ref{sec:widths}).

\subsection{Filaments as hierarchical structures}\label{sec:hierarchy}

% Self-similar
The self-similar nature of molecular clouds \citep[e.g.][]{Falgarone1991} is imprinted in its internal {\it filamentarity}. Filaments at different scales are identified within the same region depending on resolution and density thresholds  \citep[e.g.][]{Schneider2011}. Orion A may serve as an example (Fig.~\ref{fig:hierarchy}).
At large-scales, Orion A appears as an elongated $\sim$~90~pc-long cloud \citep{Grossschedl2018} of more than 10$^5$~M$_\odot$ \citep[e.g.][]{Lombardi2014,Stutz2015}, comparable to some of the largest Giant Filaments detected in Galactic Plane surveys.
At intermediate scales, this cloud resolves into dozens of parsec-size filaments with masses between $\sim$~10$^2$ and 10$^4$~M$_\odot$ detected by single-dish observations of diffuse-gas tracers \citep[e.g.][]{Nagahama1998}, including the famous Integral Shape Filament \citep{Bally1987,Johnstone1999}. Denser molecular tracers show most of these filaments are heavily structured into sub-parsec-like sub-filaments with a few M$_\odot$ in mass at interferometric resolutions \citep[][]{Wiseman1998,Hacar2018,Monsch2018,Suri2019}.

As a result, the choice of scale may result in diverging interpretations of the same region. In Fig.~\ref{fig:hierarchy} (left panel) we display the position in the $M-L$ plane of different filamentary structures reported in Orion A (cyan), B213-L1495 (red), and Musca (green) using a variety of observational techniques and resolutions (see references in the figure caption).
When considered at parsec-scales, filamentary clouds (squares) and large filaments (plus symbols) systematically exhibit line masses $m\sim 50-300$~M$_\odot$~pc$^{-1}$ suggesting them to be super-critical ($f\gg1$) and therefore gravitationally unstable. Smaller bundles of fibers (triangles), sub-filaments (pentagons), and individual fibers (circles) within the same region show line masses closer to m$_{crit}$ ($f\sim1$) and appear to remain radially stable.
This observed hierarchy allows us to connect the near and far observational results (see \S\ref{sec:MandL}). According to the left panel of Fig.~\ref{fig:hierarchy}, the massive filaments reported in Galactic Plane surveys are likely resolved out into smaller (trans-) critical filaments similar to those detected in nearby clouds once observed at higher resolutions \citep[see also][]{Schisano2014}, questioning the interpretation of their stability in terms of their global line mass. Examples of this behaviour are already present in the literature \citep[][among others]{Henshaw2014,Chen2019,Shimajiri2019}.

Filaments at different scales effectively sample distinct gas densities in the ISM.
Fig.~\ref{fig:hierarchy} (right panel) compares the filament lengths with their effective gas densities, where the latter is calculated as \mbox{$\langle n \rangle$ = $m/(\pi (\frac{FWHM}{2})^2)$} in those cases where estimates of the filament FWHM were available. Overall, the mean gas density in molecular filaments increases with decreasing length. Filaments identified at parsec scales have average densities of $\sim10^3-10^{4.5}$~cm$^{-3}$ compared to the $\gtrsim10^5$~cm$^{-3}$ seen in sub-parsec filaments and fibers.
The relation between filament density and length for molecular filaments (black solid line, right panel in Fig.~\ref{fig:hierarchy}) is given by a least-square fit
\begin{equation}
  \langle n \rangle \simeq (1.2\pm 0.3)\times10^4 \left(\frac{L}{{\rm pc}}\right)^{-1.1\pm 0.1}\,{\rm cm}^{-3}
  \label{eq:dens_length}
\end{equation}
despite a relatively large ($\sim1$~dex) dispersion due to the choice of different tracers and mass thresholds.

The observed filament $M-L$ distribution shows a sub-linear relation
\begin{equation}
    L \propto M^{0.5 \pm 0.2}
\label{eq:ml-fit}
\end{equation}
(black, solid line, left panel). Qualitatively speaking this dependence is expected if large-scale filaments are actually networks of smaller structures (see §\ref{sec:m-l} for details).
Indeed, nearby filaments are 
characterized by some authors as multi-fractal structures embedded within the mono-fractal cloud gas \citep{Ossenkopf-Okada2019,Robitaille-2019,Robitaille2020,Yahia2021}. 
Beyond their morphological identification, this nested structure of {\it filaments within filaments} should be considered for the interpretation of both observations and simulations.

In general, this hierarchical structure could arise from both top-down fragmentation, as in \citet{Hacar2013}, and from a bottom-up process 
whereby gravitational collapse assembles larger filaments from 
gas containing pre-existing filamentary structure, 
as in \citet{Smith2014b}. In reality both large and small filaments may be formed simultaneously. Hierarchical structure formation would move the distribution diagonally downwards to the left of Fig.~\ref{fig:hierarchy} (left panel; see arrows) as opposed to other mechanisms such as longitudinal collapse (moving down), ablation (moving left) or stretching/shear (moving to the top left).

While individual clouds tend to follow a relatively tight $M-L$ relation (see color-coded symbols for each cloud), the normalization of the individual relations varies between clouds. The normalization is set by the line mass of the largest hierarchical structure, and it may be determined by the specific cloud environment (see \S\ref{sec:m-l}-\ref{sec:normalization}).

Ridges and hubs (crosses in Fig.~\ref{fig:hierarchy}) appear to depart from the mass-length and density-length correlations observed in most other filamentary clouds, showing approximately an order of magnitude larger masses and effective densities than filaments with comparable lengths.
The large contraction timescales (\S\ref{sec:timescales}) and small velocity gradients observed at parsec scales (\S\ref{sec:velfields}) rule out the longitudinal collapse of larger filaments as the main origin for these structures (i.e. a vertical displacement in the left panel of Fig.~\ref{fig:hierarchy}). Instead, local accretion and/or merging of multiple filaments \citep[see][]{Myers2009a,Kumar2020} or cloud-cloud collisions \citep{Inoue2013,Inoue2018,Fukui2021} will all move filaments rightward in the plot.
In combination with homologous collapse  \citep[see][]{Peretto2013}, this appears to be the more plausible mechanism for the origin of these massive filamentary clouds.

The origin of the mass-size relation $L\propto M^\alpha$ has been classically attributed to the hierarchical nature of molecular clouds imprinted by turbulence and fragmentation \citep[e.g.][]{Larson1981,Myers1983,Stutzki1998,Kauffmann2010b,Roman-Duval2010}. Clouds are expected to follow a mean mass-size relation such as \mbox{$L\propto M^{0.52}$} (or $M\propto L^{1.9}$), equivalent to the third Larson's relation \citep[\mbox{$n \propto L^{-1.1}$};][]{Larson1981}, also found in simulations \citep[e.g.][]{Beattie2019a}.
Shallower distributions are found in different cloud surveys \citep[$\alpha=0.27-0.4$;][]{Elmegreen1996} while steeper correlations and different normalizations are found in resolved clouds and clumps \citep[see][$\alpha\sim 0.5-0.7$]{Kauffmann2010a,Kauffmann2010b,Barnes2021}. %although part of these variations could be
These variations can be at least partly attributed to observational biases such as sensitivity and superposition \citep[e.g.][]{BallesterosParedes2019}. Our filament survey shows a similar global mass-size relation with approximately $L\propto M^{0.5}$ while individual clouds may present slightly steeper values.

\subsection{Magnetic fields}
\label{sec:bfields}

The properties of the magnetic field within and around filamentary structures are most commonly measured via polarized dust emission or stellar polarization, which both trace the magnetic field orientation as projected on the plane of the sky (POS). Our catalog contains estimates of the POS magnetic field strength, $B_\textrm{POS}$, derived from observations of polarized dust emission via the Davis-Chandrasekhar-Fermi (DCF) method \citep{Davis1951,Chandrasekhar1953}. This method relies on measuring the dispersion of polarization angles, as well as the density and the velocity dispersion of the gas, and has a number of limitations \citep[e.g.,][]{Ostriker2001,Heitsch2001b,Cho2016,Liu2021,Skalidis2021}. Furthermore, for several objects, the DCF results are obtained on a different scale (both larger and smaller) than the filament width. 
In addition, although magnetic fields have been mapped in many star-forming regions, the magnetic field strength is typically not measured for individual filaments.

\begin{figure}[ht]
    \centering
    \includegraphics[width=\linewidth]{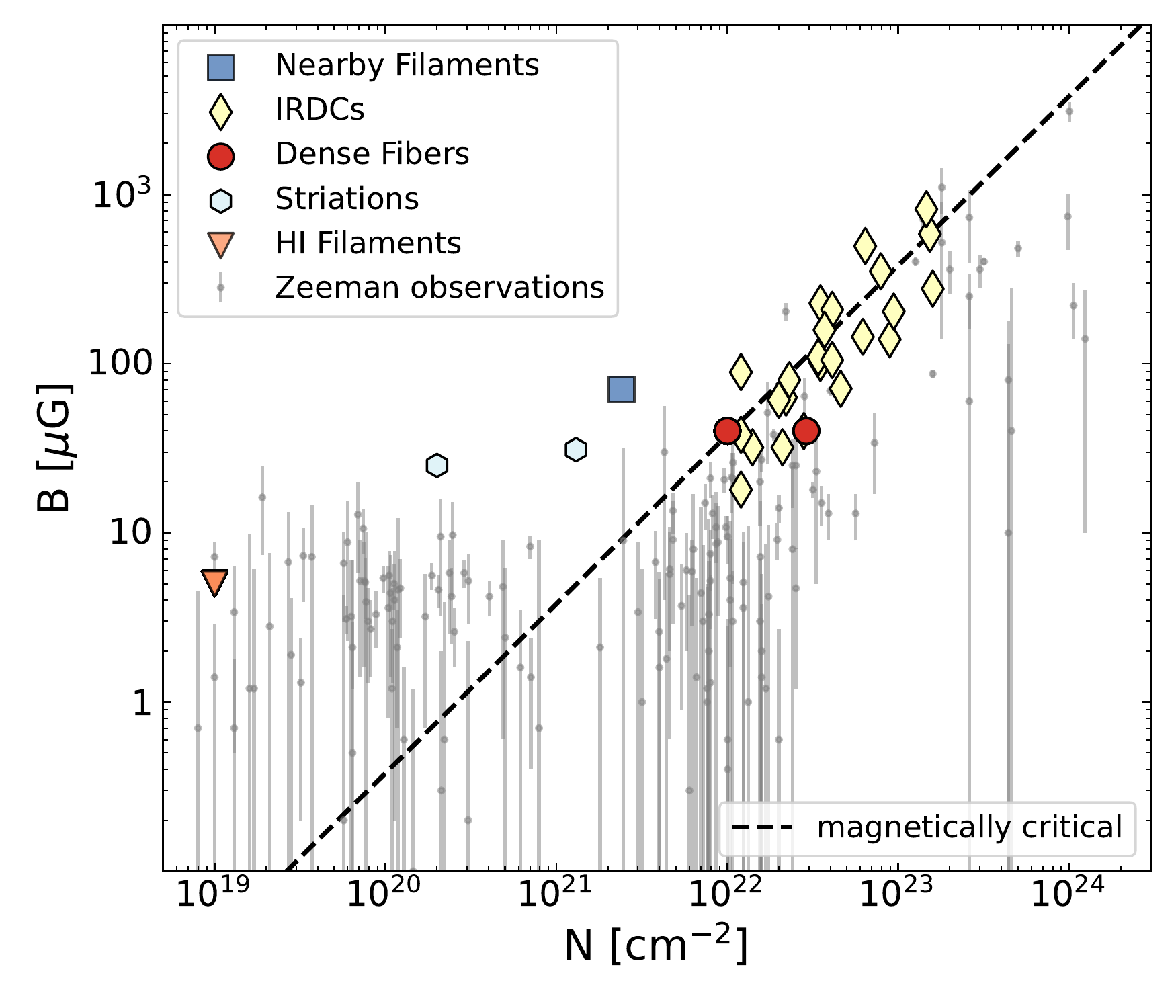}
    \caption{\small Magnetic field strength versus filament column density. The field strength in our catalog (colored, large symbols) is obtained via the DCF method (except the data point for HI). The dashed black line corresponds to a mass-to-flux ratio equal to the critical one. The data from our catalog seem to be broadly consistent with data from \citet{Crutcher2012} (gray dots and error bars).}
    \label{fig:B}
\end{figure}

Fig.~\ref{fig:B} shows the magnetic field strength against the column density of filaments in our catalog. 
Most of the measurements in the catalog refer to a single cloud from \citep[][yellow diamonds]{Arzoumanian2021}. For comparison, we also show the magnetic field measurements from \citet{Crutcher2012} (grey dots), which have been obtained using observations of the Zeeman effect (probing the line-of-sight magnetic field). The black dashed line indicates where the ratio of the mass to the magnetic flux %, $B/N$, 
is equal to the critical value \citep{Spitzer68,Mouschovias76}, which roughly matches the data points at high $N$ ($\gtrsim 10^{22}$~cm$^{-2}$). While the data in the filament catalog seem consistent with the Zeeman observations, more statistics are needed to quantify the agreement.

In addition to the field strength, the orientation of the magnetic field with respect to the filament axis can also inform us about the effect of the field on filaments \citep[e.g.][]{Nagai1998,Li2013,Palmeirim2013,Malinen2016,Panopoulou2016a,Planck2016XXXVSoler,Soler2017a,Jow2018,Fissel2019,Soler2019}. Low-column density \hi filaments (\S\ref{sec:hi}) tend to be aligned parallel to the local magnetic field orientation \citep{McClure-Griffiths2006,Clark2014,Clark2015}. 
%A statistical evaluation
Statistical evalulations of the relative orientation between the magnetic field and filamentary structures \citep{Soler2013,Jow2018} reveal 
a strong dependence with column density: magnetic fields are preferentially parallel to column density structures at low $N$ but can become preferentially perpendicular to filaments at high $N$. The transition in relative orientation appears at \mbox{$N \simeq 10^{21-22}$ cm$^{-2}$}, which approximately coincides with the column density at which the magnetic field strength starts to increase \citep[Fig.~\ref{fig:B} and][]{Crutcher2012}.
For an interpretation of these results we refer to \S\ref{sec:formation_bfield} (see also the chapter by Pattle et al. in this book).

The number of magnetic field measurements in the literature and thus in our catalog is rather limited compared to other properties of filaments (e.g. mass, length, velocity dispersion). Given the potential wealth of information stored in the magnetic field structure and strength on the evolution of filaments (see \S\ref{sec:formation_bfield}), a more detailed investigation of the magnetic field properties in individual filamentary objects is warranted.

\begin{deluxetable}{l|c|c|c|c|c|c|c}
\scriptsize
%\tablewidth{0pt}
\tablewidth{65em}
\tablecaption{FILAMENT FAMILIES (II): DYNAMIC PROPERTIES$^{(1)}$}
\renewcommand{\arraystretch}{.7}
\tablehead{Family$^{(\#)}$ & $\sigma_\textrm{tot}$ & $\sigma_\textrm{NT}$ & $m$ & $f$ & $\nabla v_{\textrm{LSR,}\parallel}$ & $B_0$   \\
 & (km~s$^{-1}$) & (km~s$^{-1}$) & (M$_\odot$~pc$^{-1}$) & $(=m/m_\textrm{crit})$ &  (km~s$^{-1}$~pc$^{-1}$) &  ($\mu$G)   }
\startdata
Nearby Filaments & 0.28 - 1.32 & 0.21 - 1.30 & 4.4 - 18.6 & 0.2 - 0.7 & 1.02 - 1.60 & 70 - 70   \\ 
Galactic Survey Fils. & 1.02 - 1.55 & 1.00 - 1.54 & 23 - 191 & 0.8 - 6 & 0.22 - 0.55 & ?   \\ 
IRDCs & 0.40 - 0.62 & 0.35 - 0.62 & 31 - 115 & 0.9 - 2.5 & 0.20 - 0.90 & 45 - 500 \\ 
Giant Filaments & 1.1 - 3.57 & 2.43 - 3.93 & 250 - 2513 & 15 - 35 & 0.04 - 0.08 & ?   \\ 
Dense fibers & 0.26 - 0.36 & 0.15 - 0.24 & 9.3 - 33.5 & 0.3 - 1.4 & 0.35 - 3.23 & 40 - 1000$^{(2)}$  \\ 
Striations & 0.71 - 0.90 & ? & 2.5 - 5.6 & 0.1 - 0.2 & ? & 26 - 30  \\ 
HI fibers & 0.88(?) & 0.65(?) & 0.05 - 0.15 & ? & ? & 5$^{(3)}$  \\ 
\enddata
\tablenotetext{(1)}{\small Values correspond to the interquartile [Q25\%,Q75\%] range for each measured quantity. $^{(2)}$ Inferred from the surrounding medium. $^{(3)}$ Standard $B_0$ value for the diffuse medium \citep[$n<10^3$~cm$^{-3}$,][]{Crutcher2010}.
(?) Unknown or poorly constrained. $^{(\#)}$ See Table~\ref{tab:filprop1} for references.
}
\label{tab:filprop2}
\normalsize
\end{deluxetable}

\subsection{Gas kinematics}
\label{sec:velfields}

\begin{figure*}[ht]
    \centering
    \includegraphics[width=\textwidth]{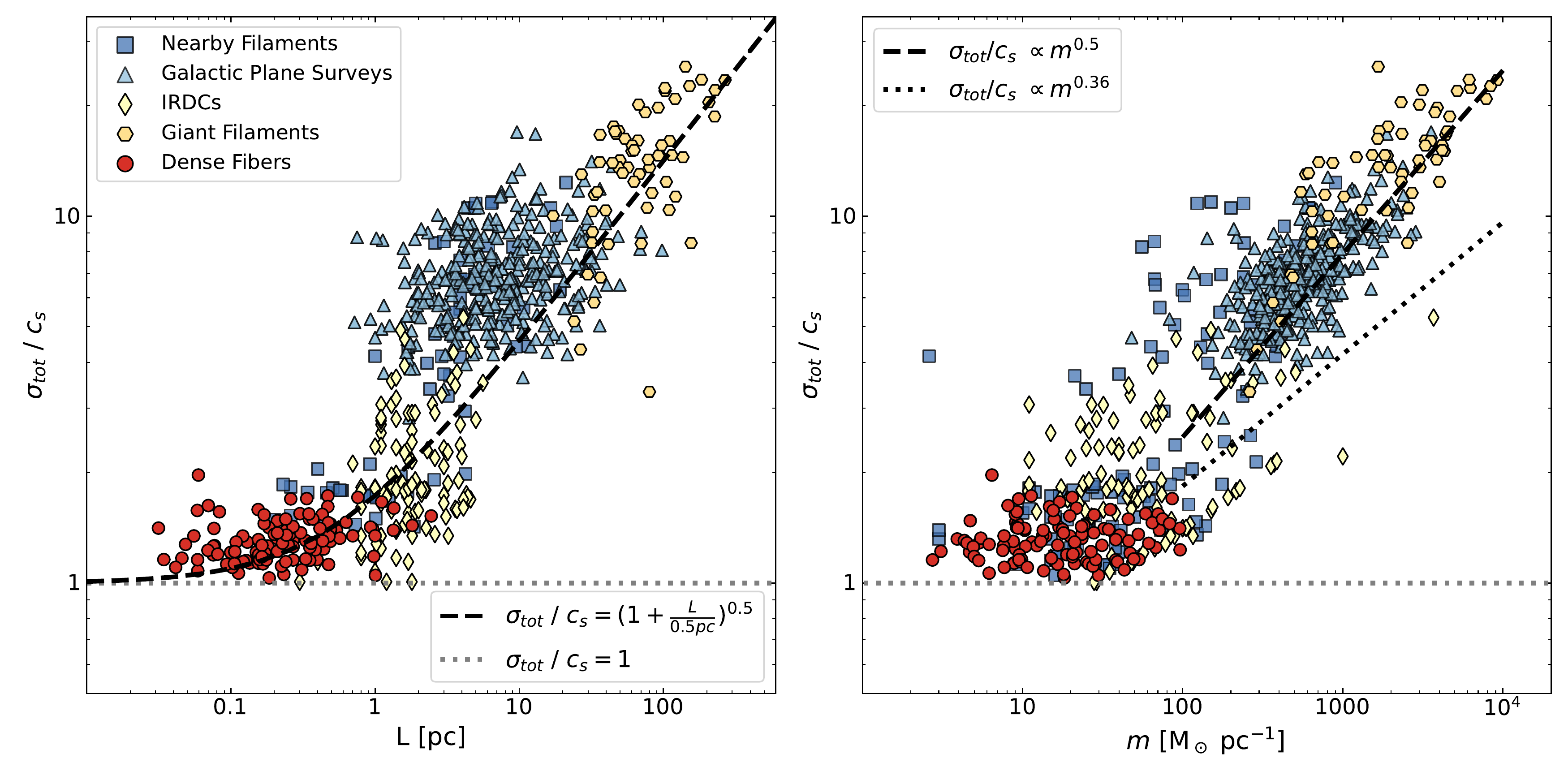}
    \caption{\small Velocity dispersion in units of the sound speed ($\sigma_\textrm{tot}/c_\textrm{s}$) as function of the filament length $L$ {\bf(left panel)} and line mass $m$ {\bf(right panel)}.
    The observed correlations of $\frac{\sigma_\textrm{tot}}{c_\textrm{s}}=\left(1+\frac{L}{0.5\mathrm{pc}} \right)^{0.5}$  (Eq.~\ref{eq:sigma}) and $\frac{\sigma_\textrm{tot}}{c_\textrm{s}}\propto m^{0.5}$  
    are indicated by dashed lines. For comparison, the previously reported $\sigma_\textrm{tot}/c_\textrm{s}\propto m^{0.36}$ by \citet{Arzoumanian2013} is also indicated in the right panel.
    In both plots, the $\sigma_\textrm{tot}/c_\textrm{s}=1$ is indicated by a horizontal line.
    \label{fig:kinematics}
    }
\end{figure*}

The gas velocity field at different scales within filaments can be characterized using both line-of-sight ($\sigma_\textrm{tot}$ and $\sigma_\textrm{nt}$) and POS ($\nabla v_\textrm{LSR}$) measurements (see definitions in \S\ref{sec:basicprop}).
We report the change of $\sigma_\textrm{tot}/c_\textrm{s}$ with the total filament length $L$ in the left panel of Fig.~\ref{fig:kinematics}. We use $c_\textrm{s}$ for each filament individually for normalisation. However, due to the small temperature variations found between individual filaments (see \S\ref{sec:pressure}), using a constant $c_\textrm{s}$ for all filaments would not change our overall findings.

Resolved nearby filaments and fibers show total velocity dispersions close to the sonic speed (\mbox{$\sigma_\textrm{tot}/c_\textrm{s} \simeq 1-2$}) \citep[e.g.][]{Hacar2011,Hacar2013,Arzoumanian2013}. On the other hand, larger Galactic Plane and Giant Filaments exhibit highly supersonic velocity dispersions  (\mbox{$\sigma_\textrm{tot}/c_\textrm{s}\gg 1$}) \citep[e.g.][]{Wang2016,Mattern2018}. The large dynamic range covered by our sample allows an empirical parametrization of the observed $\sigma_\textrm{tot}$-$L$ scaling in filaments as
\begin{equation}
    \frac{\sigma_\textrm{tot}}{c_\textrm{s}}=\left(1+\frac{L}{0.5\,\mathrm{pc}} \right)^{0.5}
\label{eq:sigma}
\end{equation} 
(dashed line in the left panel of Fig.~\ref{fig:kinematics}) with uncertainties of $\pm 0.2$~pc for the length normalization and $\pm 0.2$ for the exponent. 
The relative increase of $\sigma_\textrm{tot}$ with respect to $c_\textrm{s}$ can be attributed to an increasing contribution of non-thermal motions at larger scales
as \mbox{$\sigma_\textrm{nt}/c_\textrm{s} \propto L^{0.5}$},
reminiscent of the velocity dispersion-size relationship observed in the ISM \citep[][see also \S\ref{sec:dyn-dispersion}]{Solomon1987}. 

As shown in the right panel of Fig.~\ref{fig:kinematics}, a similar correlation is observed in terms of the line mass $m$. Filaments with $m < $~100~M$_\odot$~pc$^{-1}$ exhibit sonic-like $\sigma_\textrm{tot}$ values while increasing values are seen in targets with $m ~>$~100~M$_\odot$~pc$^{-1}$. The change of $\sigma_\textrm{tot}$ with $m$ was first investigated by \citet{Arzoumanian2013}, finding that
\mbox{$\sigma_\textrm{tot}/c_\textrm{s} \propto  m^{0.36}$} (dotted line in the right panel of Fig.~\ref{fig:kinematics}). Our new results indicate a somewhat steeper correlation with 
\begin{equation}
\sigma_\textrm{tot}/c_\textrm{s} \propto  m^{0.5}
\label{eq:ml-sigma}
\end{equation}
(dashed line). The interpretation of this $\sigma_\textrm{tot}-m$ correlation is, however, non-trivial (see \S\ref{Sec:env-accretion}) and subject to strong observational biases. While velocity dispersions in local filaments are obtained from resolved, single-velocity components of density-selective tracers, measurements in large filaments in Galactic Plane surveys typically make use of more diffuse tracers \citep[e.g. $^{13}$CO; ][]{Mattern2018}, and average large-scale motions (see below) as well as multiple velocity components blended within a beam \citep[e.g.][]{Sokolov2019}.

\begin{figure}[ht]
    \centering
    \includegraphics[width=\linewidth]{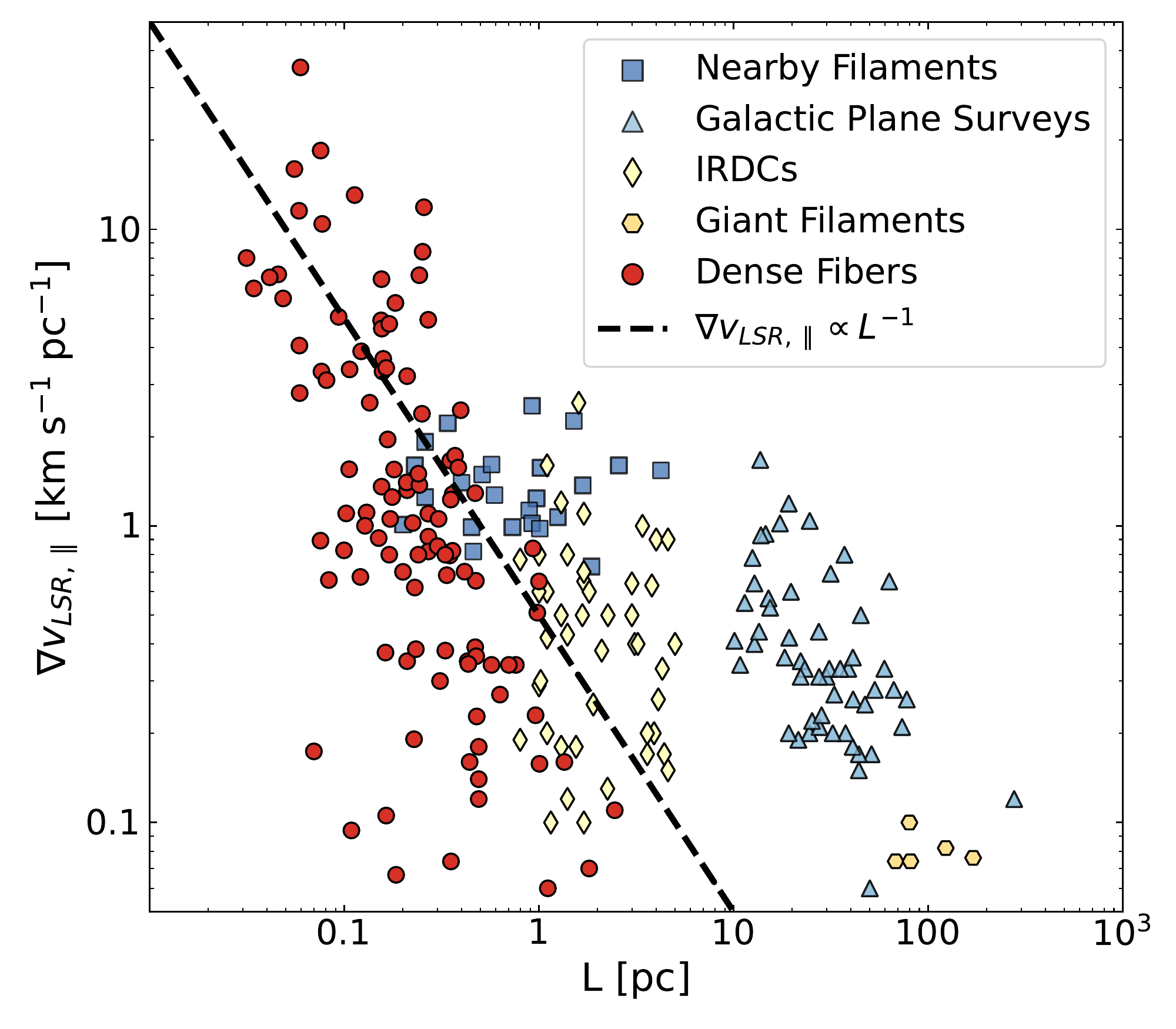}
    \caption{\small Parallel velocity gradients $\nabla v_{\textrm{LSR,}\parallel}$ as function of the filament length $L$. The velocity gradient follows roughly a $L^{-1}$ relation (black dashed line).
    % commented out as not necessary as given in legend.
    %Symbols are identical to Fig.~\ref{fig:Main_prop}.
    \label{fig:gradients}}
\end{figure}

Velocity gradients $\nabla v_\textrm{LSR}$ can be employed to quantify the bulk gas motions inside filaments. We find systematic variations of $\nabla v_\textrm{LSR}$ with $L$ as shown in Fig.~\ref{fig:gradients}. Nearby parsec-scale filaments show typical longitudinal gradients along their main axis of \mbox{$\nabla v_{\textrm{LSR,}\parallel} \sim$ 1-2 km s$^{-1}$ pc$^{-1}$} \citep[][for additional references see also \S\ref{sec:hubs}]{Loren1989b,Bally1987,Schneider2010,Kirk2013,Lu2018,Hu2021}.  While cloud-size IRDCs \citep[e.g.][]{Tackenberg2014,Sokolov2019} and large-scale filaments \citep{Ragan2014} show similar or even smaller $\nabla v_{\textrm{LSR},\parallel}$ values, shorter fibers (at scales $<$~1~pc) show larger gradients up to \mbox{$\nabla v_{\textrm{LSR},\parallel} \gtrsim$ 10 km s$^{-1}$ pc$^{-1}$} \citep{Lee2013,Hacar2018,Dhabal2018,Chen2020a,Chen2020c} following an approximate $\nabla v_{\textrm{LSR},\parallel} \propto L^{-1}$ relationship (dashed line).
For an interpretation of the observed correlation we refer to \S\ref{sec:coherence}.

We find no clear correlation between $\nabla v_{\textrm{LSR},\parallel}$ and $m$ (not shown) suggesting that most filaments do not  show a global homologous collapse (see also \S\ref{sec:timescales}) with only few potential cases reported in the literature \citep[e.g.][]{Zernickel2013}. Similarly, gradients associated with edge effects \citep{Burkert2004} appear to be an exception \citep{Dewangan2019}.
Local collapse can nonetheless become dominant, as converging velocity gradients at scales of $\sim$~1~pc are seen in hub-like structures \citep[e.g.][see also \S\ref{sec:hubs}]{Peretto2013,Kirk2013,Hacar2017a}

%%%%%%%%%%%%%%%%%%%%%%%%%%%%%%%%%%%%%%%%%%%%%%%%%%%%%%%%%%%

\section{FILAMENTS AS DYNAMIC STRUCTURES}\label{sec:theory}

\begin{figure*}
    \centering
    \includegraphics[height=0.42\textwidth]{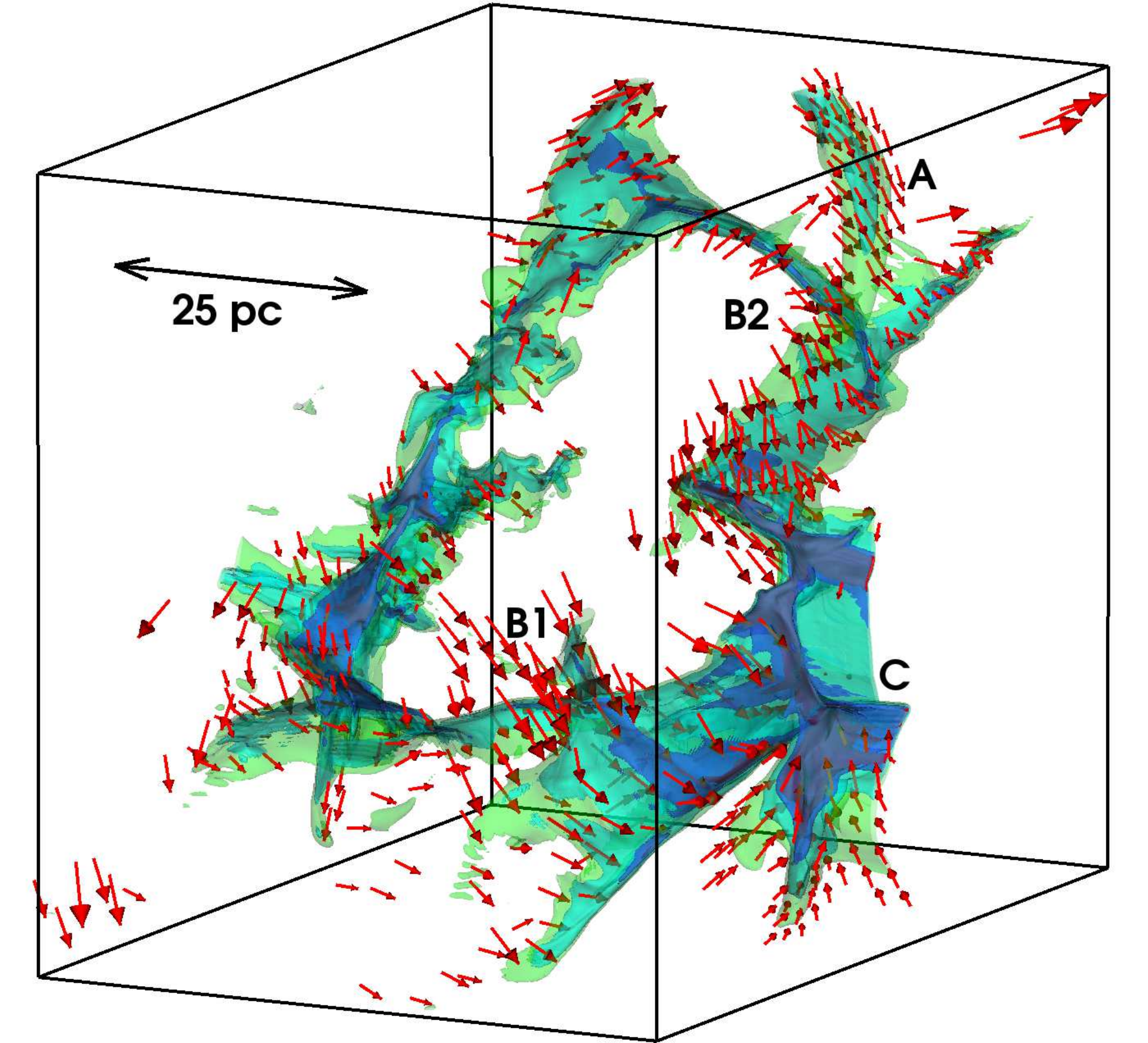}\hspace{0.07\textwidth}
    \includegraphics[height=0.42\textwidth]{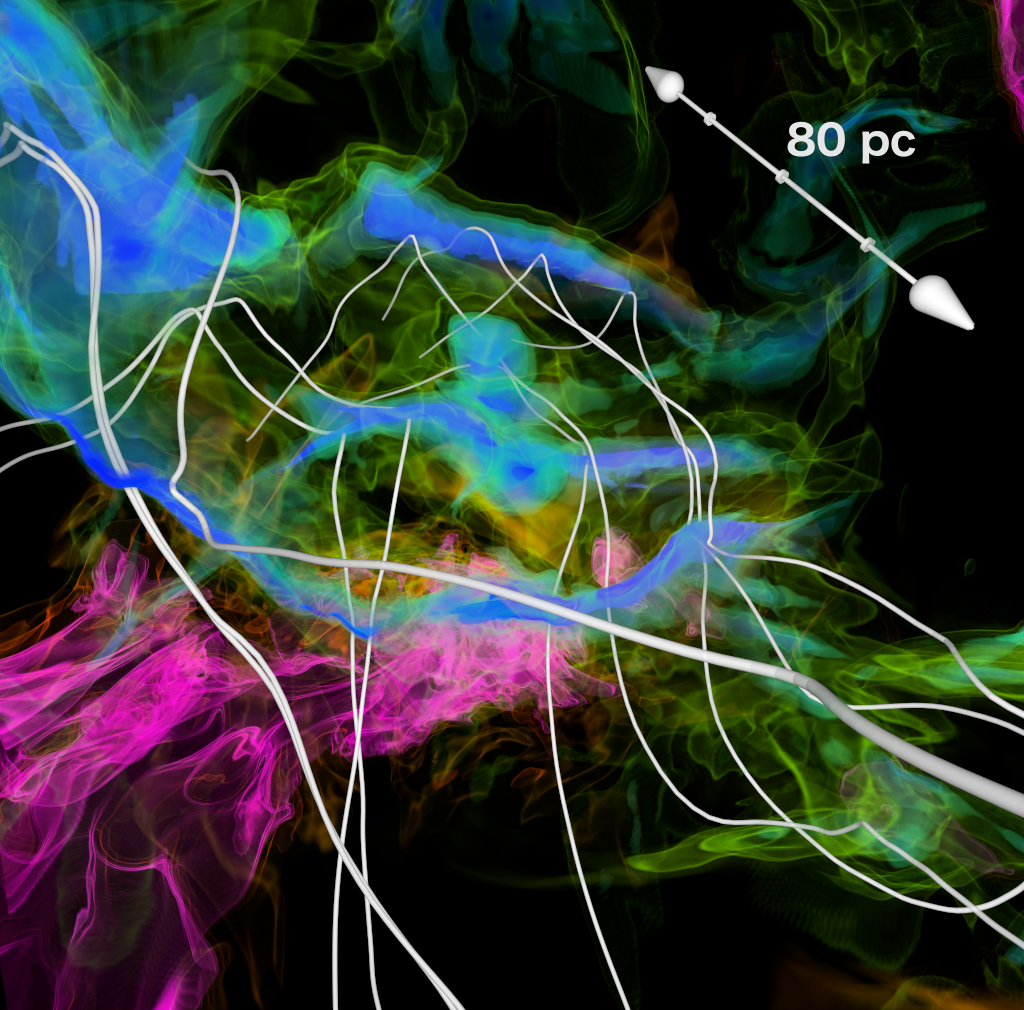}
    \caption{\small {\bf Left panel:} Filamentary density structure (depicted with various isocontours between $\sim$100 and 2000~cm$^{-3}$) of a simulated molecular cloud. The red arrows show the complex velocity field in the vicinity of the filamentary structures suggesting features like accretion along filaments (A), accretion onto filaments (B1, B2) and accretion onto a hub-like system (C). 
    {\bf Right panel:} Volume rendering of the density structure in a simulated molecular cloud showing cold and dense filaments (blue-green) as well as warm and diffuse gas in its surrounding (pink).
    Both the density structure as well as representative fields lines associated with filaments show the internal complexity of these objects, e.g. a kinked/bow-like field structure towards the spine of filaments.
    Simulations are taken from \citet{Seifried2017SILCC,Seifried2020}
    }
    \label{fig:simulations}
\end{figure*}

Measured line masses (\S\ref{sec:MandL}) and internal non-thermal motions (\S\ref{sec:velfields}) indicate that filaments are highly dynamical structures, significantly different from the hydrostatic case outlined in \S\ref{sec:basicprop}.
Furthermore, their hierarchical nature (\S\ref{sec:hierarchy}) suggests that 
filaments and their interstellar environments both exhibit a complex 3D structure that greatly departs from idealized cylinders, which complicates the interpretation of POS-projected observations. 
As illustrated in Fig.~\ref{fig:simulations}, simulations provide unique insights into the gas velocity field, internal mass distribution, and magnetic field structure of filaments in space and time. 
This section reviews the status of numerical simulations in shaping our current interpretation of the filamentary structure of the ISM.

\subsection{Filament formation mechanisms}
\label{sec:formation}

Filaments are definitionally anisotropic, which suggests that their formation is at least affected -- if not controlled -- by anisotropic agents, such as large-scale gas motions sweeping up material (\S\ref{sec:formation_sheet}), shock interactions induced by turbulence (\S\ref{sec:formation_turb}), magnetic fields (\S\ref{sec:formation_bfield}), feedback (\S\ref{sec:formation_feedback}), or global environments such as Galactic spiral arms (\S\ref{sec:formation_dynamics}). The abundance of anisotropic physics across a wide range of physical scales %agents
may explain why filamentarity is ubiquitous in the ISM.
The relative unimportance of thermal pressure at least on large scales in molecular clouds gives credence to gravitational contraction as a key ingredient in filament evolution \citep{VazquezSemadenietal2019}. 
Moreover, many mechanisms create networks of filaments rather than single objects, which contributes to the hierarchy of filaments throughout the ISM.

\subsubsection{Classical Formation via sheets}
\label{sec:formation_sheet}

In the classical picture, filaments can form in infinite sheets due to gravitational instabilities \citep[equivalent to the Jeans instability in 3D;][]{Tomisaka1983,Larson1985,Miyama1987a,Miyama1987b,Nagai1998,VanLoo2014}, once the wavelength of a perturbation is above a few times the thickness of the sheet. 
For a sheet in pressure equilibrium with a thickness of approximately the Jeans length, this results in filaments with near-critical line masses \citep{Miyama1987a}.

However, under realistic conditions, the planar shock fronts that form in the ISM (e.g. due to cloud-cloud collisions, stellar feedback-driven bubbles, or cascading turbulent motions) will not be infinite in their lateral extension. The behaviour of a finite sheet is markedly different \citep{Li2001,Burkert2004,Hartmann2007}. In general, the sheet will undergo global collapse, and gravitational focusing will lead to a concentration of mass at the edge of the sheet rather than fragmentation inside the sheet. Elongated sheets will collapse into filamentary structures with density enhancements at either end.

\subsubsection{Turbulence-induced formation} \label{sec:formation_turb}

In the absence of gravity, interactions of sheets can lead to the formation of filaments. This has been seen for more than two decades in simulations of turbulent boxes intended to represent parts of molecular clouds \citep[e.g.][and many more]{Porter1994, Padoan1999,Heitsch2001a,Padoan2001,Pudritz2013,Hennebelle2013,LiKlein2019,Federrath2016,Federrath2021}. 
%In this picture t
The turbulent motions must be both supersonic and super-Alfv\'enic in the magnetised case. 
They create shock planar fronts (sheets), which can collide obliquely to form a filament at the intersection with a turbulent wake behind it \citep{Pudritz2013}.

Turbulence is a natural mechanism for forming networks 
of filaments that trace the underlying gas velocity field. The turbulent energy cascade generates an internal hierarchy of filaments from the driving scale downwards, as seen in observations (see \S\ref{sec:hierarchy}). A turbulent origin has also been proposed to explain the intertwined fibers inside filaments, although their exact origin remains under debate: In a top-down scenario, the large-scale coherence and parallel organization of fibers with respect to the main filament reported in regions such as B213-L1495 suggest that these structures might form out of their parental filament \citep[fray and fragment scenario;][]{Tafalla2015,Clarke2017}. 
Alternatively, some simulations favour a bottom-up approach, with small subsonic filaments forming in a turbulent medium first, and then being gathered by collapse and shear flows \citep[fray and gather scenario;][]{Smith2016}. 

\subsubsection{Magnetic field-assisted formation}
\label{sec:formation_bfield}

Since magnetic fields exert anisotropic forces, they can assist the formation of filaments in various ways depending on the relative strength of the magnetic field with respect to the gas motions.

\textit{Shock compression of a magnetised clump:} If a magnetised clump is hit by a shock wave, the clump is swept up and becomes deformed in a kink-like fashion together with the magnetic field threading the clump. The resultant kinked magnetic field structure focuses the gas flow in the post-shock gas towards the convex point of the deformed shock, leading to the formation of an elongated structure \citep{Inoue2013,Vaidya2013,Inoue2018,   Fukui2021}. \citet{Abe2021} argue that this mechanism is dominant when
the unshocked gas is supersonic and super-Alfv\'enic, but becomes sub-Alfv\'enic behind the shock.

\textit{Stretching of overdensities:} 
Magnetic fields may affect filament formation most strongly in the diffuse atomic gas, and in the lower-density molecular gas as traced by e.g. $^{12}$CO, where gas motions are sub- to trans-Alfv\'enic. 
In such regions, magnetic fields introduce anisotropy because motions perpendicular to the magnetic field are impeded by the Lorentz force, while gas can move freely along the field lines. Thus overdensities will be stretched out by (turbulent) shear flows preferentially along field lines \citep{Hennebelle2013,Inoue2016,Xu2019}.
This mechanism may explain the observed predominance of
filaments that are parallel to the local magnetic field at column densities below $\sim10^{21-22}$ cm$^{-2}$ (\S\ref{sec:bfields}).
This is particularly pronounced for the \hi filaments, which are well aligned with the Galactic magnetic field (\S\ref{sec:hi}). These possibly originate from 
overdensities which are created by thermal instability and are subsequently stretched out \citep{Inoue2016,Wareing2021}.

\textit{Converging flow-assisted formation:} 
Around $N \simeq 10^{21-22}$ cm$^{-2}$ the preferred relative orientation of filaments and magnetic fields switches from parallel to perpendicular (see \S\ref{sec:bfields}),
a trend also seen in a number of simulations including turbulence and strong magnetic fields \citep[][but see also \citealt{Hennebelle2019} for a recent review]{Heitsch2001a,Ostriker2001,Li2004,Nakamura2008,Collins2011,Hennebelle2013,Soler2013,Chen2015,Li2015,Chen2016,Zamora2017,Mocz2018,LiKlein2019,Chen2020b,Seifried2020,Dobbs2021}.
The observation that high-column density filaments are preferentially perpendicular to the ambient magnetic field admits multiple interpretations for how magnetic fields, gas flows, and gravity might interact in the filament formation process.

\citet{Soler2017a} argue that the transition from a parallel to a perpendicular field orientation
indicates an anisotropic and converging (accretion) flow. The anisotropy is introduced by a dynamically important magnetic field on larger scales (causing aligned magnetic fields and filaments); the converging flow can be caused by either shocks \citep{Kortgen2020} or gravitational collapse \citep{Seifried2020}.
For the latter case,
the \textit{relative orientation} of the magnetic field is related to its \textit{relative strength} with respect to gravity and in the end to the formation of filamentary structures: for high (column) densities (\mbox{$n$ $\simeq$ 10$^2$ - 10$^3$ cm$^{-3}$}, \mbox{$N$ $\simeq$ 10$^{21-22}$ cm$^{-2}$}) gravity starts to dominate over the magnetic force \citep{Seifried2020,Ibanez2021}.
For the converging flow mechanism to work
the motions are globally sub-Alfv\'enic but become super-Alfv\'enic on smaller scales (if they were globally  super-Alfv\'enic, we would end up with the turbulence-induced formation mechanism described in \S\ref{sec:formation_turb}).
\citet{Chen2016} argue that locally super-Alfv\'enic motions alone are sufficient to create filaments perpendicular to the magnetic field. More recent simulations find that super-Alfv\'enic motions are necessary but not sufficient
\citep{Soler2017a,Seifried2020}.

A common conclusion of these analyses is that the geometrical configuration of the magnetic field with respect to the filament can provide insights into its formation mechanism. An example is
the bow-like structure of magnetic fields around filamentary structures, which is caused when a dynamically weak field is dragged along with the filament (see right panel of Fig.~\ref{fig:simulations}). This configuration is found in both simulations \citep[e.g.][]{Hartmann2002,Inoue2018,LiKlein2019,Reissl2021}, and observations \citep{Heiles1997,Heiles2009,Tahani2018,Tahani2019,Tahani2022}.
Helical magnetic fields, 
suggested to explain the radial stability of filaments \citep{Fiege2000}, have -- to the authors' knowledge -- to date not been found in any simulation and would also be difficult to distinguish from a bow-like structure \citep[][see also \citealp{Tahani2019} for a comparison between different field geometries around Orion]{Reissl2021}.

\subsubsection{Feedback-induced filaments}\label{sec:formation_feedback}

Filaments can also be shaped (and enhanced) by the pressure effects of mechanical and radiation feedback on pre-existing density inhomogeneities on the cloud surfaces and edges. Two types of filaments occur in the context of feedback. 
Filaments orthogonal to the wind direction will show asymmetric radial profiles \citep{Peretto2012} that can be enhanced by the effects of radiation \citep{Suri2019}.
On the other hand, pillar-like filaments elongated along the wind direction exhibit signatures of head compression \citep[e.g.][]{Bally2018} and Kelvin-Helmholtz instabilities \citep{Berne2010}. In both cases magnetic fields can play a crucial role in shaping the filaments \citep{Ntormousi2017}.
While initially constructive, the continuous action of these feedback mechanisms eventually destroys these filamentary structures via photo-dissociation, erosion, and ablation of their molecular material over time \citep[e.g.][]{Goicoechea2016}. 

These wind-shaped filaments are typically low-mass, are found at parsec scales within clouds, and are located in the vicinity of the high-mass stars that supply winds and radiation \citep[e.g.][]{Suri2019}. Examples of filaments sculpted by externally-applied forces have been classically reported in connection to HI expanding shells \citep[e.g. L204;][]{Heiles1988} and stellar winds \citep[e.g. Oph Streamers;][]{Loren1989b}. %The advent of 
High-sensitivity atomic \citep[{[C{\sc ii}]};][]{Pabst2020} and molecular observations \citep[CO;][]{Suri2019} %reveals the prevalence of
have identified these feedback-induced filaments at the walls of H{\sc ii} regions.

\subsubsection{Shear, Differential rotation and Galactic Dynamics}
\label{sec:formation_dynamics}

A common feature of the above models is the presence of shear due to turbulence, which enhances the filamentarity of gas structures.
However, the mechanisms discussed in \S\ref{sec:formation_turb} and \S\ref{sec:formation_bfield} usually only consider turbulent motions within molecular clouds. Simulations have shown that on larger scales ($\sim$100 pc or more) Galactic rotation can play an important role in forming the Giant Filaments discussed in \S\ref{sec:giantfilaments} \citep[e.g.][]{Wada2002,Smith2014a}. For example, \citet{Duarte2017} zoomed into the Giant Filament population of an isolated galaxy simulation to show that Giant Filaments form predominantly through galactic shear. 

Giant Filament formation is aided by instabilities such as the Parker instability \citep[e.g.][]{Kim2002,Machida2013,Kortgen2018}, and the thermal instability  which rapidly drives gas from the warm to the cold phase \citep{Hennebelle2007,Kim2010}.
These filaments are then further elongated by differential rotation as they move between arms as, due to the their length ($\gtrsim$~100~pc), the ends of the filaments are at different galactic radii \citep{Smith2020}. Within spiral arms, dense filamentary clouds known as `bones' such as the Nessie filament are proposed to form from gas compressed as it falls into the Milky Way spiral potential \citep{Goodman2014,Zucker2015,Zucker2019}.

\subsection{Environment}\label{sec:environment}

Filaments do not exist in isolation. Fig.~\ref{fig:thermal} (see also \S\ref{sec:pressure}) suggests that -- depending on the line mass -- the pressure of the ambient gas (\S\ref{Sec:env-pressure}) or accretion (\S\ref{Sec:env-accretion}) driven by the filament's gravitational potential might play a role in filament structure and evolution.

\subsubsection{External Pressure}\label{Sec:env-pressure}

Whether pressure confinement is required to prevent the filament from expanding depends on the effective equation of state \citep{Toci2015}, on the line mass \citep{Fischera2012a}, the global gravitational potential \citep{BallesterosParedes2009} and on the magnetic field geometry \citep{Fiege2000}. The Galactic midplane pressure of $\sim 2\times 10^4$~K~cm$^{-3}$ \citep{Cox2005,Blitz2006} provides a lower limit for the external pressure. Magnetic pressure \citep[][see also \citealp{Kalberla2016} for H{\small{I}} filaments]{Hennebelle2013} and ram pressure \citep{Audit2005,Duarte2017,Chira2018} of the ambient gas can provide additional pressure confinement.

Ram pressure confinement suggests dynamical or even transient filaments, since it entails a mass inflow leading to accretion \citep[][\S\ref{sec:formation}]{Heitsch2013b}. Ram pressure exerted by turbulent flows \citep{Audit2005,Duarte2017,Chira2018} in combination with magnetic stresses \citep{Hennebelle2013} can provide a confining pressure. 

Assuming a truncated Ostriker filament, the role of pressure confinement can be quantified by the steepness of the column density profile \citep{Fischera2012a} as a function of the criticality parameter $f$ (Eq.~\ref{eqn:deff}). Beyond the inner radius $R_\textrm{flat}$ (Eq.~\ref{eq:rflat}), filaments show column density profiles consistent with power-law slopes of $1.5<p<2.5$, though steeper slopes approaching $p=4$ have been measured (see \S\ref{sec:radialprof} and reference therein). It remains to be seen -- for example by correlating the power-law slope $p$ with the estimated central pressure -- whether flatter profiles indeed indicate stronger confinement as predicted by \citet{Fischera2012a}. The effective equation of state \citep{Toci2015}, turbulence \citep{Gehman1996b} or magnetic fields \citep{Fiege2000} may also play a role for shaping the filament profiles.

Pressure confinement can also be interpreted as radial compression of a filament at constant line mass. How a filament responds to an external pressure increase depends on the effective equation of state and magnetization. Isothermal filaments cannot be squeezed into collapse, since the internal and gravitational energy increase with the same dependence on the radius \citep{Fiege2000}, yet, pressurization affects the critical and maximum wave number for longitudinal fragmentation \citep[Eq.~\ref{eq:lambdacrit};][]{Nagasawa1987}. For equations of state softer than isothermal (\mbox{$\gamma_\textrm{eff} < 1$}), increasing the external pressure leads to collapse \citep{Toci2015} and affects the longitudinal fragmentation and resulting shape of cores \citep{Heigl2018a,Hosseinirad2018,Anathpindika2021,Motiei2021}.

\subsubsection{Accretion onto filaments}\label{Sec:env-accretion}

Line observations of the medium surrounding filaments show velocity gradients often roughly perpendicular to the main axis of the filament, suggestive of gas accretion onto the filament \citep{Schneider2010,Kirk2013,Palmeirim2013,Beuther2015,Dhabal2018,Williams2018,Shimajiri2019,Bonne2020,Chen2020a,Gong2021}. Such ongoing accretion onto filaments is also evident in 3D simulations of filamentary molecular clouds (see positions labelled B1 and B2 in the left panel of Fig.~\ref{fig:simulations}). Estimates of the associated accretion rates, $\dot{m}$, range from a few $10$ to a few $100$~M$_\odot$~Myr$^{-1}$~pc$^{-1}$ \citep{Kirk2013,Palmeirim2013,Schisano2014,Bonne2020,Gong2021}. These estimates are somewhat higher but still in rough agreement with simulations of filaments located inside molecular clouds giving accretion rates of a few %1 to a few 10~M$_\odot$~Myr$^{-1}$~pc$^{-1}$
to a few 10 M$_\odot$~Myr$^{-1}$~pc$^{-1}$\citep{Gomez2014,Chira2018}.
Variations in the observed accretion rates can in part be attributed to environmental effects, as $\dot{m}$ is found to correlate with the column density \citep{Heitsch2013a,Gomez2014}. This is supported by the observation that higher-mass filaments are preferentially embedded in environments with a higher background column density \citep{RiveraIngraham2016}.

The above accretion rates lead to accretion timescales
\begin{equation}
  \tau_\textrm{acc} \equiv \frac{m}{\dot m},
  \label{eqn:acctime}
\end{equation}
on which a filament would double its mass, as short as 0.1~Myr. 
The implications of such short accretion timescales for the evolution of filaments will be discussed further in \S\ref{sec:turb-frag} and \S\ref{sec:timescales}.

The geometry of accretion flows and their relation to the ambient magnetic field are still under discussion. \citet{Shimajiri2019} argue that the accretion in B211/B213 in Taurus occurs within a sheet-like structure \citep[see][for a simulation counterpart]{Chen2020c}, consistent with a formation scenario via sheets
(see \S\ref{sec:formation_sheet}).
\citet{Palmeirim2013} suggest that accretion occurs along the striations which are oriented perpendicular to the main filaments B211/213 but parallel to the local magnetic field \citep[see also e.g.][]{Hennemann2012}.
This relative orientation of magnetic fields and filaments was first proposed by \citet{Nagai1998} and subsequently confirmed in numerous observational and theoretical works (see \S\ref{sec:formation_bfield} and the review article of Pattle et al. in this book). This geometry suggests a magnetic field strong enough to channel the gas flow.

Mass accretion can drive turbulence in molecular clouds \citep{Klessen2010,Goldbaum2011}. While 3D simulations show that mass accretion leads to turbulence within a filament \citep{Seifried2015,Clarke2017,Heigl2018b}, %,Heigl2020},
\citet{Heigl2020} point out that the resulting turbulent pressure does not contribute to the stability of the filament since the turbulent pressure profile is flat. If accretion is driven by gravity, invoking accretion-driven turbulence as a stabilization mechanism against gravitational collapse conflicts with energy conservation. Turbulence driven by (gravitational) accretion implies that kinetic energy flows from the outside in, and is 
dissipated on a crossing timescale \citep{MacLow1998,Klessen2010}. If accretion-driven turbulence were to support the filament against collapse, the turbulent kinetic energy ($m\sigma_\textrm{nt}^2$) would have to grow faster than the gravitational energy ($|m^2G|$), a proposition difficult to reconcile with energy conservation, since it is the gravitational potential that drives the accretion in the first place. Moreover, the conversion efficiency between the kinetic energy of the inflow and the resulting turbulent energy is only a few percent \citep{Klessen2010}, resulting in trans- to mildly supersonic turbulence \citep{Seifried2015,Clarke2017,Heigl2018b,Heigl2020} in agreement with observations \citep[][see  \S\ref{sec:velfields}]{Hacar2011,Hacar2013,Arzoumanian2013}. Therefore, accretion cannot result in filament stabilization, but it may induce fragmentation (\S\ref{sec:fragmentation}).

\subsection{Filament Dynamics} \label{sec:dynamics}

\subsubsection{Coherence}
\label{sec:coherence}

One of the characteristic features of filaments is their coherence. This is illustrated in two ways 1) in the POS velocity gradients along their length, and 2) in the LOS velocity dispersion where there is a transition to sonic motions at small scales
(see \S\ref{sec:dyn-dispersion}). 

We first consider the large-scale longitudinal coherence shown in 
Fig.~\ref{fig:gradients} in the observed velocity gradients of our filament families. Two distinct classes emerge. On scales larger than $\sim1$ pc, average POS velocity gradients of only \mbox{1 km s$^{-1}$ pc$^{-1}$} or less are observed, independent of the filament's length. This is particularly apparent for the Giant Filaments on 100~pc scales. 
This large-scale longitudinal coherence is likely due to two factors. First, it is often an explicit condition in the algorithm used to identify filaments, particularly in the case of Giant Filaments \citep[e.g.][]{Ragan2014,Wang2015,Zucker2015,Wang2016}. Second, coherency may simply be a necessary feature of survival, as any larger gradient may lead to a rapid destruction of the filament. Simulations of galactic-scale filaments \citep[e.g.][]{Duarte2016,Smith2020} exhibit similar coherency in large-scale structures.

However, velocity gradients rise on smaller scales. Fig.~\ref{fig:gradients} shows that below scales of 1~pc gradients can exceed values of \mbox{$10$ km s$^{-1}$ pc$^{-1}$} and there is a tentative indication that the dispersion scales inversely with the filament length, \mbox{$\nabla v_{\textrm{LSR},\parallel} \propto L^{-1}$}. Such an effect could be due to a number of reasons. 
First, larger gradients along the filament could be a sign of global collapse, possibly also connected to fragmentation.
Second, feedback effects such as outflows may disrupt filaments locally on small scales. \citet{Smith2020} investigated the velocity gradients along filamentary cloud networks formed as a result of galactic-scale dynamics (their Figure~12) and found a similar distribution to that seen in Fig.~\ref{fig:gradients}. In this case the gradients at small scales were high in both i) a massive filament system undergoing significant collapse and fragmentation and ii) a highly turbulent filament system undergoing disruption by supernovae - suggesting that both mechanisms play a role in the velocity dispersion scaling behaviour.

An additional contribution to the increasing velocity gradient at small scales could arise from projection effects. 
Considering a hypothetical filament with length $L$ and constant velocity gradient $g$, one obtains \mbox{$L_\textrm{obs} = L \cos\alpha$} and \mbox{$g_\textrm{obs} = g \tan\alpha$}, where $\alpha$ is the angle between the filament and the POS. It follows that \mbox{$\alpha$ = arccos($L_\textrm{obs}/L$) = arccos($x$)} and thus $g_\textrm{obs}$~= $g$~tan(arccos($x$))~= $g$~$\frac{\sqrt{1-x^2}}{x}$ which, for \mbox{$x \ll 1$}, gives 1/$x$. Hence, the observed gradient would scale as 1/$x$, i.e. \mbox{$\nabla v_{\textrm{LSR},\parallel} \propto$ $L^{-1}$}.

\subsubsection{The $\sigma_\textrm{tot}-L$ relation and the sonic regime} \label{sec:dyn-dispersion}

Coherence is also observed in the velocity dispersion of filaments, but only on small scales (\mbox{$\leq$ 1 pc}) were \mbox{$c_\textrm{s} < \sigma_\textrm{tot} <2 c_\textrm{s}$} (Fig.~\ref{fig:kinematics}).
Such behaviour is reminiscent of the transition to sonic coherence seen in star forming cores at $\sim 0.1$ pc \citep[e.g.][]{Goodman1998}. Despite the complications in measuring filament widths (\S\ref{sec:radialprof}, \S\ref{sec:widths}), it is clear that the FWHM of filaments with $L < 1$ pc is typically less than $0.2$~pc (see \S\ref{sec:widths}), and hence in the sonic regime. Subsonic filaments are a natural consequence 
of supersonic turbulent shocks (see \S\ref{sec:formation_turb}). This would imply that the transition to coherence in molecular clouds occurs \textit{before} star-forming cores are formed, i.e. at the filament-formation stage \citep{Hacar2011}. Hence, the cores simply inherit their subsonic nature when they are formed by filament fragmentation (\S\ref{sec:fragmentation}).

However, $\sigma_\textrm{tot}$ becomes increasingly supersonic for larger filament lengths (Fig.~\ref{fig:kinematics}, Eq.~\ref{eq:sigma}). The scaling $\sigma \propto L^{0.5}$ is reminiscent of the second Larson relation \citep{Larson1981}. For a constant column density $\Sigma \propto R_\textrm{flat} $, such an exponent would be consistent with an energy equipartition of \mbox{$E_\textrm{k}/|E_\textrm{p}| = \sigma^2 R_\textrm{flat} / 2 G M \sim 1/2$} as expected from the Virial theorem.

Non-thermal motions are commonly interpreted as evidence for internal turbulence, potentially driven by large-scale accretion flows. Interpretation of such non-thermal motions is further complicated as different tracers are used (e.g. N$_2$H$^+$ for small, nearby filaments, and $^{13}$CO for large, distant filaments) resulting in a larger column of gas and more diffuse material being included in the measurement of larger structures. Given the hierarchical nature of the filamentary ISM, this can lead to superposition effects. %Nearer clouds which are more easily resolved,
Better-resolved nearby clouds break into \textit{networks} of filaments. If a network of small trans-sonic fibers  with different centroid velocities were averaged into a single beam, the 
measured linewidth would be supersonic \citep{Hacar2016b}.

\subsubsection{Filament rotation and angular momentum} \label{sec:perp-gradients}

In addition to their longitudinal gradients $\nabla v_{\textrm{LSR},\parallel}$ (\S\ref{sec:velfields} and \S\ref{sec:coherence}), orthogonal velocity gradients inside filaments are also reported for both fiber-like \citep{Fernandez-Lopez2014,Dhabal2018} and nearby filaments \citep{Palmeirim2013} with values of $\nabla v_{\textrm{LSR},\perp} \lesssim$~10~km~s$^{-1}$~pc$^{-1}$. These gradients could represent a continuation of the accretion onto the filaments from larger scales (\S\ref{Sec:env-accretion}). 
Organized, large-scale velocity gradients are observed in some cloud-size filaments %\citep[$\nabla v_{\textrm{LSR},\perp} \sim$~2~km~s$^{-1}$ pc$^{-1}$][]{Alvarez2021}
\citep[$\nabla v_{\textrm{LSR},\perp} \sim2~\mathrm{km}~\mathrm{s}^{-1} \mathrm{pc}^{-1}$,][]{Alvarez2021} 
and massive IRDCs %($\nabla v_{\textrm{LSR},\perp} >~20~\mathrm{km}~\mathrm{s}^{-1} \mathrm{pc}^{-1}$;, \citet{Beuther2015}). 
\citep[$\nabla v_{\textrm{LSR},\perp} > 20 $~km~s$^{-1}$ pc$^{-1}$; ][]{Beuther2015}.

%Such
Turbulence simulations also show filament velocity gradients \citep{Smith2016,Chen2020c}.
Orthogonal gradients may be related to filament formation mechanisms such as self-gravity or turbulent shocks, which can be distinguished by comparing the transverse kinetic and gravitational energies at different radii \citep{Chen2020c}.  
Velocity gradients across filaments could also be interpreted as rotation along the long axis of the filament \citep{Levshakov2016,Alvarez2021}.
Rotation can provide additional support against collapse \citep{Recchi2014}
and may play an important role in determining the angular momentum of cores formed within them \citep{Hsieh2021}. Alternative explanations for perpendicular velocity gradients include magnetic reconnection events from clump collisions with misaligned magnetic fields \citep{Kong2021}.

\subsubsection{Accretion flows and velocity oscillations along filaments} \label{sec:dyn-acc}

The observed longitudinal velocity gradients discussed in \S\ref{sec:velfields} and \S\ref{sec:coherence} often extend coherently along large parts of the filament, with values from \mbox{$\sim$0.5 km s$^{-1}$ pc$^{-1}$} to a few \mbox{10 km s$^{-1}$ pc$^{-1}$}.
In particular in hub systems, accretion flows appear directed towards the intersection point, supporting their gravitational origin (see \S\ref{sec:hubs} and the location labelled C in the left panel of Fig.~\ref{fig:simulations}).
Together with the radial extent and the local density of the filament, velocity gradients
enable estimation of the accretion rates at which cores accumulate mass. Accretion rates estimated  
from the aforementioned observations cover a wide range from 10$^{1}$ -- 10$^{4}$ M$_\odot$~Myr$^{-1}$.

The mass growth of cores via accretion along filaments was first proposed by \citet{Gehman1996a} and studied subsequently in a number of numerical and theoretical works 
\citep{Balsara2001,Banerjee2006,Myers2009b,Smith2011,Smith2016,Gomez2014,Gomez2018,LiKlein2019}. 
Such an accretion flow along a filament is also visible in the left panel of Fig.~\ref{fig:simulations} at position A.
\citet{Gomez2014} suggest that the accretion flow \textit{along} the filament is fed by the accretion \textit{onto} the filament 
(such an accretion flow conversion might occur between location B1 and C in the left panel of Fig.~\ref{fig:simulations}). Due to mass conservation, the accretion rate increases along the filament towards the central core reaching values up to 10$^{-4}$~M$_\odot$~yr$^{-1}$. 
Evidence for similar accretion flow patterns has been seen in observed filaments \citep{Schneider2010,Kirk2013,Chen2020a}.

Besides coherent velocity gradients over the entire length of a filament, velocity gradients may take the form of oscillations along the filament spine \citep{Heiles1988,Loren1989b,Hacar2011,Henshaw2014,Hacar2017b,Barnes2018,Liu2019,Sokolov2019}. 
According to \citet{Henshaw2020}, velocity oscillations are present at different scales in the ISM, from sub-parsec to kpc scales.
Indeed, theoretical studies by \citet{Gritschneder2017} have shown that long-lived, stable oscillations can be driven in hydrostatic filaments due to the action of gravity upon small bends within the filament. 
Large-scale velocity oscillations have been proposed to also produce stellar ejections via a slingshot mechanism under the presence of helical magnetic fields \citep{Stutz2016}.

In a few tantalising cases velocity gradient and density oscillations appear to have the same wavelength $\lambda$ but be systematically shifted by $\lambda/4$ \citep{Hacar2011,Henshaw2016}. This could be evidence for accretion flows within filaments onto cores, which in turn suggests gravitational instabilities as the origin for these motions \citep{Gehman1996a,Heigl2016}.
However, such a ``dephased'' behaviour between the velocity and density perturbations is not observed in all cases \citep{Tafalla2015}. As stated by \citet{Gehman1996a}, this could be due to turbulent motions which superimpose a different velocity pattern on the gravitational flow thus breaking the correspondence of velocity and density oscillations.

\subsection{Fragmentation}
\label{sec:fragmentation}

Accretion onto and along filamentary structures both increase the gas density, either locally (e.g. at the intersection points of filaments in hub systems) or globally along the entire length of the filament. This directly influences the instantaneous fragmentation properties by decreasing the Jeans length of the structures. 
Furthermore, as discussed in \S\ref{sec:dynamics}, filaments are turbulent, and supersonic compression creates local over-densities and thus seeds fragmentation 
before the turbulent energy decays \citep{MacLow1998,MacLow1999}. Hence, fragmentation most likely does not occur in the perfectly smooth hydrostatic filaments as described in \S\ref{sec:basicprop}, but instead within a hierarchy of structures which have complex motions.
The end point of this fragmentation process is the formation of dense cores that may go on to form stars (see also \S\ref{sec:starformation}). Does filamentary fragmentation determine the properties of these cores?

\subsubsection{Edge fragmentation}
\label{sec:edge-frag}

Theoretical studies on the evolution of quiescent filaments of finite length have shown that gravitational focusing causes an enhanced collapse at either end of the filament \citep{Bastien1983,Burkert2004,Pon2011,Pon2012,Toala2012,Clarke2015,Hoemann2021}.
The timescale for the filament to fully collapse along its long axis is proportional to the free-fall time, \mbox{$\tau_\textrm{ff} = \sqrt{3 \pi/(32 G \rho_0)}$}, modified by the aspect ratio \mbox{$A~\equiv L/ \textrm{FWHM}$} of the filament \citep{Pon2012,Toala2012},
\begin{equation}
  \tau_\textrm{long}=\frac{\sqrt{32A}}{\pi}\tau_\textrm{ff}=1.9 A^{1/2}\left(\frac{n_0}{10^3 \, \mathrm{cm}^{-3}}\right)^{-1/2}\,\mathrm{Myr}.
  \label{eqn:tau_edge}
\end{equation}
The actual onset of edge collapse, however, should be observable much earlier \citep{Seifried2015}. 

However, observationally there are only a few filaments observed with clear indications for a pure edge fragmentation \citep{Zernickel2013,Friesen2016,Kainulainen2016,Dewangan2019,Yuan2020,Cheng2021,Gong2021}.
Indeed, moderate density perturbations, supercritical line masses \citep{Seifried2015}, or localised radial inflow \citep{Heigl2022} can effectively mask this fragmentation mode by causing fragmentation inside the filament proceeding on a comparable or even faster timescale than at the edges. Moreover, any tapering of the edge will reduce the extreme accelerations caused by gravitational focusing  \citep{Li2001}. The paucity of observational evidence for edge fragmentation is thus not surprising.

In summary, given the turbulent nature of the ISM, it would thus be difficult to find a filament that shows edge collapse exclusively, 
specifically if the presence of protostars is used as an indicator. 
Line-of-sight velocity information may provide a clearer diagnostic tool \citep{Heitsch2013a}.

\subsubsection{Fragmentation spacing}
\label{sec:frag-spacing}

Isothermal cylinders can fragment for perturbations at wavelengths 
$\lambda$ exceeding $\lambda_\mathrm{crit}~\simeq 4R_\mathrm{flat}$ (Eq.~\ref{eq:lambdacrit}). Thus \mbox{(sub-)critical} filaments ($f\le1$) are expected to gravitationally fragment to form equally spaced cores with typical masses of \mbox{$M_\mathrm{core} \simeq m_\mathrm{crit} \lambda_\mathrm{crit}$}
or \mbox{$\sim M_\mathrm{Jeans}$}. The separation of these cores is set by the fastest growing unstable mode $\lambda_\mathrm{max}$, which is equivalent to about 2 times the critical wavelength \citep{Nagasawa1987,Larson1985,Inutsuka1992}. The $e$-folding time for the fastest growing mode at $\lambda_\mathrm{max}$ is
\begin{equation}
  \tau_\mathrm{frag} = \frac{3}{2\sqrt{\pi G\rho_0}} = 1.66\left(\frac{n_0}{10^3\,\,\mathrm{cm}^{3}}\right)^{-1/2} \,  \mathrm{Myr}.
  \label{eqn:longfrac}
\end{equation}
The growth rate $\propto \tau_\mathrm{frag}^{-1}$  vanishes for 
perturbations with increasing $\lambda$
for an isothermal, unmagnetized filament, but it can remain finite for a non-isothermal equation of state \citep{Gehman1996b,Hosseinirad2018,Motiei2021}. Axial magnetic fields increase the factor in Eq.~\ref{eqn:longfrac} from $1.66$ to $1.84$ \citep{Nagasawa1987}. A supercritical filament will collapse faster radially than it can fragment longitudinally, unless it is seeded with sufficiently high perturbations \citep[][their Fig.~11]{Inutsuka1997}.

Observed core spacings rarely align well with these idealised models. Frequently filaments do not exhibit regular spacing at all \citep[e.g.][]{Kainulainen2017,Mattern2018Nessie}, or -- when regular spacing is observed -- it is not at $\lambda_\mathrm{max}$ but at $\lambda_\mathrm{crit}$ \citep{Tafalla2015} or shows a significant scatter \citep{Beuther2021}. First-order diagnostics such as the median core-to-core separation may omit information on the clustering of fragmentation and can give varying results depending on the statistical method \citep[e.g.][]{Clarke2019}. Indeed there is evidence that the fragmentation spacing may be hierarchical, with smaller chains of regularly spaced cores consistent with Jeans fragmentation embedded within clumps whose spacings are determined by the large-scale, gravitationally-unstable modes of the filament \citep{Teixeira2016} or the dominant turbulent mode \citep[][but see also \S\ref{sec:turb-frag}]{Seifried2015,Clarke2017}. \textit{Herschel} filaments have median spacings similar to the filament width \citep[e.g.][]{Zhang2020, Konyves2020}, while in other sources they only become compatible with idealised models if an additional source of support, such as turbulence, is included beyond the thermal case \citep{Jackson2010,Busquet2013,Lu2014}. 
Analysing core separations requires care, as in addition to the potential presence of more than one characteristic fragmentation length scale, a significant number of cores ($N \geq 20$) is required to obtain statistically meaningful results \citep{Clarke2019}.

\subsubsection{Fragmentation in a turbulent environment}
\label{sec:turb-frag}

The %afore-
discussed results depict a complex picture of filament fragmentation. This could indicate an effect of accretion flows on the fragmentation behaviour of filaments, in particular as the timescales for fragmentation of (Eq.~\ref{eqn:longfrac}) and accretion onto a filament (Eq.~\ref{eqn:acctime}) can be comparable. Indeed, numerical simulations show that ongoing accretion of material onto the filament affects gravity-induced fragmentation \citep{Clarke2016}, which deviates significantly from %those of
fragmentation of initially quiescent, pre-existing filaments \citep{Inutsuka1992}. Moreover, due to the turbulent energy cascade, the accreted material is expected to be moderately turbulent as shown by the filament velocity dispersions (see \S\ref{sec:velfields} and \S\ref{sec:dyn-dispersion}). Simulations show that in the trans- to mildy supersonic regime, fragmentation is shifted from a gravity-dominated to turbulence-dominated process, where 
fragment locations are seeded by turbulent motions \citep{Seifried2015,Clarke2017}. Both these studies show that the observed fragmentation spacing strongly depends on the dominant turbulent length scale. Furthermore, \citet{Chira2018} show that filament fragmentation might set in %already
before the critical line mass is reached, further emphasising the importance of turbulent motions and moderate density enhancements \citep[see also][]{Seifried2015} during the formation phase of filaments. 
Consequently, core formation can take place at any point along the filament, in agreement with observations (\S\ref{sec:frag-spacing}).

%%%%%%%%%%%%%%%%%%%%%%%%%%%%%%%%%%%%%%%%%%%%%%%%%%%%
\section{TOWARDS A PHYSICAL INTERPRETATION OF FILAMENTS}\label{sec:phydes}

\subsection{The Mass - Length ($M-L$) scaling relation}\label{sec:mass_size}
\label{sec:m-l}

Observed filaments show a wide range of values for the line mass $m$ (Fig.~\ref{fig:Main_prop}). Close inspection of the data indicates that there might not be a single powerlaw \mbox{$L \propto M^{\alpha}$} describing the entire distribution
(left panel of Fig.~\ref{fig:hierarchy} and Eq.~\ref{eq:ml-fit}).
In the following we speculate on the origin of the observed $M-L$ relation.

\subsubsection{The scaling and scatter of the $M-L$ relation}\label{sec:origin-ml}

The observed $M-L$ relation may originate from the hierarchical structure of the filamentary ISM (\S\ref{sec:hierarchy}).
To build intuition, we 
consider how a parent filament may populate the $M-L$ plane with 
filamentary substructures.
Simple longitudinal 
segmentation of the parent filament (neglecting longitudinal contraction of the fragments, see \S\ref{sec:edge-frag}), in combination with mass conservation, necessitates $L \propto M$, as each sub-filament will by definition have the same line mass. This is significantly steeper than the observed scaling relation $L \propto M^{\alpha}$ with \mbox{$\alpha \simeq 0.5$} (Eq.~\ref{eq:ml-fit}).

Assuming a top-down (fray and fragment) filament-formation scenario \citep{Tafalla2015}, we consider the following toy model to explain the observed scaling. We note that a similar argument can be made for a bottom-up (fray and gather) scenario \citep{Smith2016}.
Picture a straight filament of length $L_0$ and mass $M_0$ embedded in a turbulent cloud of the same size. Turbulent motions will bend and stretch the filament. For self-gravitating filaments, any bends in the filament will enhance local gravitational forces, resulting in fragmentation. Assuming the filament's length $L_0$ does not change, and equating the turbulent bending with a random walk process, the filament will fragment into $N$ sections of length $L=L_0/\sqrt{N}$ each. If the filament does not accrete an appreciable amount of mass during this random walk process, each fragment has a mass of $M=M_0/N$.
Eliminating $N$ thus leads to
\begin{equation}
 L \simeq L_0\left(\frac{M}{M_0}\right)^{1/2} \, ,
\label{eq:randomwalk}
\end{equation}
similar to the relation found in Fig.~\ref{fig:hierarchy}. Hence, 
the random walk toy model appears 
to be a viable first-order
approximation of the substructure generated by turbulent motions.

We add, however, some words
of caution: First, this picture certainly oversimplifies the hierarchical, filamentary structure inside the parental object.
Second, if random column density perturbations,
e.g. due to observational noise or choices in the filament identification algorithm \citep[e.g.][and \S\ref{sec:algorithms}]{Panopoulou2014}, dominate the identification of sub-filaments, this could also mimic a random walk behaviour.
Thirdly, not all of the mass $M_0$ might be assigned to the smaller-scale filaments, e.g. due to different density dependences of observational tracers, causing the relation in Eq.~\ref{eq:randomwalk} to become shallower.

\begin{figure}[ht!]
    \centering
   \includegraphics[width=0.92\linewidth]{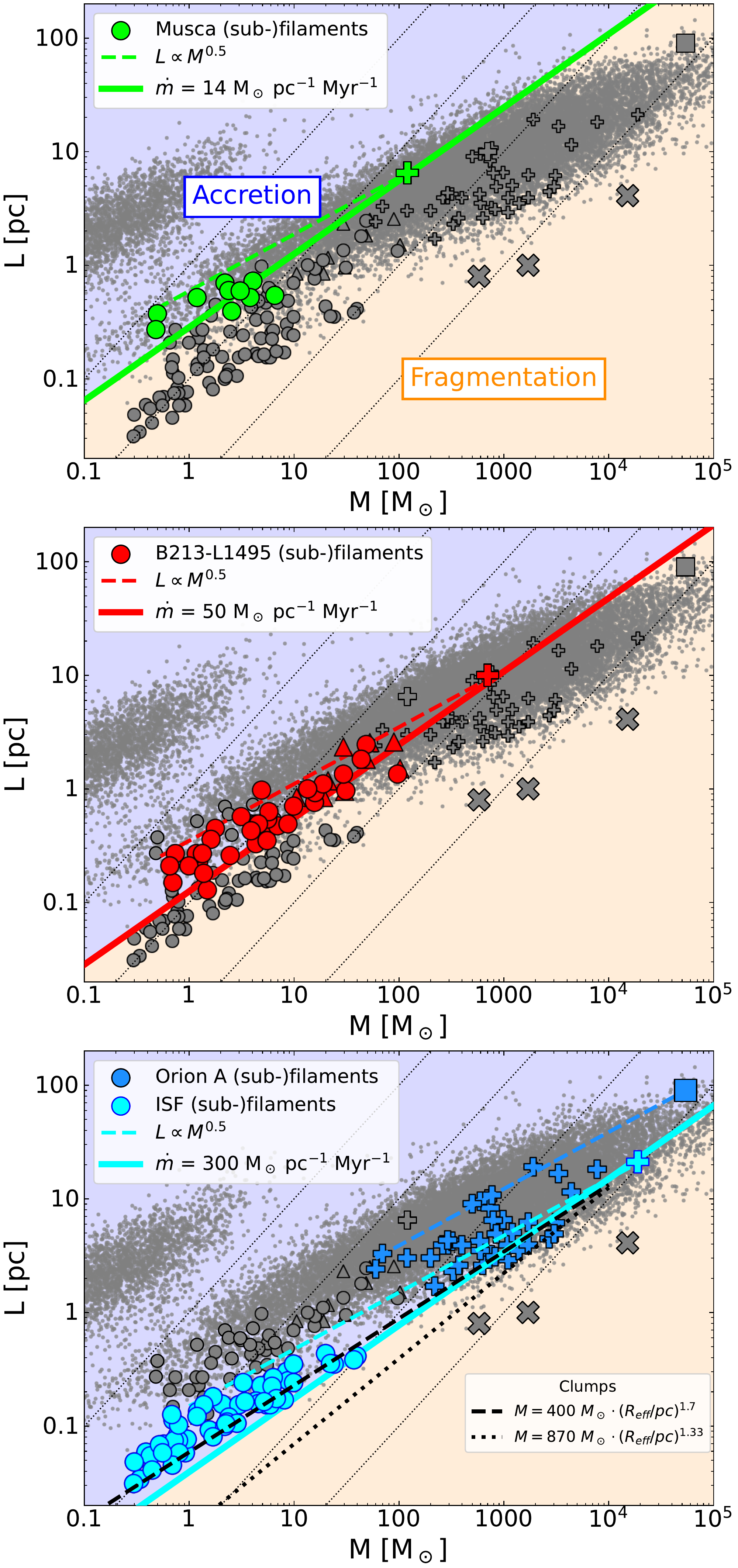}
    \caption{\small Top to bottom: Regimes of dominant timescales in the $M-L$ plane for Musca, L1495-B213 (Taurus) and Orion~A, compared to the full filament population in our sample (grey dots, see also Fig.~\ref{fig:hierarchy}). Dotted black lines mark values of constant $m$.
    %For each panel, f
    Fragmentation (and possibly longitudinal collapse) dominates over accretion below the solid colored line (equating Eqs.~\ref{eq:acc2} and~\ref{eq:frag2}).
    Coloured dashed lines show Eq.~\ref{eq:randomwalk} originating from the corresponding main filament (cross) towards associated sub-filaments (coloured circles), indicating a hierarchical structure.
    For Orion (bottom panel) we include two hierarchical levels:   
    from the Orion A cloud (dark blue square) to its parsec-scale filaments (dark blue crosses) and from the ISF (cyan crosses) to its fibers (cyan circles). %In case of
    %For the latter, we take the 
    $L$ and $M$ values for the ISF are from the ISF-extended region \citep[cyan cross;][]{Nagahama1998}, but 
    are similar to values for the ISF-alone filament \citep[][]
    {Bally1987}. 
    The lower panel also includes the mass-size relations (dotted/dashed black lines) previously reported for clumps (see \S\ref{sec:normalization}).
    }
    \label{fig:timescaleslm}
\end{figure}

We emphasise that $L_0$ and $M_0$ in Eq.~\ref{eq:randomwalk} are set by the conditions on the largest scale and can thus be different for different objects. This motivates Fig.~\ref{fig:timescaleslm}, where we plot the $M-L$ relation for separately Musca, Taurus, and Orion~A.
Turbulent effects in combination with fragmentation will cause filaments to evolve down along the colored, dashed lines, which represent hierarchical fragmentation originating at the %position of the harbouring structure
cloud-scale $M_0, L_0$ (Eq.~\ref{eq:randomwalk}). %In addition, since 
Because a filament might continue accreting or contracting longitudinally, its sub-filaments are not expected to exist above the dashed line. Indeed, for all three cases the sub-filaments sit below the corresponding line, suggesting that accretion does play a role.
For Orion~A, we additionally plot the observed dense fibers, suggesting an additional level of hierarchy.

Though the entire distribution (neglecting the \hi filaments) roughly follows a \mbox{$L \propto M^{1/2}$} relation, for a given $M$ there is a significant scatter in $L$ of about \mbox{0.5 - 1 dex}. A number of factors may contribute to this scatter.  
First of all, environmental effects for individual objects 
(Fig.~\ref{fig:timescaleslm}) seem to introduce a vertical scatter in the overall distribution.
Also inclination can account for a small part of the vertical scatter (\mbox{$\sim 0.1$ dex}), as can be derived from geometrical considerations, similar to the effect of blending of unresolved sub-filaments along the LOS \citep{Juvela2012,Moeckel2015}.
The identification algorithm (\S\ref{sec:algorithms}) can influence the determination of the length and filament radial extent \citep[e.g.][]{Schisano2014}, thus introducing scatter in both $M$ and $L$.
As longitudinal collapse would make the $M-L$ relation steeper (in contradiction to the observations), we speculate that this process is subdominant, but may contribute 
to the intrinsic scatter at a given $M$.
Accretion from the ambient medium can introduce a horizontal shift in $M$ for constant $L$.
Further study is needed to isolate the relative importance of each of these processes.

\subsubsection{The lower and upper envelopes of the $M-L$ distribution}\label{sec:ML_edges}

In a first attempt to understand the envelopes of the global $M-L$ distribution, we compare the distribution to equilibrium filaments. 
We replace the total velocity dispersion (Eq.~\ref{eq:sigmatot_definition}) 
used to determine $m_\textrm{vir}$ (Eq.~\ref{eq:Mcrit_tot}) with the relation inferred from our catalog
(Eq.~\ref{eq:sigma}), yielding:
\begin{equation}
 m_\textrm{vir} = \frac{M}{L} \simeq \frac{2 c_\textrm{s}^2}{G}\left(1 + \frac{L}{\textrm{0.5 pc}} \right) \, .
\label{eq:MLscaling}
\end{equation}
In Fig.~\ref{fig:Main_prop} we show this relation using \mbox{$T$ = 10 K} (red solid line).
It divides the $M-L$ plane into two regions. Above the line, a combination of thermal and non-thermal pressure dominates. The latter may be caused by turbulence or infall. Below the line, gravity dominates, leading to collapse and fragmentation (\S\ref{sec:timescales}). The line approximately traces the lower envelope of the filament distribution, suggesting that both pressure components together set the maximum achievable line mass.

The distinction into these two regions is not exact, for at least three reasons. First, decreasing the
scaling length of 0.5~pc in Eq.~\ref{eq:MLscaling} or increasing $T$ would shift the red curve slightly downwards/to the right, but would not affect the overall scaling behaviour. Second, also inclination effects might shift the curve somewhat downwards.
Third, Eq.~\ref{eq:MLscaling} could be modified to include the contribution from a radially isotropic magnetic pressure (e.g. by a field parallel to the filament's main axis) by adding the term $(v_\textrm{A}/c_\textrm{s})^2$ inside the bracket. By assuming a scaling relation of the magnetic field strength with density of \mbox{$B = 10$ $\mu$G $\cdot$ ($n$/300 cm$^{-3}$)$^{\beta}$} and the specific case of \mbox{$\beta = 0.5$} \citep{Tritsis2015}, $v_\textrm{A}$ becomes constant, i.e. independent of $n$.
This term acts to slightly shift the solid line towards lower $L$ for a given $M$. A more in-depth investigation of the magnetic contribution to this $M-L$ relation, however, would also have to take into account the effect of magnetic field geometry:
a perpendicular field orientation with respect to the filament and typically observed field strengths  (Fig.~\ref{fig:B})
would increase $m_\mathrm{crit}$ (Eq.~\ref{eq:mcrit_therm})
% and~\ref{eq:mvir_bfield}) 
by a factor or $\sim2-3$ only, compared to a factor of $\sim10$ for a parallel orientation \citep{Tomisaka2014,Seifried2015,Hanawa2017,Kashawagi2021}.

An upper limit for the distribution in the $M-L$ plane -- at least for molecular filaments -- could be caused by the requirement of a sufficient shielding against UV radiation to form molecular gas (here we assume CO). The column density of a filament available for shielding is \mbox{$N_\textrm{shield} \simeq (m/\mu m_\mathrm{p})/R_\textrm{flat}$}.
Assuming that the central density (required to calculate $R_\textrm{flat}$; Eq.~\ref{eq:rflat}) is 10 times the average density (Eq.~\ref{eq:dens_length}) yields the relation
\begin{equation}
  L \simeq 1.5 \left(\frac{M}{M_\odot}\right)^{0.65}\left(\frac{N_\textrm{shield}}{10^{21}\,\mathrm{cm}^{-2}}\right)^{-0.65}\,\mathrm{pc} \, .
  \label{eq:Nshield}
\end{equation}
We note that this represents only a rough estimate as also the material in the environment of the filament contributes to the shielding and as it is based on the assumption of a filament in hydrostatic equilibrium. Despite this simplification,
this relation \citep[red dashed line in Fig.~\ref{fig:Main_prop} using \mbox{$N_\textrm{shield}$ = 10$^{21}$ cm$^{-2}$}, \mbox{$A_\textrm{V} \simeq$ 1$^{mag}$;}][]{vanDishoeck1988} roughly matches with the upper envelope of the data distribution. This could indicate that the upper envelope of the distribution of (molecular) filaments is set by 
a shielding column density typically required for molecular gas to form. Consequently, this line also roughly separates the \hi filaments from the rest.

Overall, the solid and the dashed red lines in Fig.~\ref{fig:Main_prop} bracket the distribution of data points in our catalog.
The physical parameterization introduced in Eqs.~\ref{eq:MLscaling} and~\ref{eq:Nshield} could be used to predict the expected $M-L$ distribution of molecular filaments in different Galactic environments depending on  $\sigma_\textrm{nt}$, $B$, $c_\textrm{s}$, and $N_\textrm{shield}$.

\subsection{Timescales and the $M-L$ relation} 
\label{sec:timescales}

Timescales of dominant physical processes can identify characteristic domains in the $M-L$ plane. We consider three characteristic timescales, namely the accretion timescale (Eq.~\ref{eqn:acctime}), the (linear) fragmentation timescale (Eq.~\ref{eqn:longfrac}), and the timescale for longitudinal (edge) collapse (Eq.~\ref{eqn:tau_edge}). The latter two timescales are variants of the free-fall time where
$\tau_\mathrm{frag}$ is always shorter than $\tau_\mathrm{long}$ for aspect ratios $A > 1$. For this reason, in the following we focus on the accretion and fragmentation time scales, but note that, in particular for filaments with a low $A$ and hub-like filaments, longitudinal collapse might become important as well. To place the timescales in the $M-L$ plane, we use the density-length scaling (Eq.~\ref{eq:dens_length}) inferred from Fig.~\ref{fig:hierarchy} resulting in 
\begin{eqnarray}
  \tau_\textrm{acc}&=& \frac{M/L}{\dot{m}} \label{eq:acc2}\\%  = \frac{\left(\frac{M/L}{M_{\odot} \mathrm{pc}^{-1}}\right)}{\left(\frac{\dot{m}}{M_{\odot} \mathrm{pc}^{-1} \mathrm{Myr}}\right)} \, \mathrm{Myr} \,\mathrm{Myr} \label{eq:acc2}\\
  \tau_\textrm{frag}& \simeq & 0.5 \cdot \left(\frac{L}{\mathrm{pc}}\right)^{0.55} \,\mathrm{Myr}%\left(\frac{L}{\mathrm{pc}}\right)^{3/4}\,\mathrm{Myr}
  \label{eq:frag2}
\end{eqnarray}

These timescales delineate two
regions in the $M-L$ plane (Fig.~\ref{fig:timescaleslm}).
Both regions are dominated by one process, the upper by accretion, the lower by fragmentation (and in parts longitudinal collapse). However, none
of the processes act exclusively:
rather, all processes will compete, one having the shortest timescale.  
Each panel uses the observed accretion rates (14~M$_\odot$~pc$^{-1}$~Myr$^{-1}$ for Musca \citep{Bonne2020}; upper limit for Taurus at 50~M$_\odot$~pc$^{-1}$~Myr$^{-1}$ \citep{Palmeirim2013}; estimate for Orion at 300~M$_\odot$~pc$^{-1}$~Myr$^{-1}$). For all main filaments, accretion dominates, while sub-filaments straddle the boundary between accretion and fragmentation or longitudinal collapse. In general, for long, massive filaments, accretion tends to dominate -- longitudinal collapse will be present, but will not control filament evolution. 

Main and sub-filaments in the individual panels in Fig.~\ref{fig:timescaleslm} are bounded toward lower $L$ by the transition from accretion to fragmentation
(solid lines), and toward higher $L$ by their hierarchical structure (Eq.~\ref{eq:randomwalk}, dashed line).
We thus suggest the following \textit{evolutionary picture for filaments}:
    (i) Filaments and their hierarchical sub-structure(s) ``move'' diagonally downwards,
(ii) then accrete mass and move horizontally to the fragmentation-dominated regime and
(iii) finally form cores rapidly and get dispersed (\S\ref{sec:starformation}).
Therefore, the majority of filaments are expected lie above the solid line, which indeed seems to be the case for the three examples shown. Our results tentatively suggest that the distribution in the $M-L$ plane could be used to estimate the accretion rate in the system.
Comparisons with synthetic filament evolutionary tracks for different simulations in the $M-L$ plane could shed more light on the interpretation of the observed $M-L$ relation in clouds.

The narrowness of the filament distribution between the (dashed)  fragmentation line and the (solid) accretion limit in Fig.~\ref{fig:timescaleslm} may place a limit on filament evolution timescales.
\citet{RiveraIngrahametal2017} estimated timescales of $1-2$~Myr for nearby filaments in the Galactic Cold Core \textit{Herschel} sample, while \citet{Gong2021} find an age of $\sim 2$~Myr for the Serpens filament based on timescales for C$^{18}$O freeze-out. These timescales are consistent with the spatial correlation between gas and YSOs \citep{Hacar2017a}, a characteristic property of nearby (resolved) star-forming regions \citep{Hartmann2001}.

Short characteristic timescales are also expected for massive and compact HFS such as NGC1333 or Mon-R2 (see grey crosses in Fig.~\ref{fig:timescaleslm}). After a likely high accretion phase in earlier stages, their current position in the $M-L$ phase-space suggests that these filamentary clouds could undergo a global collapse in %timescales of
$\sim$~1~Myr (see Eq.~\ref{eqn:tau_edge}). This rapid evolution may explain large velocity gradients and high infall rates reported in these objects (see \S\ref{sec:hubs}). Additional analysis of different HFS are needed to confirm these conclusions.

\subsection{Normalization}\label{sec:normalization}

The origin of the mass-size relation (sometimes referred to as mass-radius relation for clumps and cores) has been widely discussed in the literature \citep[e.g.][]{Elmegreen1996}.
In addition to the slope variations (see \S\ref{sec:hierarchy}), the comparison of the mass-size relations obtained in resolved clouds reveals systematic offsets between regions \citep{Kauffmann2010a,Kauffmann2010b}.
The combination of hierarchical fragmentation (Eq.~\ref{eq:randomwalk}) and differing accretion rates (Eq.~\ref{eqn:acctime}) provides a potential interpretation of these different normalization values as seen in regions such as Musca, Taurus, and Orion (see \S\ref{sec:timescales}).
Fig.~\ref{fig:timescaleslm} (bottom panel) also shows the mass-size relation obtained for the most massive clumps in the solar neighborhood, \mbox{$M = 400 \, \mathrm{M}_\odot\cdot (R_\mathrm{eff}/\mathrm{pc})^{1.7}$} with $R_\mathrm{eff}$ corresponding to the object's effective radius, and where we assume $L=2R_{eff}$ \citep[black dashed line;][]{Kauffmann2010a}.
For these local regions our filamentary models predict a maximum accretion rate  of \mbox{ $\sim$ 300 M$_\odot$ pc$^{-1}$ Myr$^{-1}$} similar to Orion.

Our analysis could also shed some light on the origin of high-mass stars in our Galaxy. High-mass stars are found in regions with \mbox{$M>870 \, \mathrm{M}_\odot \cdot (R_\mathrm{eff}/\mathrm{pc})^{1.33}$}, implying a minimum mass limit for the formation of massive stars \citep{KauffmannPillai2010}. When displayed in Fig.~\ref{fig:timescaleslm} (black dotted line in the bottom panel), this threshold appears to separate most of the ``standard'' filaments from the more massive and compact HFS.  The location of HFS in the $M-L$ parameter space tentatively suggests that they form under high accretion rates with \mbox{ $\dot{m} >$ 500 M$_\odot$ pc$^{-1}$ Myr$^{-1}$}. This would reinforce the connection between the origin of high-mass stars and the dynamic evolution of massive filamentary hubs (see \S\ref{sec:hubs}). 

\subsection{Revisiting Larson's relations}\label{sec:larson}

The scaling relations derived in this work bear resemblance to the classical relations first described by Larson \citep{Larson1981,Solomon1987,Heyer2009}. We can examine the similarities and differences between the scaling relations, and further, re-interpret Larson's relations by taking into account the filamentary geometry. Importantly, Larson assumes spherical geometry. Cylindrical geometry naturally allows varying geometries via aspect ratio. Thus, different filament types or families may play a role in setting the scaling relations. 

The first similarity occurs for the scaling of our $\sigma_\mathrm{tot}-L$ relation of filaments (Eq.~\ref{eq:sigma}) and that
of galactic clouds \citep[first Larson's relation: $\sigma_{tot}\propto L^{0.5}$; see e.g.][]{Solomon1987}: both show the same power-law exponent. Secondly, the power-law exponent of our $n-L$ relation (Eq.~\ref{eq:dens_length}) is consistent with that of the third Larson's relation: $n \propto L^{-1.1}$.

Larson's $M \propto L^{1.9}$ relation \citep[contained in the data but not explicitly shown in][]{Larson1981} combined with a spherical geometry leads to the aforementioned $n-L$ relation, which suggests a constant column density. In contrast, for a filamentary picture, i.e. under the assumption of cylindrical geometry, our observed scaling of $M \propto L^2$ does not directly lead to $n \propto L^{-1.1}$. This is because, in addition to $M$ 
and $L$, the radius (FWHM), or alternatively the average column density, must be determined independently.
Furthermore, the column density of filaments in our catalog is not constant (indeed it covers almost three orders of magnitude in dynamical range) and is uncorrelated with $L$. It is thus remarkable that 
the classical $n-L$ relation appears to hold in the view of a filamentary ISM.

An analysis using the first ($\sigma-L$) and second ($\sigma-M$) Larson's relations in a spherical configuration is interpreted as a manifestation of energy equipartion independent of scale \citep[][]{Larson1981}. In the case of filaments the $\sigma-m$ relation (Eq.~\ref{eq:ml-sigma}) would imply energy balance in the radial direction only. Fragmentation could still occur in the longitudinal direction, and of course, energy equipartition is consistent with a range of scenarios from virial equilibrium to free-fall collapse.

Finally,  under  the  assumption  of  cylindrical  geometry, our observed $n-L$ and $M-L$ relation directly necessitate that the two geometrical scales of filaments are linked, i.e. that the filament width $FWHM \propto L^{\gamma}$ with $\gamma$ close to unity. Given the large dynamical range of $L$ (see Fig.~\ref{fig:Main_prop}) this is an independent indication that the width of filaments can in general not be constant.

A definition issue should be emphasised when considering filament widths over the wide dynamic ranges the scaling relations span. It is not clear how to reconcile the concept of width for a hierarchical structure. Current filament-definition approaches only consider filaments at one given scale (\S\ref{sec:algorithms}). For example, giant filaments whose widths have been measured in parsecs \citep[][]{Zucker2018}, harbor — or \emph{are composed of} — filaments whose widths are much smaller \citep[][]{Mattern2018Nessie}.  A better description and understanding of the filament hierarchy is necessary also for the interpretation of the scaling relations.

\subsection{Filament widths: intrinsic properties vs observational artefacts}
\label{sec:widths}

\begin{figure*}[h!]
    \centering
    \includegraphics[width=\textwidth]{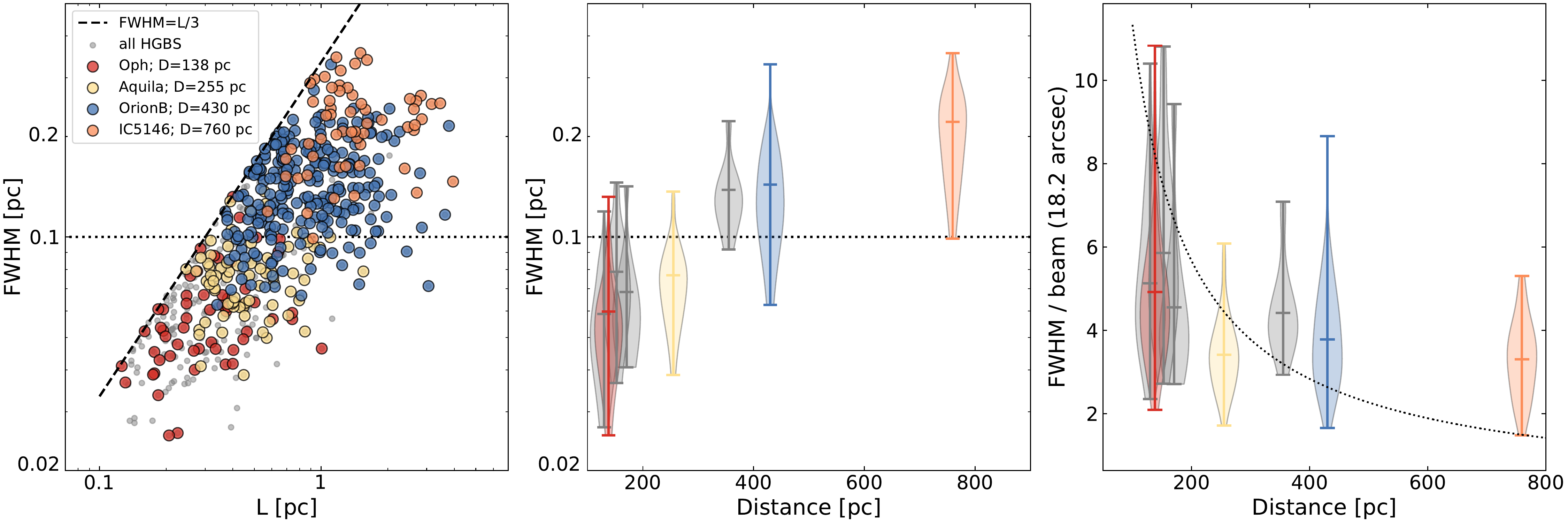}
    \caption{\small {\bf Left panel:} Observed filament widths, FWHM (in parsec), as function of the filament length, L, in 7 nearby clouds as seen by \textit{Herschel} \citep[from][]{Arzoumanian2019}. The data for Ophiuchus (red), Aquila (yellow), Orion B (blue), and IC5146 (magenta) are highlighted in all panels. The distances to the clouds have been updated based on the latest 3D dust extinction maps, see text and \citet{Panopoulou2021}.
    {\bf Middle panel:} Observed FWHM as a function of the cloud distance, summarised for the various clouds in the form of a violin plot. The plot includes the maximum, minimum and mean of each FWHM distribution. 
    {\bf Right panel:} Same as in the middle panel, but now the FWHM is in units of the beamsize (18.2 arcsec). In all panels, the value of 0.1~pc is shown by a dotted line.
    }
    \label{fig:widths}
\end{figure*}

One of the main findings from \textit{Herschel} is the narrow distribution of filament widths measured in nearby clouds with the \textit{Herschel} Gould Belt Survey. \cite{Arzoumanian2011} first found that the distribution of widths is centered around $\sim$0.1~pc, which is inconsistent with the Jeans length. Theoretical models have attempted to explain this observation \citep{Fischera2012a,Fischera2012b,HennebelleAndre2013,Federrath2016,Auddy2016}, \citep[ some have since been excluded][]{Hennebelle2013,Ntormousi2016} while none of the remaining models are applicable to the wide range of filament column densities in the sample of \citet{Arzoumanian2011}. On the other hand, an increasing number of filament widths significantly different from 0.1~pc are reported in a variety of regions with the use of different gas tracers and observational techniques (\S\ref{sec:radialprof}).
Here, we re-examine the understanding of filament widths, in light of existing measurements of filament properties. 
We briefly summarize three factors that may affect the filament width:
(a) physical processes, (b) observational tracers, (c) biases in the measurement method.

A variation of the filament width across environments and over time is theoretically expected due to various physical processes  \citep[e.g. accretion, gravitational collapse, ambipolar diffusion, varying magnetic field geometry, cosmic ray ionization rate, sonic Mach number, see][]{Heitsch2013a,Smith2014b,VanLoo2014,Seifried2015,Federrath2016,Ntormousi2016,Seifried2017}.
Some evidence for environmental dependence and time evolution has been found \citep{Hacar2018,Schmiedeke2021}. However, observations are hampered by biases from chemistry \citep{Tritsis2016b,Seifried2017,Priestley2020,Schuller2021}, radiative transfer \citep{Juvela2012,Howard2019}, resolution \citep{Schisano2014} and the exact method of measuring filament widths \citep[averaging, confusion, choice of fitting function and range, see][]{Malinen2012,Smith2014b,Panopoulou2017,Arzoumanian2019,Suri2019,Priestley2020}.
Complementary approaches have been developed to characterize filamentary structures, which do not suffer from the same biases as profile fitting
(e.g. power spectrum, wavelet analysis, see \S\ref{sec:algorithms} and \S\ref{sec:hierarchy}).

By examining data in our catalog we have found that filament width measurements are still affected by a more classical problem: the distance determination.
There is a correlation between the width and length of filaments in the data from \citet{Arzoumanian2019}. This
becomes even more apparent in the left panel of our Fig.~\ref{fig:widths}, when using more recent distance estimates \citep{Panopoulou2021}, which are based on 3D dust extinction \citep{Zucker2019,Zucker2020,Leike2020}.
This correlation is directly linked to a distance issue: filaments in more distant clouds have, on average, larger FWHMs (Fig.~\ref{fig:widths}, middle panel). 
We show the observed FWHM as opposed to the reported `deconvolved' FWHM by \citet{Arzoumanian2019} \citep[as `deconvolution' assuming a Gaussian profile does not correct for beam effects][]{Panopoulou2021}. The FWHM and $L$ have been rescaled to the new estimated distance  \citep[from][]{Panopoulou2021}.
The mean FWHM per cloud are equal to 3-5 times the \textit{Herschel}-SPIRE beam of 18.2\arcsec~\citep[the resolution of the $N$-maps used in][]{Arzoumanian2019}. As seen in Fig.~\ref{fig:widths} (right panel), a constant width of 0.1 pc (black, dotted line) does not describe the data well. The scaling of mean width with distance suggests that filament widths are resolution-dependent \citep{Panopoulou2021}.

We note that \citet{Arzoumanian2019} did investigate the effect of distance and concluded that the mean width (for the entire ensemble of clouds) is not significantly affected by the choice of distance. However, the trend is clear when we break up the data in separate clouds: the width of the entire filament distribution remains similar, but the mean width per cloud clearly increases with distance. 

We observe an almost linear scaling of FWHM $\propto ~L^{0.97}$ (fit to the data in Fig. \ref{fig:widths}, left), consistent with the expectation from \S\ref{sec:larson}. At the same time, a scaling of width with distance, similar to what we find, has also been observed for Hi-GAL filaments \citep{Schisano2014} and for \textit{Herschel} cores \citep{RiveraIngraham2016} and has been connected to resolution effects. 
Thus, disentangling intrinsic variation of filament widths from possible resolution biases is an issue that requires further scrutiny.

\subsection{Initial conditions for star formation}
\label{sec:starformation}

Star-forming cores are associated with filaments \citep{Schneider1979,Hartmann2002,Andre2010,Molinari2010}. For example, in Aquila 75\% of the \textit{Herschel} core population lies within gravitationally super-critical filaments \citep{Konyves2015}.
As discussed in \S\ref{sec:fragmentation}, parsec-size filaments can gravitationally fragment into dense star-forming cores. %For example, c
Cores identified on and off filaments may even have different properties. \citet{Polychroni2013} find that cores in Orion A located in filaments (71\% of total) have a different mass function and peak mass (4~M$_\odot$) compared to those outside filaments (0.8~M$_\odot$). \citet{Schisano2014} studied the properties of cores %on and off filaments 
observed in the Galactic Plane, and find that cores exceeding the theoretical thresholds for massive star formation \citep[$\sim$~1~g~cm$^{-3}$,][]{Krumholz2008} are only located on filaments, and not outside them. %Thus,
These results suggest that the properties of filaments are crucial in determining the properties of star-forming cores.

\subsubsection{Core Mass Function}
\label{sec:cmf}

In addition to the spacing of the fragments (\S\ref{sec:frag-spacing}), one can also consider the effect of filament fragmentation on the resulting mass distribution of cores. \citet{Inutsuka2001} used the Press-Schechter function to predict the Core Mass Function (CMF) that would arise from a turbulent power spectrum of fluctuations within a filamentary geometry over time, and showed that it evolves towards a dependence of $\frac{\mathrm{d}N}{\mathrm{d}M}\propto M^{-2.5}$ within approximately one free-fall time. More recently, steady-state models by \citet{Lee2017} examined the expected masses of fragments of the observed filament population, noting that a realistic CMF could only arise from a \textit{superposition} of multiple filaments with a characteristic $m$ distribution. Early observational works suggest a tentative mass dependence for the filament mass function (FMF) of \mbox{$\frac{\mathrm{d}N}{\mathrm{d}M}\propto M^{-2.1}$} for parsec-scale filaments \citep{Nagahama1998}.
Resolved filaments in nearby star-forming regions have a power-law like distribution of both the FMF, \mbox{$\frac{\mathrm{d}N}{\mathrm{d}M} \propto M^{-2.6}$}, and filament line mass function (FLMF), \mbox{$\frac{\mathrm{d}N}{\mathrm{d}m} \propto m^{-2.4}$}, respectively, which will naturally lead to a power-law distribution of core masses \citep{Andre2019}. 
The first complete surveys of entire filament populations inside clouds \citep[e.g.][]{Orkisz2019} appear to be a promising avenue for investigating the connection between the filament and core mass distributions.

While Jeans-like cores are likely formed in fragmenting filaments, the origin of more massive supra-Jeans cores inside filaments is less clear. Massive cores are observed within highly super-critical parsec-scale filaments \citep[e.g.][]{Shimajiri2019,Konyves2020}. This could simply be due to the fact that cores of similar size in a higher-column density medium will be more massive, or it could be due to the most massive cores being preferentially at junctions (\S\ref{sec:hubs} and references therein) where cores may merge or grow by accretion (\S\ref{sec:dyn-acc}). Additional high-resolution (e.g. ALMA) observations of massive filaments and hubs are needed to explore these different scenarios \citep[e.g.][]{Hacar2018,Louvet2021,Motte2021}.

\subsubsection{Environmental variations}

Although resulting in the same mass distribution, the initial conditions for the formation of cores inside filaments may depend on the environment. According to Fig.~\ref{fig:timescaleslm}, all filamentary regions roughly follow the same \mbox{$L \propto M^{0.5}$} dependence, but with a different normalization  (\S\ref{sec:origin-ml}). 
Shorter filaments are seen in regions with higher accretion rates (such as Orion) in comparison to more quiescent clouds (such as Taurus). This effect is likely produced by the enhanced fragmentation of filaments with higher accretion rates. 
This trend could extend into more massive ridges, hubs, and filament networks (\S\ref{sec:hubs}), suggesting that filament properties may vary with local accretion rates.

We speculate that some of these environmental variations may also determine the densities and rates in which cores and stars originate inside filaments.
As a result of a vertical displacement in the $M-L$ plane (shortening $L$), filaments with the same mass are expected to exhibit higher central densities $n_0$ in regions with higher accretion rates (see also Fig.~\ref{fig:hierarchy}, right panel).
Higher accretion rates would thus translate into shorter dynamical timescales $\tau_\textrm{long}$ (Eq.~\ref{eqn:tau_edge}) and $\tau_\textrm{frag}$ (Eq.~\ref{eqn:longfrac}), as well as shorter fragmentation spacing $\lambda_\textrm{crit}$ (Eq.~\ref{eq:lambdacrit}). Also shortening the free-fall timescales once a core is formed ($\tau_\textrm{ff}\propto n_0^{-1/2}$), stars
inside filaments showing high accretion rates would form at higher star formation rates \mbox{($SFR \propto 1/\tau$)} and surface densities \mbox{($\Sigma_{SF} \propto 1/\lambda_\textrm{crit}$)}. This scenario may explain the origin of stellar clusters in dense fiber networks \citep{Hacar2018}. Detailed measurements of the local accretion rates onto filaments and their comparison with their core and stellar populations are needed to further explore this hypothesis. 
If confirmed, the environmental properties of the filamentary ISM would uniquely shape the initial conditions for the formation of cores, disks, and stars explored in other chapters of this series.

\section{CONCLUSIONS \& OUTLOOK}\label{sec:conclusions}

\subsection{Conclusions}

We have reviewed the main advances in studying filamentary structures in the ISM, focusing on the progress since Protostars and Planets VI. 
Bringing together the latest data provided by the community has enabled us to take a major step forward: 
we present a comprehensive census and meta-analysis of 22,704 filamentary structures gathered from 41 major studies in the literature. This facilitates a new, holistic view on filaments in the ISM, ultimately leading to new insights on the observed properties of filaments.

New observational studies have identified a plethora of filamentary structures over a wide range of conditions and scales in the ISM (\S\ref{sec:surveys}, \S\ref{sec:census}). These studies range from detailed observations of individual filaments at high resolution to large, Galaxy-scale surveys. Overall, these works have given rise to “filament families”, (e.g., fibers, nearby filaments, and giant filaments). These families are primarily operationally defined and any physical distinction between them remains largely unknown. Each of the families is subject to distinct selection effects and biases, hampering direct comparisons between some studies. 
We review the basic properties of the different families, including their mass distribution, internal kinematics, and magnetic field structure.
Despite the heterogeneity of the data, some distinctions and trends seem to prevail, allowing us to build physical insight across the spectrum of observed structures.

The formation of filaments remains a crucial open question. While progress has been made in proposing and describing various scenarios, observational constraints for such scenarios remain few.
Anisotropic mechanisms are particularly important for filament formation. 
Magnetic fields may play a key role in filament formation, and during the stages in which the gas is atomic and/or relatively diffuse. The role of magnetic fields during filament fragmentation and collapse seems less significant. 

The basic filament properties give rise to scaling relations that span wide ranges of filament lengths, masses, densities, and kinematic properties. The most striking of these is the $M-L$ relation that shows how filaments %follow 
are well-described by an $L \propto M^{0.5}$ scaling over eight orders of magnitude in mass. This relation, however, must immediately be accompanied by the notion that filaments show \emph{hierarchy}. A tens-of-parsec-scale filamentary cloud can harbour parsec-scale filaments within, and those may further break down into smaller and smaller filaments. Crucially, our meta-analysis shows that this hierarchy \emph{within} clouds follows the $L \propto M^{0.5}$ scaling, with the intercept of the relation set by the mass and length at the largest scales (Figs. \ref{fig:hierarchy} and \ref{fig:timescaleslm}). Understanding the hierarchical nature of these filamentary networks may provide a strong constraint for cloud and filament formation and evolution scenarios.

Particularly interestingly, our meta-analysis points out 
that the measured filament widths depend on the distance from the Sun, especially when we use the latest, arguably most accurate distances for the clouds. Determining whether this dependence reflects resolution effects, physical differences, or both, requires further study.

A large number of recent observational and numerical works -- and our meta-analysis -- promote a highly dynamical, non-idealized picture of filament evolution. At the heart of this dynamical picture lies the fact that filaments do not form, nor live, in isolation. 
Accretion from the surrounding media onto filaments seems inevitable and significant. This is likely to have a strong effect on filament evolution due to the short accretion timescales. Similarly, accretion \emph{along} the filament seems common, making filament dynamics complex. This complexity increases at length-scales smaller than roughly 1~pc, as best evidenced by increased plane-of-the-sky velocity gradients below that scale. However, there is also a transition to coherence below roughly 1~pc, as measured by the 
trans-sonic velocity dispersion below that scale. Crucially, this suggests that the transition to coherence occurs \emph{before} star-forming cores are formed, and that cores inherit their sub-sonic nature from the parental filament. Clearly, these kinematic complications together severely hamper the applicability of evolution (fragmentation) analyses arising from hydrostatic models. Indeed, observations of fragment spacing rarely align well with such models, showing more complex patterns.

The above hierarchical and dynamical nature of filamentary structures leads us to propose a new scenario for reconciling the observed mass-length scaling of filaments (\S\ref{sec:m-l} and \S\ref{sec:timescales}). We propose that the observed hierarchy of filamentary structures can be understood 
in terms of turbulent fragmentation in which 
one filament fragments into further filaments. This leads to a $L \propto M^{0.5}$ scaling between the %fiducial
large-scale filament and its sub-filaments and gives a new perspective on the observed $M-L$ relation of \emph{all} molecular filaments: it originates in effect from the \emph{observable lifetime} of a filament. At one end, the lifetime is limited by the formation of molecules and thus the appearance of the structure as a molecular filament. At the other end, the lifetime is limited by the radial and longitudinal collapse and/or fragmentation. These limits 
bound the space in the $M-L$ plane in which the observable molecular filaments gather: the hierarchy necessitates the $L \propto M^{0.5}$ scaling of individual filaments, and the lifetime considerations limit the intercepts to the observed range. As a result, the observed $M-L$ relation emerges. Further, comparing and contrasting the scaling relations of $L$, $M$, $N$, $R$ and $\sigma$ with Larson's relations open a door to  new physical interpretations (\S\ref{sec:larson}). Finally, this scenario has an important consequence: the location of the filaments in the $M-L$ plane depends on the accretion rate, and hence, it should be possible to \emph{predict} the accretion rate of the filament based on its observed substructure. This prediction could be tested by determining the internal hierarchy of more filaments (as in Fig.~\ref{fig:timescaleslm}) and --  independently of that -- their accretion rates.

%---------------------------------
\subsection{Future directions}
%---------------------------------

The above results give rise to some key development points for the future, which in equal measure apply to observational and theoretical
studies. First, our understanding of the \textit{hierarchical nature} of filamentary structures is limited. 
Our current observational descriptions of the filament hierarchy are largely incomplete, as are theoretical frameworks necessary for its interpretation. This 
hampers our understanding of the formation and early evolution of filaments.

The hierarchical nature of filaments also calls for rethinking the concept of ``environment” of filamentary structures: in some cases, filaments at larger scales provide the environment of the filaments at smaller scales. This hierarchy is connected via gas accretion, as the structures at small scales accrete from their parental structures.
Only at their largest scales do filaments blend into what can be considered an ambient environment. Thus, more studies focusing on the density, temperature, kinematics, and magnetic fields of the large-scale or diffuse gas surrounding filamentary networks are crucial. Similarly, studies focusing on the early, possibly starless, phase of filament evolution are key for understanding the build-up of the filament hierarchy. It is unclear if the properties of evolved, star-forming filaments carry observable information on the filament formation process. From the theoretical perspective, the study of evolutionary tracks of simulated filaments in a phase-space such as the $M-L$ plane (\S\ref{sec:timescales}) appears to be one key to decoding the information embedded in the filament hierarchy. 

Finally, our ability to study filament hierarchy is hampered by the inability of the commonly-used analysis techniques to consider filaments as hierarchical or multi-scale objects (\S\ref{sec:algorithms}). This also applies to numerical simulations that need to span wider ranges of physical scales to capture this hierarchy, e.g. by dedicated zoom-in studies. Most existing tools restrict us to consider filaments (only) as isolated, singular objects. Progress is needed in developing more tools to describe filamentary networks -- or more generally the structure of the media -- in a manner that enables some description of its hierarchy. 
We encourage broader recognition of the inevitably biased nature of \emph{all} detection algorithms, as concretely evidenced by the heterogeneity of the filament data in the literature (\S\ref{sec:surveys}).   

We identified filament formation as a chronically open, crucial question. Theoretical works have firmly established that filamentary networks can form via a variety of processes (\S\ref{sec:formation}), but it remains unclear what properties of that network are meaningful diagnostics of the different formation processes. What are the fingerprints of the filament formation scenarios? How do they manifest themselves in observable quantities? These key questions need to be more systematically addressed using numerical experiments to make progress.

Another key future direction emerges from the fact that our detailed knowledge of internal filament properties is still limited to the Galactic environment of the Sun. While Galactic plane surveys have described some bulk properties of filaments in larger sections of the Galaxy, their resolution does not enable the study of the internal structure of the detected filaments. Further -- arising again from the hierarchical nature of filaments -- it is unclear how the properties of structures identified at different distances should be reconciled together. As a result, the key question about the dependence of filament properties on the Galactic environment remains open. There is ample evidence, both Galactic and extragalactic, that molecular cloud properties depend on the Galactic environment -- does this dependence also extend to the cloud substructure, i.e., the filaments?

Being limited to the nearby Galactic environment translates to being limited to low-mass star formation (with some well-known exceptions like Orion, NGC6443, and DR21). How do filamentary structures manifest themselves in clustered, high-mass star-forming regions and what is their role for star formation in such objects? Are filamentary structures the dominant morphology throughout the spectrum of star formation? Further, which of the end-products of star formation (stellar masses, SFR, or clustering) depend on the properties of the parental filamentary network remains a major open question in the field that can only be addressed by more comprehensive observations. 
Following the existing Galaxy-wide surveys in continuum (e.g. {\it Herschel}), a similar community effort is needed to statistically investigate filaments' internal kinematics and magnetic field structure at different scales. Indeed, new surveys taking steps to that direction are in the making, building a base for statistical studies in the future (\S\ref{sec:surveyfilaments}, \S\ref{sec:hi}).

Filaments have been one central topic of the ISM and star formation studies over the past decade. Their study has focused on identifying filaments -- \emph{extracting} them from the surrounding media -- and analysing their intrinsic properties. But despite these fruitful works, perhaps not all important physics are encapsulated in the detailed properties of isolated filaments. Our review and analysis proposes an intriguing, alternative perspective: filaments can be seen as an inevitable byproduct of the physics involved in the holistic, multi-scale process of gas accumulation, collapse, and recycling. Filaments are unavoidable, “transitory” morphological features of the gas on its way to stars. Adopting this difference in perspective immediately elevates the role of the environment and evolution, and properties such as accretion or multi-scale mass flows, as the central concepts of filamentary networks. To date, we have gained little knowledge on these; refocusing on them, especially using the emerging generation of surveys, may be one trigger of fundamental advances during the coming decade. 

%%\newpage
\bigskip
\noindent\textbf{Acknowledgments}
The authors thank the anonymous referees for their thorough read and constructive comments..
The authors thank D. Arzoumanian, E. Schisano, M. E. Putman, and E. Y. Chung for kindly sharing their data for this analysis as well as S. Zellmann for support for Fig.~\ref{fig:simulations}. 
This project has received funding from the European Research Council (ERC) under the European Union’s Horizon 2020 research and innovation programme (Grant agreement Nos. 851435 and 639459).
SEC acknowledges support by the National Science Foundation under Grant No. 2106607. 
GVP acknowledges support by NASA through the NASA Hubble Fellowship grant No. \#HST-HF2-51444.001-A  awarded  by  the  Space Telescope Science  Institute,  which  is  operated  by  the Association of Universities for Research in Astronomy, Incorporated, under NASA contract NAS5-26555.
DS acknowledges funding by the Bonn-Cologne Graduate School, which is funded through the German Excellence Initiative, as well as by the Deutsche Forschungsgemeinschaft (DFG) via the Collaborative Research Center SFB 956 “Conditions and Impact of Star Formation” (subproject C6).
RJS gratefully acknowledges an STFC Ernest Rutherford fellowship (grant ST/N00485X/1). 
This research made use of Astropy (http://www.astropy.org), a community-developed core Python package for Astronomy \citep{astropy2013,astropy2018}. The present study has made use of data from the Herschel
Gould Belt survey (HGBS) project (\url{http://gouldbelt-herschel.cea.fr}).
%\bigskip
This publication utilizes data from Galactic ALFA \hi (GALFA-\HI) survey data set obtained with the Arecibo L-band Feed Array (ALFA) on the Arecibo 305m telescope. The Arecibo Observatory is operated by SRI International under a cooperative agreement with the National Science Foundation (AST-1100968), and in alliance with Ana G. M\'endez-Universidad Metropolitana, and the Universities Space Research Association. The GALFA-\hi surveys have been funded by the NSF through grants to Columbia University, the University of Wisconsin, and the University of California.
The Parkes Radio Telescope is part of the Australia Telescope National Facility which is funded by the Australian Government for operation as a National Facility managed by CSIRO. The EBHIS data are based on observations performed with the 100-m telescope of the MPIfR at Effelsberg. EBHIS was funded by the Deutsche Forschungsgemeinschaft (DFG) under the grants KE757/7-1 to 7-3.
This publication utilizes observations obtained with Planck (http://www.esa.int/Planck), an ESA science mission with instruments and contributions directly funded by ESA Member States, NASA, and Canada.

%%%%%%%%%%%%%%%%%%%%%%%%%%%%%%%%%%%%%%%%%%%%%%%%%%%%
%\noindent\textbf{REFERENCES}
%\bigskip
\parskip=0pt
{\small
\baselineskip=11pt
\bibliographystyle{pp7.bst}
\typeout{}
\bibliography{PP7_filaments.bib}
%PP7_filaments.bib,fh_refs.bib}
}

\end{document}